\documentclass[prd,jmp,reprint,onecolumn,superscriptaddress,12pt]{revtex4-2}

\usepackage{bm}
\usepackage{amsfonts}
\usepackage{latexsym}
\usepackage[latin1]{inputenc}
\usepackage{graphicx}
\usepackage{amsmath}
\usepackage{subfigure}
\usepackage[mathscr]{eucal}
\usepackage{enumerate}
\usepackage{palatino}
\usepackage{mathpazo}
\usepackage{textcomp}
\usepackage{float}
\usepackage{booktabs}
\usepackage{dcolumn}
\usepackage{ragged2e}
\usepackage{hyperref}
\hypersetup{colorlinks,citecolor=blue}
\hypersetup{colorlinks=true,linkcolor=red,filecolor=magenta,    urlcolor=blue}
\usepackage{amsmath}
\usepackage{xcolor}
\usepackage{commath}
\usepackage{comment}

\begin{document}

\title{Observing Black Hole Phase Transitions in Extended Phase Space and Holographic Thermodynamics Approaches from Optical Features}

\author{Chatchai Promsiri} \email{chatchaipromsiri@gmail.com} 
\affiliation{Quantum Computing and Information Research Centre (QX), Faculty of Science, King Mongkut's University of Technology Thonburi,  Bangkok 10140, Thailand}

\author{Weerawit Horinouchi} 
\email{wee.hori@gmail.com}
\affiliation{High Energy Physics Research Unit, Department of Physics, Faculty of Science, Chulalongkorn University, Bangkok 10330, Thailand}

\author{Ekapong Hirunsirisawat}
\email{ ekapong.hir@kmutt.ac.th}
\affiliation{Quantum Computing and Information Research Centre (QX), Faculty of Science, King Mongkut's University of Technology Thonburi, Bangkok 10140, Thailand}
\affiliation{Learning Institute, King Mongkut's University of Technology Thonburi, Bangkok 10140, Thailand}

\begin{abstract}

The phase transitions of charged Anti-de Sitter (AdS) black holes are characterized by studying null geodesics in the vicinity of the critical curve of photon trajectories around black holes as well as their optical appearance as the black hole images.  In the present work, the critical parameters including the orbital half-period \(\tau\), the angular Lyapunov exponent \(\lambda_L\), and the temporal Lyapunov exponent \(\gamma_L\) are employed to characterize black hole phase transitions within both the extended phase space and holographic thermodynamics frameworks. Under certain conditions, we observe multi-valued function behaviors of these parameters as functions of bulk pressure and temperature in the respective approaches. We propose that \(\tau\), \(\lambda_L\), and \(\gamma_L\) can serve as order parameters due to their discontinuous changes at first-order phase transitions. To validate this, we provide detailed analytical calculations demonstrating that these optical critical parameters follow scaling behavior near the critical phase transition point. Notably, the critical exponents for these parameters are found to be \(1/2\), consistent with those of the van der Waals fluid. Our findings suggest that static and distant observers can study black hole thermodynamics by analyzing the images of regions around the black holes. 

\end{abstract}

\maketitle

\section{Introduction} \label{Intro}

Thermodynamics of black holes (BHs) has emerged as an active research area that offers deep insights into the interplay among gravity, thermodynamics, quantum mechanics, and information theory.
Initiated by Bekenstein \cite{PhysRevD.7.2333} and Hawking \cite{Hawking:1975vcx}, the energy $E$, entropy $S$ and temperature $T$ of a BH can be written in the geometric description as follows
\begin{eqnarray}
    E=M, \ \ \ S=\frac{A}{4G_N}, \ \ \ T=\frac{\kappa}{2\pi},
\end{eqnarray}
where $M, A$, and $\kappa$ are the mass of BH, the surface area of event horizon, and surface gravity at the event horizon, respectively.
With such geometric identification,
the laws of BH mechanics~\cite{Bardeen:1973gs} and the laws of thermodynamics exhibit a surprising
mathematical analogy, leading us to consider BHs as thermal objects.


The discussion of BH thermodynamics has been significantly enriched by the discovery of holographic descriptions. Given that the symmetry group of $\text{AdS}_{n+1}$ and the conformal group in $n$-dimensional spacetime are both isomorphic to $\text{SO}(n, 2)$, the AdS/CFT correspondence emerged as a duality between the gravitational system in the bulk AdS space and the conformal field theory (CFT) on the boundary~\cite{Maldacena:1997re, Witten:1998qj, Witten:1998zw, Gubser:1998bc}. Specifically, thermal radiation and BHs within an $(n+1)$-dimensional AdS space respectively correspond to the $n$-dimensional confining phase at zero temperature and thermal states within the deconfining phase of the large-$N$ gauge theory. Consequently, the confinement-deconfinement phase transition in gauge theory has the gravitational description in the bulk through the Hawking-Page phase transition of AdS-BHs~\cite{Hawking:1982dh}.


Despite significant advancements, several issues in BH thermodynamics remain unresolved. The first issue involves the Smarr formula~\cite{PhysRevLett.30.71}, which expresses the relation
between different geometric quantities of a BH in the same fashion as the Euler equation in conventional thermodynamics. However, there is an additional term in the Smarr formula
when considering BHs with a nonzero cosmological constant $\Lambda$. The Smarr formula for AdS-BH is given by:
\begin{eqnarray}
    M=\frac{n-1}{n-2}TS+\frac{n-1}{n-2}\Omega J+\Phi Q-\frac{1}{n-2}\frac{\Theta \Lambda}{4\pi G_N},\label{Smarr 1}
\end{eqnarray}
where $\Omega$ is the angular velocity, $J$ is the angular momentum, $\Phi$ is the electric potential and $Q$ is the electric charge of BH.
Note that $\Theta$ has been suggested to be defined as the proper volume weighted locally by a Killing vector~\cite{Jacobson:2018ahi}. 
Despite this formulation, there continues to be significant debate over the appropriate thermodynamic interpretation of $\Lambda$ and its conjugate variable $\Theta$.
The second issue arises from the observation that the first law of BH thermodynamics does not include the $pd\mathcal{V}$ term. As a result, one may wonder whether the holographic description can provide a gravitational description of the mechanical work term dual to that of the conformal matter.

Remarkably, there are two recent progresses that can be applied to resolve these important issues as follows.
\begin{itemize}
\item \textit{Black hole chemistry} or \textit{extended phase space approach} is an extension of thermodynamic phase space for AdS-BHs by identifying the negative cosmological constant $\Lambda$ and its conjugate variable as a \textit{bulk pressure} $P$ and \textit{thermodynamic volume} $V$ via
\begin{eqnarray}
    P=-\frac{\Lambda}{8\pi G_N} \ \ \ \text{and} \ \ \ V=-\Theta, \label{extended phase}
\end{eqnarray}
where $V=\frac{4}{3}\pi r_+^3$ for spherical symmetric BHs \cite{Kastor:2009wy,Dolan:2011xt,Dolan:2010ha,Cvetic:2010jb}, for good reviews see \cite{kubizvnak2017black,2014Galax...2...89A} and reference therein.
In this way, the first law of BH thermodynamics has $V dP$ additional work term, so the BH's mass should be identified as an \textit{enthalpy} rather than the internal energy of the BH.
Notably, the extended phase space approach gives a more consistent gravitational analog of the first law of thermodynamics in conventional matter. 
For example, the Hawking-Page phase transition in Schwarzschild AdS can be interpreted as solid-liquid phase transition \cite{Kubiznak:2014zwa}, isotherm curves in $P-V$ plane of charged AdS-BH in canonical ensemble undergoes a family of Small-Large BHs first order phase transition ending at a critical point of second order phase transition corresponds to the liquid-gas phase transition in the van der Waals (vdW) fluid \cite{kubizvnak2012p,Xu:2015rfa,2017PhLB..773..203M}
and the Small-Large-Small BHs phase transition occurs in rotating and nonlinear electrodynamics AdS-BHs analogous to the reentrant phase transition of multicomponent liquids \cite{2013PhRvD..88j1502A,2012JHEP...11..110G}.
Moreover, this framework introduces what can be referred to as the black hole's molecules for black hole thermodynamics~\cite{PhysRevLett.115.111302,2019PhRvL.123g1103W,Wei:2019yvs}. These entities can serve similarly to the microscopic constituents in statistical mechanics, providing a means to describe thermodynamic behavior.
\item {\it Holographic thermodynamics} has been proposed to address concerns about the consistency of the extended phase space approach with the AdS/CFT correspondence. Although the extended phase space approach provides a plausible interpretation of $\Lambda$ and its conjugate $\Theta$ for describing the thermodynamics of AdS-BHs, its consistency with the AdS/CFT correspondence remains debated by some researchers~\cite{johnson2014holographic,dolan2014bose,caceres2015holographic,kastor2014chemical,karch2015holographic}. 
Before discussing progress in holographic thermodynamics aimed at resolving these consistency issues, let us first clarify two important points within the extended phase space approach that continue to be questioned. They are as follows:
\begin{enumerate}[(i)]
    \item The equation of state for conformal fluid with energy $E$, pressure $p$ and volume $\mathcal{V}$ is described by 
    \begin{eqnarray}
    E=(n-1)p\mathcal{V}. \label{eos CFT}
    \end{eqnarray}
    According to the AdS/CFT dictionary, $E$ of $n$-dimensional conformal fluid is holographically dual to the mass $M$ of BH in $n+1$-dimensional spacetime. 
    It is evident that upon replacing $p$ and $\mathcal{V}$ in the above equation with $P$ and $V$ in Eq.~\eqref{extended phase} respectively, the result indicates that $E$ is not equal to $M$ of the BH. This implies that the bulk pressure $P$ and the volume $V$ for BH are not equivalent to the fluid's pressure $p$ and volume $\mathcal{V}$. In other words, the Smarr relation of BHs in the bulk does not correspond to the Euler equation of large-$N$ gauge theories on the boundary. 
    \item In fact, the AdS/CFT correspondence suggests that $\Lambda$ should be dual to the number of colors $N_c$ in the large-$N_c$ gauge theory via the relation \cite{karch2015holographic}
    \begin{eqnarray}
    k\frac{L^{n-1}}{16\pi G_N}=N_c^2 = \mathcal{C}, \label{dictionary}
    \end{eqnarray}
    where $\mathcal{C}$ is the central charge and $k$ is the constant depending on the details of a particular system.
    These relations suggest that the variation of $L$ implies a variation of $N_c$, which is the rank of the gauge group.
    Therefore, varying $\Lambda$ of the bulk leads to changing from one field theory to another.
    Moreover, the variation of $L$ also corresponds to variation in the volume $\mathcal{V}$ of the dual gauge theory since
    the geometry of dual field theory depends on $L$ of the bulk.
\end{enumerate}
Visser~\cite{Visser:2021eqk} proposed an approach to overcome this issue, demonstrating that the Euler equation of the dual field theory requires the $\mu_\mathcal{C}\mathcal{C}$ term but lacks the \(p\mathcal{V}\) term, while the \(pd\mathcal{V}\) term appears in the first law of thermodynamics. 
Here, $\mathcal{C}$ and $\mu_\mathcal{C}$ are the central charge and its chemical potential, respectively.
On the gravity side, introducing \(\mathcal{C}\) and \(\mu_\mathcal{C}\) as thermodynamic variables in the boundary theory is equivalent to allowing for variations in both the Newton constant \(G_N\) and the AdS radius \(L\) in the bulk theory.  Using Eq.~\eqref{dictionary}, it becomes possible to vary both \(L\) and \(G_N\) while keeping \(\mathcal{C}\) fixed at the boundary, ensuring the field theory remains unchanged to another. 
Thus, one can more appropriately study the variations in the volume \(\mathcal{V}\) of the dual field theory as \(L\) changes. As suggested in \cite{Visser:2021eqk}, the terms \(\Lambda\) and \(\Theta\) in Eq.~\eqref{Smarr 1} can be rewritten in terms of \(\mathcal{C}\) and \(\mu_\mathcal{C}\). This leads to consistency between the Smarr formula and the Euler equation within the framework known as holographic thermodynamics.
Recently, the thermodynamic behaviors resulting from holographic thermodynamics have been extensively studied \cite{Cong:2021jgb,Qu:2022nrt,Alfaia:2021cnk,Bai:2022vmx,2023JHEP...06..105G,2023JHEP...08..142A,2023JHEP...06..115Z}. For a comprehensive recent review, see \cite{Mann:2024sru}.
\end{itemize}

While BH thermodynamics has been developed with intriguing arguments as discussed above, determining the most valid approach remains challenging. It is crucial to support theoretical claims with observations or at least establish a connection between them.  It makes sense to argue that the thermal properties of BHs should manifest in specific observational signatures.  A natural question that arises is how we can obtain proper observational signatures of a BH to identify its thermodynamic properties and phase transitions.
Unfortunately, the observational confirmation of BH thermodynamics remains difficult, although recent advancements that have provided evidence of BH existence through the gravitational wave (GW) signal emitted by a binary black hole (BBH) \cite{LIGOScientific:2016aoc} and the images of two supermassive BHs, i.e., M87* and SgrA* \cite{ETH1,*ETH2,*ETH3,*ETH4,*ETH5,*ETH6,*EventHorizonTelescope:2022wkp, *EventHorizonTelescope:2022apq, *EventHorizonTelescope:2022wok, *EventHorizonTelescope:2022exc, *EventHorizonTelescope:2022urf, *EventHorizonTelescope:2022xqj}.
More specifically, the thermodynamic quantities of BHs are typically defined via their event horizons, which are difficult to detect directly from observers at asymptotic infinity.

Although the AdS-BH seems to be interested only in the theoretical aspect, it is valuable to find ways to connect its thermodynamic behaviors with observational signatures.  Numerous studies on AdS-BH have attempted to relate the BH phase transition to some signals, which can be observed at asymptotic infinity. 
Namely, the quasinormal modes (QNMs), which can be characterized by the signature in the GWs emitted from a BH during the ringdown stage~\cite{He:2010zb,Liu:2014gvf,Chabab:2016cem,Zou:2017juz} and null geodesics of test particles moving near the BH \cite{Wei:2017mwc,Zhang:2019tzi,Zhang:2019glo,Belhaj:2020nqy,Du:2022quq}.
Recently, the geodesic instability of test particles, specifically focusing on the temporal-Lyapunov exponents of both massless and massive particles, serving as an order parameter to investigate phase transitions in numerous BH solutions in asymptotically AdS spacetime \cite{2022JHEP...08..153G,yang2023lyapunov,lyu2023probing,kumara2024lyapunov,hale2024optical}.  


Strong gravitation near a BH can cause photons traveling close to the critical curve to orbit the BH multiple times before reaching a distant observer, 
forming a narrow band on the observer's screen known as the photon ring. This ring can be characterized by three critical parameters: the orbital half-period ($\tau$), the angular-Lyapunov exponent ($\lambda_L$), and the temporal-Lyapunov exponent ($\gamma_L$), 
which govern the dynamics of unstable null geodesics and reveal universal properties of the 
photon ring independent of the
distance between the light source and observer from the BH~\cite{Johnson:2019ljv,Gralla:2019drh} as well as the feature of the emitting light source. In this study, 
we focus on these observable parameters to decode information about the phase transitions 
of BHs, particularly charged AdS-BHs, within the framework of extended 
phase space and holographic thermodynamics. Moreover, we propose that the differences in these three critical 
parameters at first-order phase transitions could serve as an order parameter for studying 
scaling behavior near the critical point, and we also provide a mathematical derivation for 
the corresponding scaling law.

This paper is organized as follows: In section~\ref{section 2}, we provide a comprehensive review of BH thermodynamics, focusing on three approaches: standard phase space, extended phase space, and holographic thermodynamics. This section sets the stage for our analysis by detailing the theoretical foundations and the distinctions between these approaches.
In section~\ref{section 3}, we introduce three critical parameters of the photon ring region. We discuss the orbital half-period \(\tau\), the angular Lyapunov exponent \(\lambda_L\), and the temporal Lyapunov exponent \(\gamma_L\), explaining their significance in probing BH phase transitions with horizon-scale observations.
Section~\ref{section 4} examines the phase transitions of BHs through optical features in the extended phase space and holographic thermodynamics approaches. We analyze how the critical parameters vary with changes in the pressure and temperature, and demonstrate their potential as order parameters indicating phase transitions. 
In section~\ref{sec 5_orderparameter}, we investigate the scaling behavior of the optical parameters near the critical point within both the extended phase space and holographic thermodynamics approaches. We explore the critical exponents and their implications for understanding BH phase transitions. The results of this investigation support the use of these optical critical parameters as order parameters for characterizing BH phase transitions. 
Finally, section~\ref{sec 6_conclusion} concludes the paper by summarizing our findings and discussing the broader implications of our study for BH thermodynamics and observational astrophysics.

\section{Many facets of black hole thermodynamics} \label{section 2}

In this section, we provide a comprehensive review of the thermodynamics of charged AdS-BHs using three approaches: standard phase space, extended phase space, and holographic thermodynamics. The insights gained here will inform our analysis of null geodesics near the critical curve, which may reflect the phase transitions of black holes within each thermodynamic framework discussed in section~\ref{section 4}.

\subsection{Thermodynamics of charged AdS-BH within standard phase space}
The action for Einstein-Maxwell gravity in $(n+1)$-dimensional AdS spacetime is given by
\begin{eqnarray}
    S=\frac{1}{16\pi G_N}\int d^{n+1}x\sqrt{-g}\left[R-\mathcal{F}^2+\frac{n(n-1)}{L^2}\right], \label{action}
\end{eqnarray}
where $F$ is the $U(1)$ gauge field strength tensor and $\Lambda =-\frac{n(n-1)}{2L^2}$ is the negative cosmological constant with the length scale of the AdS space $L$. 
A spherical symmetric BH solution from this action is 
\begin{eqnarray}
    ds^2&=&-f(r)dt^2+\frac{dr^2}{f(r)}+r^2d\omega^2_{n-1}, \label{metric} 
\end{eqnarray}
where
\begin{eqnarray}
    f(r)&=&1+\frac{r^2}{L^2}-\frac{m}{r^{n-2}}+\frac{q^2}{r^{2n-4}}.
\end{eqnarray}
Note that
$d\omega^2_{n-1}$ is the metric of unit $n-1$ sphere.
The parameter $m$ is related to the mass $M$ of BH as follows
\begin{eqnarray}
    M=\frac{(n-1)\omega_{n-1}}{16\pi G_N}m,
\end{eqnarray}
where $\omega_{n-1}$ is the volume of the unit $n-1$ sphere.
The electric charge of BH $Q$ can be written in the form of the parameter $q$ as
\begin{eqnarray}
    Q=\frac{(n-1)\omega_{n-1}}{8\pi G_N}\eta q,
\end{eqnarray}
where $\eta$ is given by
\begin{eqnarray}
    \eta =\sqrt{\frac{2(n-2)}{n-1}}.
\end{eqnarray}
For spherical symmetric and static charged BH, one can choose the gauge potential as
\begin{eqnarray}
    \mathcal{A}=\left(-\frac{1}{\eta}\frac{q}{r^{n-2}}+\Phi \right)dt. \label{A_t}
\end{eqnarray} 
Here, $\mathcal{A}$ is fixed and set to vanish at the horizon surface $r_+$, so that the electric potential $\Phi$ becomes
\begin{eqnarray}
    \Phi =\frac{1}{\eta}\frac{q}{r_+^{n-2}}.
\end{eqnarray}
The Hawking temperature of BH and its corresponding entropy are associated with its geometrical quantities as
\begin{eqnarray}
    T&=&\frac{\kappa}{2\pi}=\frac{n-2}{4\pi r_+}\left( 1+\frac{n}{n-2}\frac{r_+^2}{L^2}-\frac{q^2}{r_+^{2n-4}}\right), \label{Hawking temp} \\
    S&=&\frac{A}{4G_N}=\frac{\omega_{n-1}r_+^{n-1}}{4G_N},
\end{eqnarray}
where $\kappa$ and $A$ denote the surface gravity and horizon area, respectively. 

Recall that the partition function in the grand canonical ensemble $\mathcal{Z}$ is the Laplace transform of the density of states $g(E,N)$ as follows
\begin{eqnarray}
    \mathcal{Z}=\int g(E,N)e^{-\beta (E-\mu N)}dEdN,
\end{eqnarray}
where $\beta ,E,\mu$ and $N$ denote the inverse temperature, internal energy, chemical potential and number of particles in the system, respectively.
To obtain the density of states, we apply an inverse Laplace transform to $\mathcal{Z}$ and then use the steepest descent method.  Thus, we have
\begin{eqnarray}
    g(E,N)\sim \mathcal{Z}e^{\beta (E-\mu N)}.
\end{eqnarray}
As the thermal entropy $S$ defined as the logarithm of the number of states, the above equation can give the grand potential, $\Omega \equiv -T \ln \mathcal{Z}$, in the form of thermodynamic variables as
\begin{eqnarray}
    \Omega = E-TS-\mu N, \label{W stat}
\end{eqnarray}
The first law of thermodynamics satisfies
\begin{eqnarray}
    d\Omega =-SdT-pd\mathcal{V}-Nd\mu .
\end{eqnarray}
Thus, we treat $T,\mathcal{V}$ and $\mu$ as independent thermodynamic variables so that we can express the grand potential as the function of these three variables, i.e., $\Omega =\Omega (T,\mathcal{V},\mu)$.
Applying the Legendre transformation to obtain the Helmholtz free energy of the canonical ensemble, $F(T,\mathcal{V},N)=\Omega (T,\mathcal{V},\mu)+\mu N$, we have
\begin{eqnarray}
    F=E-TS.
\end{eqnarray}
With the first law of thermodynamics, the infinitesimal change in $F$ reads
\begin{eqnarray}
    dF=-SdT-pd\mathcal{V}-\mu dN.
\end{eqnarray}

In gravitational physics, the partition function of BHs can be approximated from the on-shell Euclidean action $S_E$ as follows \cite{PhysRevD.15.2752}
\begin{eqnarray}
    \mathcal{Z}\sim e^{-S_E}.
\end{eqnarray}
To evaluate the Einstein-Maxwell action in Eq.~\eqref{action} with the potential $A_t$ fixed at infinity, the Gibbons-Hawking boundary term vanishes due to the strength tensor $\mathcal{F}$ becoming zero.
Consequently, the partition function $\mathcal{Z}$ can be simply calculated from the bulk action without any additional term. 
Using the subtraction method, the resulting thermodynamic potential is in the form~\cite{Chamblin:1999tk, Chamblin:1999hg, Burikham:2014gwa}
\begin{eqnarray}
    -T\ln \mathcal{Z}\equiv \Omega =M-TS-\Phi Q. \label{W BH}
\end{eqnarray}
Comparing Eqs.~\eqref{W stat} with \eqref{W BH}, one may identify $E$, $\mu$ and $N$ of the thermal systems with $M$, $\Phi$ and $Q$ of the charged BH.
It is important to note that the thermodynamic potential corresponding to the grand potential $\Omega$, in the above equation, is derived by calculating the on-shell action with potential $A_t$ fixed at the boundary of spacetime. Recall that the potential $\Phi$ is given by taking $r\to \infty$ into Eq.~\eqref{A_t}.
Alternatively, the thermodynamic potential corresponding to the Helmholtz free energy $F$ can be obtained by fixing the charge $Q$ at the spacetime boundary instead.  In this approach, the Gibbons-Hawking term no longer vanishes but becomes significant and contributes to the on-shell action $S_E$.
The resulting Euclidean action calculation with fixed $Q$ at the boundary yields the Helmholtz free energy $F$ as the corresponding thermodynamic potential:

\begin{eqnarray}
   -T\ln \mathcal{Z}\equiv F=M-TS.
\end{eqnarray}

\subsection{Extended phase space approach} \label{subsec_ext}
As discussed in the previous subsection, it is indeed possible to associate the thermodynamic variables of BH with the geometric properties of its spacetime.
However, the Euler's theorem for a homogeneous function suggests that the Smarr formula in Eq.~\eqref{Smarr 1} for BHs in the spacetime with non-zero $\Lambda$ become inconsistent with the first law of black hole thermodynamics unless variations in $\Lambda$ are taken into account.
Using Eqs.~\eqref{Smarr 1} and  \eqref{extended phase} and allowing for variations in the cosmological constant $\Lambda$, we obtain the Smarr formula and the variation of mass in the form
\begin{eqnarray}
    M&=&\frac{n-1}{n-2}TS+\Phi Q-\frac{2}{n-2}PV, \label{MMM} \\
    dM&=&TdS+\Phi dQ+VdP, \label{dM extended}
\end{eqnarray}
respectively.
Note that the computations concerning the Smarr formula and its associated first law derived from the scaling argument are detailed in Appendix \ref{App A} of the manuscript.

As presented in Eq.~\eqref{dM extended}, the BH mass $M$ is a function of $S, Q$ and $P$, thus it should be interpreted as the enthalpy $H(S,Q,P)$ rather than the internal energy $E(S,Q,V)$.
Using the Legendre transformation $E=H-PV$, treating $M$ to be enthalpy leads to the first law of black hole thermodynamics in the extended phase space of the form 
\begin{eqnarray}
    dE=TdS+\Phi dQ-PdV.
\end{eqnarray}
Significantly, the first law can now describe the change in internal energy during processes associated with changes in volume within a black hole's event horizon when it absorbs energy or emits Hawking radiation through the $PdV$ term, akin to conventional thermodynamics.

\begin{figure}[t]
    \centering
    \includegraphics[width = 6cm]{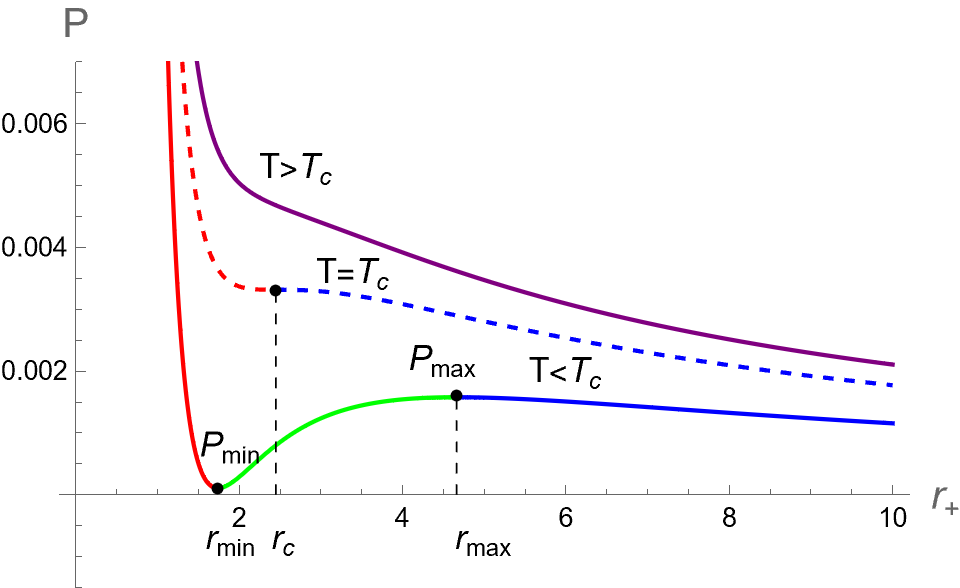}\hspace{2 cm}
    \includegraphics[width = 6cm]{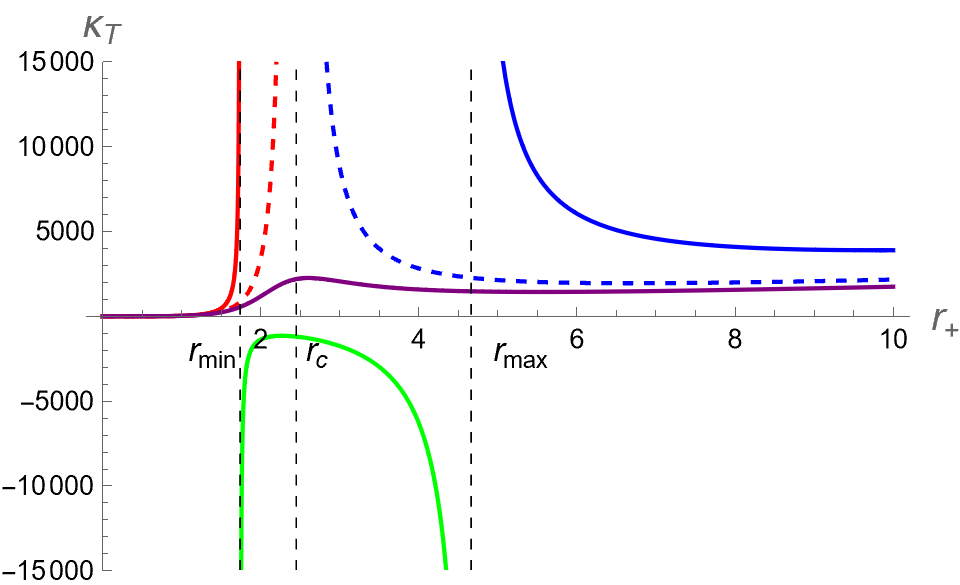}
    \caption{\textbf{Left}: Plots of isotherm curves in the $P-r_+$ plane for fixed $q=1$. The temperature increases from bottom to top as $T=0.0310, 0.0433$ and $0.0500$, where $T_c=0.0433$ is the critical temperature. \textbf{Right}: The plots of isothermal compressibility $\kappa_T$ as a function of $r_+$ correspond to those of the isotherm curves shown in the left panel.
    }  
    \label{fig: P vs rh}
\end{figure}

Let us consider black hole thermodynamics in the canonical ensemble within the extended phase space approach. 
Eliminating $L$ in Eq.~\eqref{Hawking temp} by identifying $\displaystyle P=\frac{3}{8\pi G_NL^2}$, we obtain the equation of state for a $4$-dimensional charged AdS-BH as follows~\cite{kubizvnak2012p}
\begin{eqnarray}
    P=\frac{T}{2r_+}-\frac{1}{8\pi r_+^2}+\frac{q^2}{8\pi r_+^4}. \label{eqf1}
\end{eqnarray}
By comparing the above equation with the equation of state for vdW fluids, one can find that $q$ relates to the gas's constant in the vdW equation of state~\cite{kubizvnak2012p}.
Using Eq.~\eqref{eqf1}, we plot the isotherm curves for different temperatures, as shown in the left panel in Fig.~\ref{fig: P vs rh}.  
This depiction presents the critical behavior in the $P-r_+$ plane. Note that 
the critical values $r_c, T_c$ and $P_c$ of this system satisfy the following conditions:
\begin{eqnarray}
    \frac{\partial P}{\partial r_+}=0, \ \ \ \text{and} \ \ \ \frac{\partial^2 P}{\partial r_+^2}=0.
\end{eqnarray}
Solving these two conditions, we obtain
\begin{eqnarray}\label{critical}
    r_c=\sqrt{6}q, \ \ \ T_c=\frac{\sqrt{6}}{18\pi q}, \ \ \ P_c=\frac{1}{96\pi q^2}.
\end{eqnarray}
Since $P$ should have positive values for $r_+>0$, there exists a particular value of temperature $T_0$ which is the lower bound of temperature where $P=0$ at a particular value of horizon radius $r_0$.
Namely, 
\begin{eqnarray}
    T_0=\frac{\sqrt{3}}{18\pi q}, \ \ \ r_0=\sqrt{3}q. \label{T_0}
\end{eqnarray}
At $T$ below $T_c$, the isotherm curves in the $P-r_+$ plane show the appearance of a local minimum $P_{\text{min}}$ and a local maximum $P_{\text{max}}$ at the horizon radii $r_{\text{min}}$ and $r_{\text{max}}$, respectively. Using Eq.~\eqref{critical} and fixing $q=1$, the critical parameters are given by $r_c=2.45$, $T_c=0.0433$, and $P_c=0.0033$. To illustrate the characteristics of the system, we present the curves for $T$ less than, equal to, and greater than $T_c$. As shown in Fig.~\ref{fig: P vs rh}, the curves for $T=0.0310$ (below the critical point), $T=T_c$, and $T=0.0500$ (above the critical point) are depicted. For $T=0.0310$, we find that $P_{\text{min}}=0.0001$ and $P_{\text{max}}=0.0016$, corresponding to $r_{\text{min}}=1.7391$ and $r_{\text{max}}=4.6615$, respectively.

According to the \textit{Le Chatelier principle} \cite{landau1980,callen1985,Sekerka2015,PhysRevD.105.124049}, 
the \textit{response function} associated to the isotherm curves in $P-r_+$ plane is the \textit{isothermal compressibility} $\kappa_T$, which is defined as
\begin{eqnarray}
    \kappa_T=-\frac{1}{V}\frac{\partial V}{\partial P}\Big |_T=\frac{12\pi r_+^4}{2\pi Tr_+^3-r_+^2+2q^2}, \label{compressibility}
\end{eqnarray}
where $V=\frac{4}{3}\pi r_+^3$ is thermodynamic volume.
The right panel in Fig.~\ref{fig: P vs rh} displays $r_+$ dependence of $\kappa_T$ corresponding to isotherm curves in $P-r_+$ plane in the left panel. 
The results reveal three branches of BHs below $T_c$: Small BH (red solid curve), Intermediate BH (green solid curve) and Large BH (blue solid curve) correspond to the ranges of $r_+<r_\text{min}, r_\text{min}<r_+<r_\text{max}$ and $r_+>r_\text{max}$, respectively.
Here $r_\text{min}$ and $r_\text{max}$ is the event horizon radii that $\kappa_T$ diverge.  
Note that these values of $r_+$ can be obtained by setting the denominator in Eq.~\eqref{compressibility} to be zero, namely $2\pi Tr_+^3-r_+^2+2q^2=0$, and finding its roots. 
The Small BH and Large BH (both have $\kappa_T>0$) are mechanically stable against microscopic fluctuation while the Intermediate BH ($\kappa_T<0$) is unstable.
The authors in \cite{Lan:2015bia}  have shown that the unstable Intermediate BH in the $P-V$ plane can be replaced by a straight line at $P=P_f$ obeying the Maxwell equal area law, where the Small-Large BHs first-order phase transition occurs.
Note that $P_f$ can be called as the \textit{Hawking-Page pressure}.
Defining the reduced variables as
\begin{eqnarray}
    p=\frac{P}{P_c},\ \ \ t=\frac{T}{T_c}\ \ \ \text{and}\ \ \ v=\frac{V}{V_c},
\end{eqnarray}
an analytic expression for Hawking-Page pressure is given by \cite{Lan:2015bia}
\begin{eqnarray}
    p^*=\left[1-2\cos \left(\frac{\arccos \left(1-t^{2}\right)+\pi}{3} \right)\right]^2, \label{Maxwell 1}
\end{eqnarray}
where $p^*=P_f/P_c$ denotes the reduced Hawking-Page pressure.
Considering the isotherm curve with $T=0.0310$, for example, which is less than $T_c$ as shown in Fig.~\ref{fig: P vs rh}, the corresponding reduced temperature $t=0.7159$ can be put into the above equation such that the reduced pressure at which the first-order phase transition occurs is at  $p^*=0.4384$, corresponding to $P_f=0.0014$.
Moreover, the horizon radii for Small and Large BHs at the first-order phase transition are respectively expressed as follows
\begin{eqnarray}
    \frac{r_S}{r_c}&=&\frac{2}{t}\cos^2\phi -\sqrt{\frac{4}{t^2}\cos^4\phi -\frac{\sqrt{2}}{t}\cos \phi}, \label{small r+} \\
    \frac{r_L}{r_c}&=&\frac{2}{t}\cos^2\phi +\sqrt{\frac{4}{t^2}\cos^4\phi -\frac{\sqrt{2}}{t}\cos \phi}, \label{large r+}
\end{eqnarray}
where $\phi=\frac{\pi -\theta}{3}$ and $\cos{\theta}=\frac{\sqrt{2}}{2}t$.

To investigate the global stability for charged AdS-BH associated with the isotherm curves in the $P-r_+$ plane, one needs to consider the free energy as a function of bulk pressure $P$ with the temperature $T$ held fixed.
In the fixed charge ensemble, treating \( M \) as the internal energy results in the thermodynamic potential being the Helmholtz free energy \( F \), as is typical in standard phase space. However, in this section, where \( M \) is treated as the enthalpy, the thermodynamic potential is instead the Gibbs free energy \( G \).
We express $G$ in term of $r_+$ with fixed $T$ in the following form
\begin{eqnarray}
    -T\ln \mathcal{Z}\equiv G &= & E+PV-TS \nonumber \\ 
                              &=& M-TS \nonumber \\ 
                              &=& \frac{2 q^2-\pi r_+^3 T+r_+^2}{3r_+}. \label{F extended}
\end{eqnarray}
By using $dM$ from Eq.~\eqref{dM extended} with the charge fixed, the infinitesimal change of the Gibbs free energy is
\begin{eqnarray}
    dG=dM-TdS-SdT,
\end{eqnarray}
can be changed to be in the form   
\begin{eqnarray}
    dG=-SdT+VdP.
\end{eqnarray}
This relation suggests that $G=G(T,P)$ leading to the expression for entropy and thermodynamic volume given by 
\begin{eqnarray}
    \left(\frac{\partial G}{\partial T}\right)_{P}=-S, \ \ \ \left(\frac{\partial G}{\partial P}\right)_{T}=V.
\end{eqnarray}

\begin{figure}[t]
    \centering
    \includegraphics[width = 4.cm]{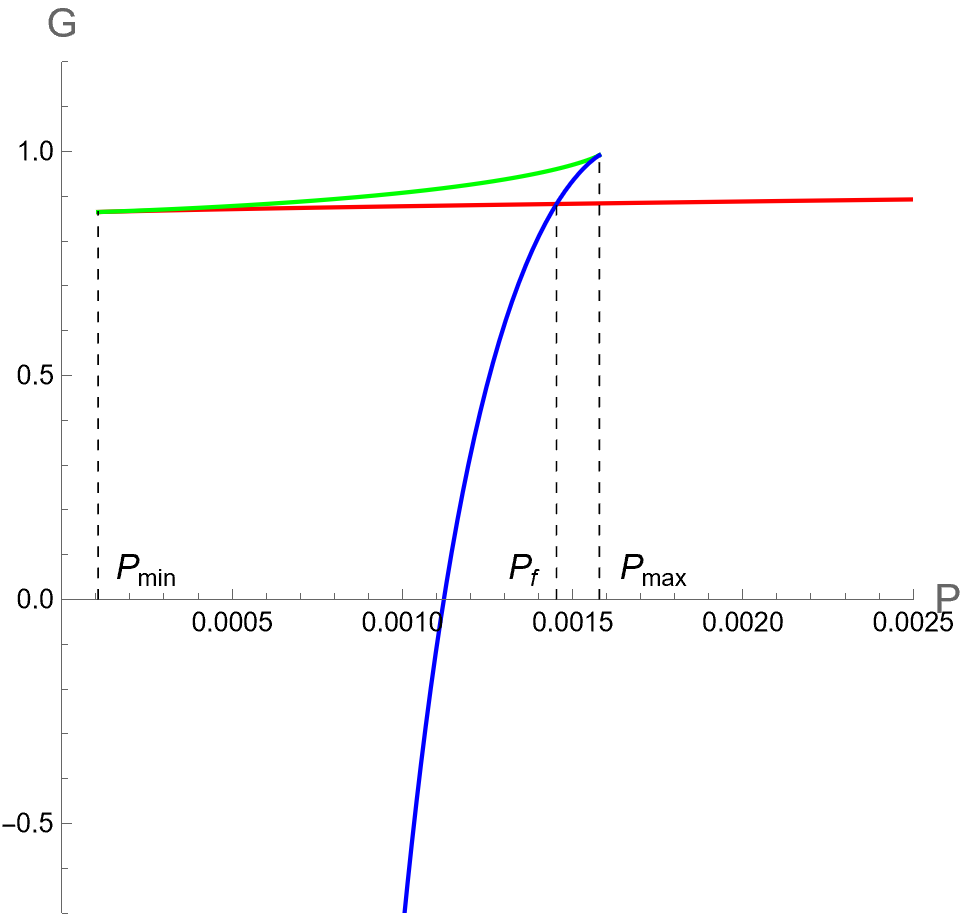}\hfill
    \includegraphics[width = 4.cm]{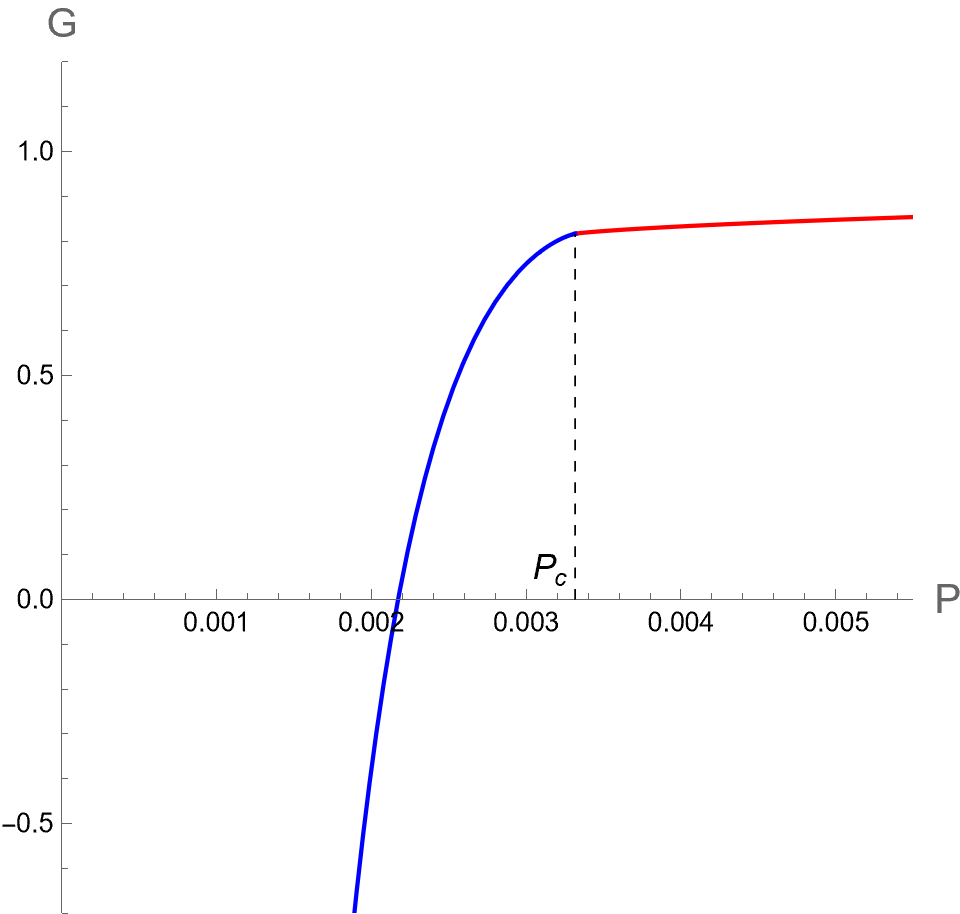}\hfill
    \includegraphics[width = 4.cm]{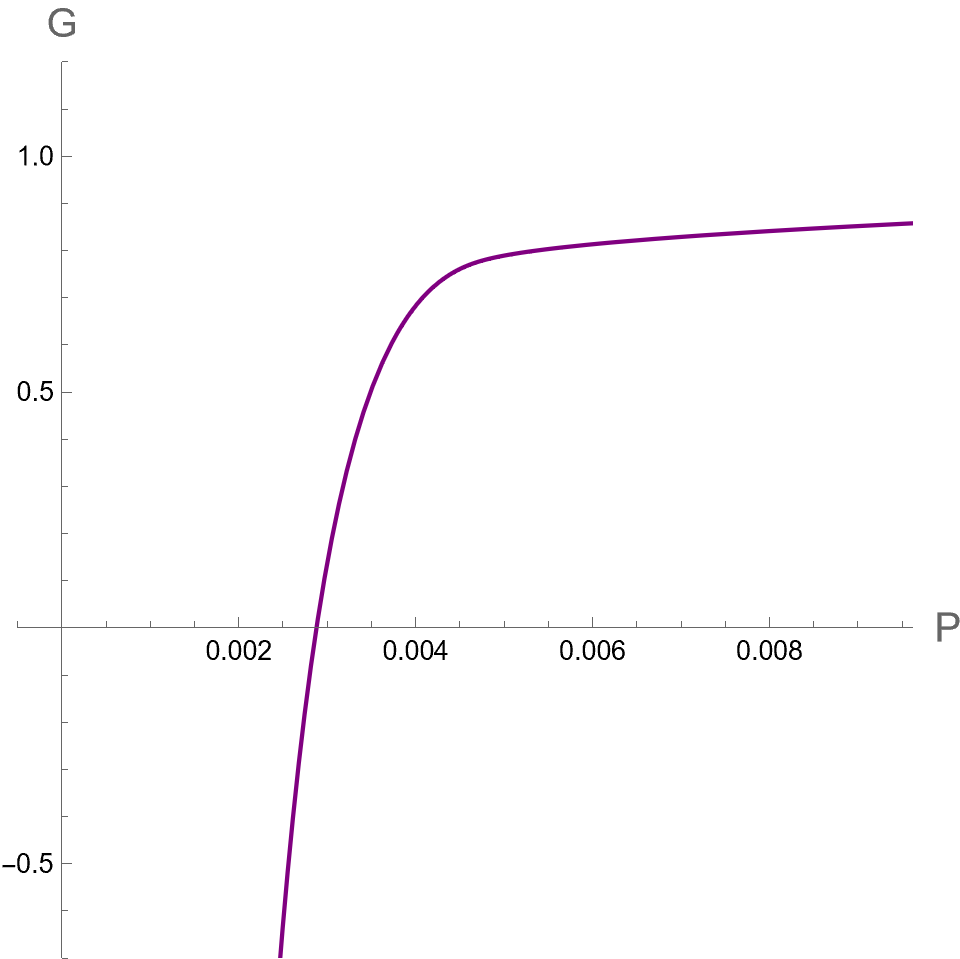}
    \caption{The isotherm curves in the $G-P$ plane for fixed $q=1$ are plotted at different temperatures, namely the temperature $0.0310$ (left), $0.0433$ (middle), and $0.0500$ (right), which represent the behaviors of the systems at the temperature smaller than, equal to, and larger than $T_c$, respectively.} 
    \label{fig: F vs P}
\end{figure}

We plot \( G \) against \( P \) for a fixed \( T \) in Fig.~\ref{fig: F vs P}. It is important to note that the point \( r_+ \) where \( \kappa_T \) becomes divergent corresponds to a cusp in the \( G(P) \) graph. This feature reflects the second-order phase transition resulting from $\kappa_T= -\frac{1}{V}\left(\frac{\partial^2 G}{\partial P^2}\right)_{T}$.
As shown in the left panel in Fig.~\ref{fig: F vs P}, the free energy in the case of $T<T_c$, exhibits the swallowtail behavior with two cusps at $P_\text{min}$ and $P_\text{max}$ corresponding to the horizon radii that $\kappa_T$ diverges $r_\text{min}$ and $r_\text{max}$, respectively. 
Considering the bulk pressure $P$ running from a low value where only Large-BH exists, Small and Intermediate BHs both emerge at $P=P_\text{min}$ with free energy larger than the Large BH. Obviously, Large BH is thermodynamically preferred in the region of $P<P_f$. It turns out that the system encounters the first-order phase transition at $P_f$ such that Small BH phase becomes thermodynamically preferred when $P>P_f$.
The middle panel in Fig.~\ref{fig: F vs P} shows the case of $T=T_c$ where the Small-Large BHs transition becomes the second-order type of phase transition at $P=P_c$. 
The right panel in Fig.~\ref{fig: F vs P} shows that only one phase of BH exists at high temperature $T>T_c$, here at $T=0.0500$, where $\kappa_T>0$ at all values of $P$.
Remarkably, by identifying Small, Intermediate, and Large BHs as liquid, metastable, and gas phases, respectively, the behaviors of the BH phase transition is in a similar way as what occurs in the vdW fluid.

\subsection{Holographic thermodynamics approach}
As mentioned in section~\ref{Intro}, Introduction, the pressure $P$ and thermodynamic volume $V$ characterizing BHs in the bulk through the extended phase space approach do not directly correspond to the pressure $p$ and volume $\mathcal{V}$ of the dual field theory on the boundary within the AdS/CFT correspondence. Additionally, the Smarr formula, which relates the mechanical quantities of BHs to thermodynamic variables, depends on the number of spacetime dimensions, this feature is typically absent in the Euler equation for ordinary matter. 
Recently, the holographic thermodynamics approach has provided insights into effectively addressing these two issues in associating the black hole thermodynamic properties with its dual field theory counterpart. 
In this section, we examine the thermodynamics of CFT, which is holographically dual to the charged AdS-BH, within the holographic thermodynamics approach. 

Introducing the central charge $\mathcal{C}$ and its chemical potential $\mu_\mathcal{C}$ as a new pair of thermodynamic variable in the large-$N$ gauge theory, the author in \cite{Visser:2021eqk} suggests that the scaling behavior of its internal energy $E$ may be written as
\begin{eqnarray}
    E(\alpha S,\alpha^0 \mathcal{V},\alpha B_i,\alpha \mathcal{C})=\alpha E(S,\mathcal{V},B_i,\mathcal{C}),
\end{eqnarray}
where $B_i$ denote the conserved quantities, such as charge and angular momentum.
The Euler equation and its corresponding first law due to the Euler scaling argument can be obtained by using Eqs.~\eqref{Euler} and \eqref{df} in Appendix \ref{App A} as
\begin{eqnarray}
    E&=&TS+\nu^iB_i+\mu_\mathcal{C} \mathcal{C}, \label{Euler1} \\
    dE&=&TdS-pd\mathcal{V}+\nu^idB_i+\mu_\mathcal{C} d\mathcal{C}, \label{first law holography}
\end{eqnarray}
where
\begin{eqnarray}
    T=\left(\frac{\partial E}{\partial S}\right)_{\mathcal{V},B_i,\mathcal{C}}, 
    p=-\left(\frac{\partial E}{\partial \mathcal{V}}\right)_{S,B_i,\mathcal{C}},
    \nu^i=\left(\frac{\partial E}{\partial B_i}\right)_{S,\mathcal{V},\mathcal{C}},
    \mu_\mathcal{C}=\left(\frac{\partial E}{\partial \mathcal{C}}\right)_{S,\mathcal{V},B_i}.
\end{eqnarray}
Note that the $p\mathcal{V}$ term is absent in the Euler equation but the $pd\mathcal{V}$ term appear in the first law.
Since the variation of $E$ in Eq.~\eqref{Euler1} should be equal to the first law in Eq~\eqref{first law holography}, i.e., $TdS+SdT+\nu^idB_i+B_id\nu^i+\mu_\mathcal{C}d\mathcal{C}+\mathcal{C}d\mu_\mathcal{C}=TdS-pd\mathcal{V}+\nu^idB_i+\mu d\mathcal{C}$, this implies the Gibbs-Duhem equation as follows
\begin{eqnarray}
    SdT+B_id\nu^i+\mathcal{C}d\mu_\mathcal{C}=-pd\mathcal{V}.
\end{eqnarray}
The grand potential is given by 
\begin{eqnarray}
    \Omega &=&E-TS-\nu^iB_i, \nonumber \\
    &=&(TS+\nu^iB_i+\mu_\mathcal{C} \mathcal{C})-TS-\nu^iB_i, \nonumber \\
    &=&\mu_\mathcal{C}\mathcal{C},
\end{eqnarray}
where we have substituted $E$ by using Eq.~\eqref{Euler1}.

Since $\mathcal{C}$ depends on both $L$ and $G_N$ due to the holographic dictionary as shown in Eq.~\eqref{dictionary}, thus varying $\mathcal{C}$ in the dual field theory should be equivalent to variation of $\Lambda$ and $G_N$ in the gravity side.
By including $\Lambda$ and $G_N$ into the mass formula of AdS-BH in Eq.~\eqref{M(A,Q,Lambda)}, 
an infinitesimal change of $M$ can be obtained from Eq.~\eqref{df} as
\begin{eqnarray}
    dM=\frac{\kappa}{8\pi G_N}dA+\Phi dQ+\frac{\Theta}{8\pi G_N}d\Lambda -(M-\Phi Q)\frac{dG_N}{G_N},
\end{eqnarray}
where an additional partial derivative of $M$ with respect to $G_N$ is $\frac{\partial M}{\partial G_N}=-\frac{(M-\Phi Q)}{G_N}$.
Substituting $\Theta$ in term of other variables by using the Smarr formula in Eq.~\eqref{Smarr 1} with $\Lambda$ expressed in term of $L$, the above equation becomes
\begin{eqnarray}
    dM=\frac{\kappa}{2\pi}d\left( \frac{A}{4G_N}\right)+\frac{\Phi}{L}d(QL)-\frac{M}{n-1}\frac{dL^{n-1}}{L^{n-1}}+\left( M-\frac{\kappa A}{8\pi G_N}-\Phi Q\right)\frac{d(L^{n-1}/G_N)}{L^{n-1}/G_N}. \nonumber \\ \label{pre 1st law}
\end{eqnarray}
Thermodynamic variables of dual field theory can be identified with geometric quantities of AdS-BH as
\begin{eqnarray}
    E=M, \ \ \ \Tilde{\Phi}=\frac{\Phi}{L}, \ \ \ \Tilde{Q}=QL, \ \ \ \mathcal{V} \sim L^{n-1}, \ \ \ \mathcal{C} \sim \frac{L^{n-1}}{G_N}.
\end{eqnarray}
Substituting them into Eq.~\eqref{pre 1st law} and comparing with Eq.~\eqref{first law holography}, we obtain
\begin{eqnarray}
    dE=TdS+\Tilde{\Phi}d\Tilde{Q}-pd\mathcal{V}+\mu_\mathcal{C} d\mathcal{C}, \label{holographic 1 st}
\end{eqnarray}
where
\begin{eqnarray}
   p=\frac{E}{(n-1)\mathcal{V}}, \ \ \ \mu_\mathcal{C} =\frac{1}{\mathcal{C}}\left(E-TS-\Tilde{\Phi}\Tilde{Q}\right). \label{p_mu_holo}
\end{eqnarray}
Notably, the former relation in \eqref{p_mu_holo} expresses the pressure $p$, in the field theory side, satisfies the equation of state in Eq.~\eqref{eos CFT}, while the latter gives the Euler equation corresponding to Eq.\eqref{Euler1} with identifying $\nu =\Tilde{\Phi}$ and $B=\Tilde{Q}$.
It is interesting to note that the CFT in the field theory side from the approach of holographic thermodynamics can live in the curved spacetime with the curvature radius $R$, which has a value not necessarily equal to the bulk AdS radius $L$. 
This can be seen by redefining the thermodynamic variables as \cite{Visser:2021eqk,Cong:2021jgb}
\begin{eqnarray}
    E=M\frac{L}{R}, \ \ \ T=\frac{\kappa}{2\pi}\frac{L}{R}, \ \ \ S=\frac{A}{4G_N}, \ \ \ \Tilde{\Phi}=\frac{\Phi}{L}\frac{L}{R}, \ \ \ \Tilde{Q}=QL. \label{rescale_holo}
\end{eqnarray}
From the resulting first law, we have $\mathcal{V}\sim R^{n-1}$ and $\displaystyle \mathcal{C}\sim \frac{L^{n-1}}{G_N}$. Thus, the volume $\mathcal{V}$ and central charge $\mathcal{C}$ of the dual CFT are now independent due to the choice of rescaling as shown in Eq.~\eqref{rescale_holo}.

It is worthwhile to review here some interesting results about the thermodynamics of CFT that are holographically dual to charged AdS-BH within the novel holographic thermodynamics approach.
For simplicity, we define the dimensionless quantities: 
\begin{eqnarray}
    x=\frac{r_+}{L}, \ \ \ y=\frac{q}{L^{n-2}}. \label{xy}
\end{eqnarray}
The thermodynamic quantities of the dual CFT, which is in $n$-dimensional spacetime of curvature radius $R$, can be written in parametric equations of $x$ and $y$ as follows \cite{Cong:2021jgb}
\begin{eqnarray}
    E&=&\frac{(n-1)\mathcal{C}x^{n-2}}{R}\left(1+x^2+\frac{y^2}{x^{2n-4}}\right), \\
    T&=&\frac{n-2}{4\pi Rx}\left(1+\frac{n}{n-2}x^2-\frac{q^2}{x^{2n-4}}, \right), \label{Hawking T} \\
    S&=&4\pi \mathcal{C}x^{n-1}, \\
    \Tilde{Q}&=&2\eta (n-1)\mathcal{C}y, \label{Q vs y} \\
    \Tilde{\Phi}&=&\frac{1}{\eta R}\frac{y}{x^{n-2}},  \\
    \mu_\mathcal{C} &=&\frac{x^{n-2}}{R}\left(1-x^2-\frac{y^2}{x^{2n-4}}\right).
\end{eqnarray}
Here, the field theory under consideration lives in the spacetime with $n=3$ dimensions.  The CFT is in the ensemble with fixed $(\Tilde{Q}, \mathcal{V}, \mathcal{C})$, corresponding to the canonical ensemble. The temperature and heat capacity of the dual CFT are given by
\begin{eqnarray}
    T&=&\frac{1}{4\pi Rx}\left( 1+3x^2+\frac{Q^2}{16\mathcal{C}^2x^2}\right), \label{temp 4D} \\
    C_{\Tilde{Q},\mathcal{V},\mathcal{C}}&=&T\left(\frac{\partial S}{\partial T}\right)_{\Tilde{Q},\mathcal{V},\mathcal{C}}
    =\frac{8\pi \mathcal{C}x^2\left(16\mathcal{C}^2 x^2 (3 x^2+1)-\Tilde{Q}^2\right)}{16\mathcal{C}^2 x^2 (3 x^2-1)+3 \Tilde{Q}^2}.
\end{eqnarray}
From Eq.~\eqref{temp 4D}, the horizon radius of extremal BH $(T=0)$ can be written as a function of $\Tilde{Q}$ and $\mathcal{C}$ as   
\begin{eqnarray} x_\text{ext}^2=\frac{-2\mathcal{C}^2+\sqrt{4\mathcal{C}^4+3
\mathcal{C}^2\Tilde{Q}^2}}{12\mathcal{C}^2}.
\end{eqnarray}
Due to the $T-x$ criticality of charged AdS-BH, thermodynamics of dual CFT also critically change in a similar way as the ratio $\Tilde{Q}/\mathcal{C}$ crosses the critical point, which can be determined from the point of inflection as follows
\begin{eqnarray}
    \left( \frac{\partial T}{\partial x}\right)=0, \ \ \ \text{and} \ \ \ \left( \frac{\partial^2 T}{\partial x^2}\right)=0.
\end{eqnarray}
Solving these two conditions, we obtain
\begin{eqnarray}
    x_c=\frac{1}{\sqrt{6}}, \ \ \ T_c=\frac{\sqrt{2}}{\sqrt{3}\pi},  \ \ \ S_c=\frac{2\pi \mathcal{C}}{3}, \ \ \ \frac{\Tilde{Q}_c}{\mathcal{C}_c}=\frac{2}{3}.
\end{eqnarray}
\begin{figure}[t]
    \centering
    \subfigure[]{\includegraphics[width = 4cm]{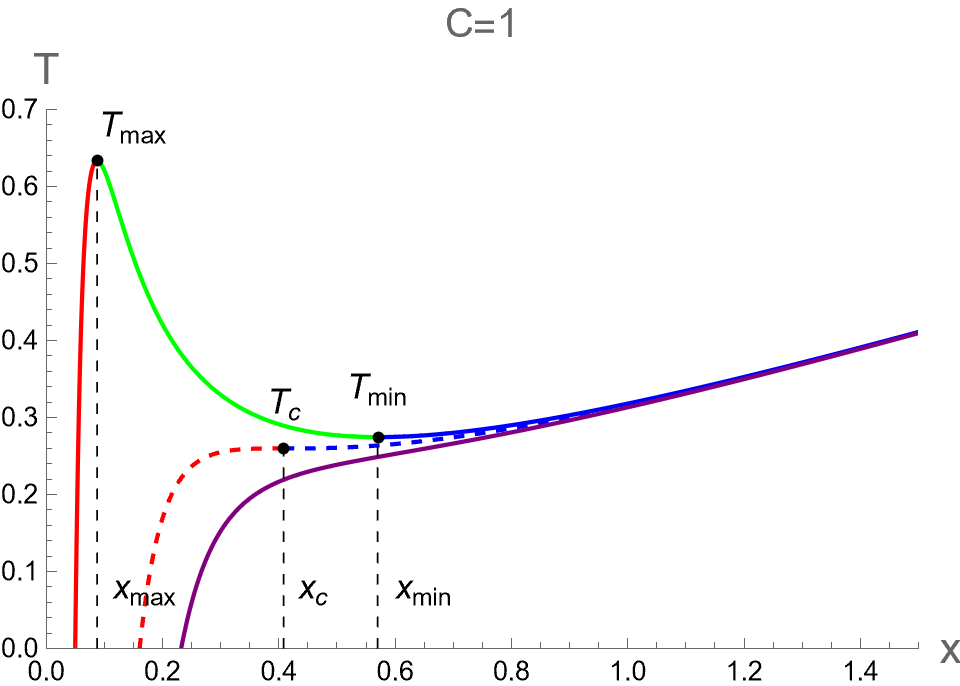}}\hspace{2 cm}
    \subfigure[]{\includegraphics[width = 4cm]{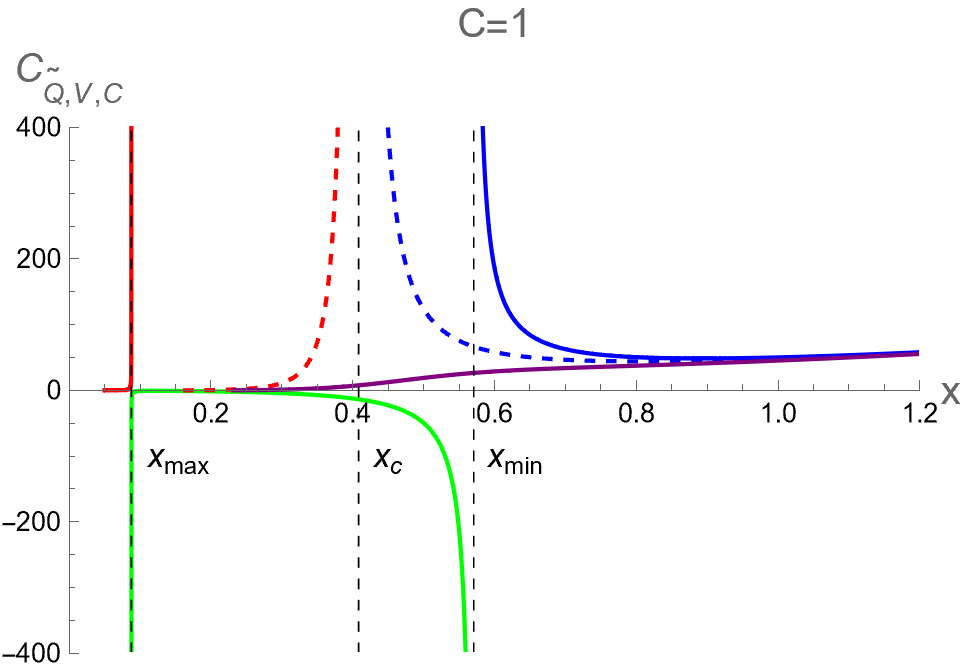}}\\
    \subfigure[]{\includegraphics[width = 4cm]{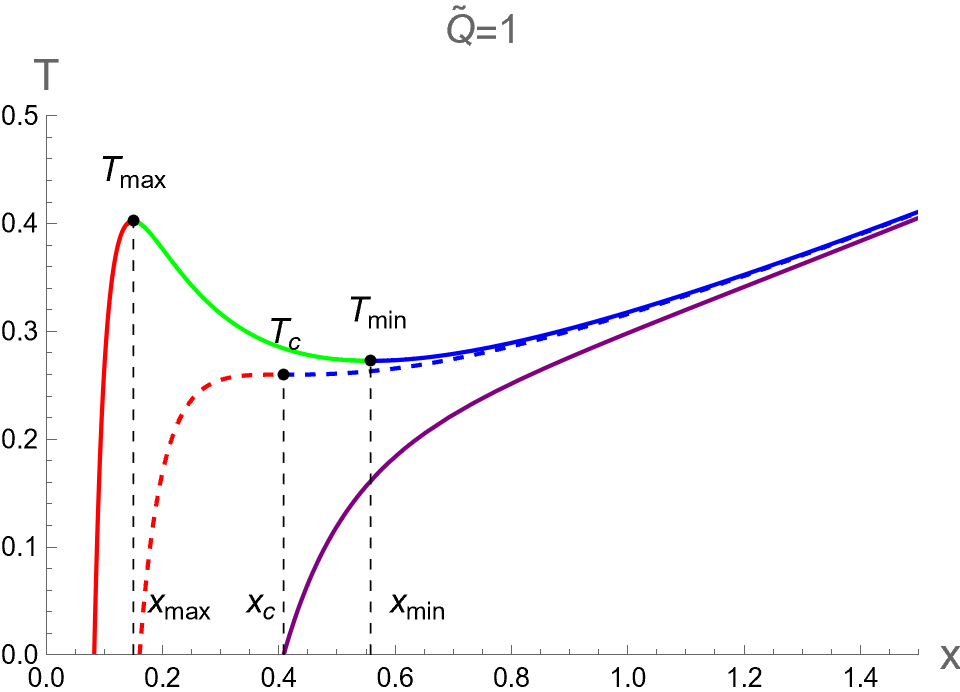}}\hspace{2 cm}
    \subfigure[]{\includegraphics[width = 4cm]{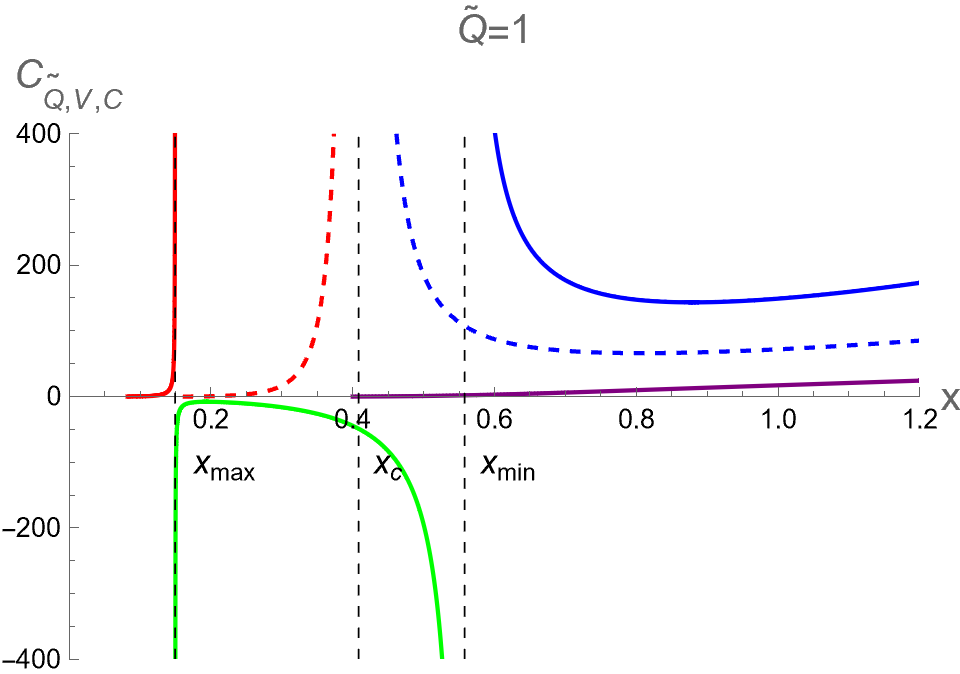}}
    \caption{\textbf{(a)} Plots of isocharge curves in the $T-x$ plane for $\Tilde{Q}=0.2,2/3,1$ with fixed $\mathcal{C}=1$. 
    \textbf{(b)} Plots of the heat capacity $C_{\Tilde{Q},\mathcal{V},\mathcal{C}}$ as a function of $x$ that corresponds to figure (a). \textbf{(c)} The dependence of $T$ on parameter $x$, while keeping $\Tilde{Q}=1$ fixed with varying $\mathcal{C}=0.5,3/2,3$. \textbf{(d)} Plots of heat capacity $C_{\Tilde{Q},\mathcal{V},\mathcal{C}}$ versus $x$ that corresponds to figure (c). Note that the critical values of electric charge and central charge are $\Tilde{Q}_c=2/3$ and $\mathcal{C}_c=3/2$, respectively.}  
    \label{fig: T and heat capacity}
\end{figure}
We examine thermal phase structures of dual CFT in two cases: (I) fixed $\mathcal{C}$ with different values of $\Tilde{Q}$ and (II) fixed $\Tilde{Q}$ with different values of $\mathcal{C}$.
We illustrate the behaviors of $T$ and $C_{\Tilde{Q},\mathcal{V},\mathcal{C}}$ as functions of $x$ for the former and latter cases in the first and second rows of Fig.~\ref{fig: T and heat capacity}, respectively. 
Note that  $x_\text{max}$ and $x_\text{min}$ are given by
\begin{eqnarray} x_\text{max,min}^2=\frac{2\mathcal{C}^2\mp\sqrt{4\mathcal{C}^4-9\mathcal{C}^2\Tilde{Q}^2}}{12\mathcal{C}^2}.
\end{eqnarray}
Two radii $x_\text{max}$ and $x_\text{min}$ indicate positions of local maximum and minimum of Hawking temperature as shown in Fig.~\ref{fig: T and heat capacity} (a) and (c), namely $T_\text{max}=T(x_\text{max})$ and $T_\text{min}=T(x_\text{min})$. 
At the critical value of $\Tilde{Q}/\mathcal{C}=2/3$, these two radii coincide at $x_c$, which corresponds to the critical temperature $T_c$.
Remarkably, there exist three thermal states of dual CFT for $\Tilde{Q}<\Tilde{Q}_c$ in case (I) and $\mathcal{C}>\mathcal{C}_c$ in case (II). These three states consist of \textbf{pCFT1} (red), \textbf{nCFT} (green), and \textbf{pCFT2} (blue), which refers to the states of positive heat capacity when $x_\text{ext}<x<x_\text{max}$, negative heat capacity when $x_\text{max}<x<x_\text{min}$,  and positive heat capacity when $x>x_\text{min}$, respectively. It is important to note that our notation might cause some confusion when $x_{\text{max}}$ is less than $x_{\text{min}}$. As defined above, $x_{\text{max}}$ and $x_{\text{min}}$ in this paper refer to the points at which the temperature $T$ is at its maximum and minimum, respectively. Please do not be confused by these terms.
\begin{table}[!htbp]
\caption{The numerical values of parameters that characterize the phase transition of charged AdS-BH in cases I and II from the holographic thermodynamics.}
\vspace{0.5cm}
\begin{tabular}{|*{7}{c|}}
  \cline{2-7}
\multicolumn{1}{c|}{} & \multicolumn{3}{c|}{Case I} & \multicolumn{3}{c|}{Case II} \\
  \cline{2-7}
 \multicolumn{1}{c|}{} & $\Tilde{Q}=0.200$ & $\Tilde{Q}_c=0.667$ & $\Tilde{Q}=1.00$ & $\mathcal{C}=0.500$ & $\mathcal{C}_c=1.50$ & $\mathcal{C}=3.00$ \\ \hline
  $x_\text{ext}$ & 0.0498 & 0.161 & 0.232 & 0.408 & 0.161 & 0.0825 \\ \hline
  $x_\text{min}$ & 0.571 & - & - & - & - & 0.558 \\ \hline
  $x_\text{max}$ & 0.0876 & - & - & - & - & 0.149 \\ \hline
  $x_c$ & - & 0.408 & - & - & 0.408 & - \\ \hline
  $T_\text{min}$ & 0.275 & - & - & - & - & 0.273 \\ \hline
  $T_\text{max}$ & 0.633 & - & - & - & - & 0.403 \\ \hline
  $T_c$ & - & 0.260 & - & - & 0.260 & - \\ \hline
  $T_f$ & 0.302 & - & - & - & - & 0.291 \\ \hline
\end{tabular}
\label{tab: Table I}
\end{table}

Using the Maxwell's equal area law, the curve associated with the unstable nCFT phase in the $T-x$ plane in cases (I) and (II) can be replaced by a horizontal line, which indicates the first-order phase transition between these pCFT1 and pCFT2.
In Appendix \ref{App B}, we elaborate on the detailed calculations using the method employed by \cite{Lan:2015bia} to obtain the Hawking-Page phase transition temperature $T_f$ within the holographic thermodynamics framework. The temperatures for cases I and II are as follows:
\begin{eqnarray}
    t^*=\frac{3+\sqrt{3(q-3)(q-1)}-q}{2\sqrt{3+\sqrt{3(q-3)(q-1)}-2q}}, \label{Tf of fix C}
\end{eqnarray}
and
\begin{eqnarray}
    t^*&=&\frac{3c+\sqrt{3(1-3c)(1-c)}-1}{2\sqrt{3c^2+c\sqrt{3(1-3c)(1-c)}-2c}},
\end{eqnarray}
respectively, where $t^*=T_f/T_c, q=\Tilde{Q}/\Tilde{Q}_c$ and $c=\mathcal{C}/\mathcal{C}_c$.
The horizon radii for Small and Large BHs of case I and II are expressed respectively as follows
\begin{eqnarray}
    \frac{x_S}{x_c}=\frac{q}{\sqrt{3+\sqrt{3(q-3)(q-1)}-2q}}, \label{Small r holo} \\
    \frac{x_L}{x_c}=\sqrt{3+\sqrt{3(q-3)(q-1)}-2q}, \label{Large r holo} 
\end{eqnarray}
and
\begin{eqnarray}
   \frac{x_S}{x_c}=\frac{1}{\sqrt{3c^2+c\sqrt{3(1-3c)(1-c)}-2c}}, \\
   \frac{x_L}{x_c}=\frac{1}{c}\sqrt{3c^2+c\sqrt{3(1-3c)(1-c)}-2c}.
\end{eqnarray}
Note that we summarize our numerical results of some important parameters in our study in Table \ref{tab: Table I}.

\begin{figure}[t]
    \centering
    \includegraphics[width = 5cm]{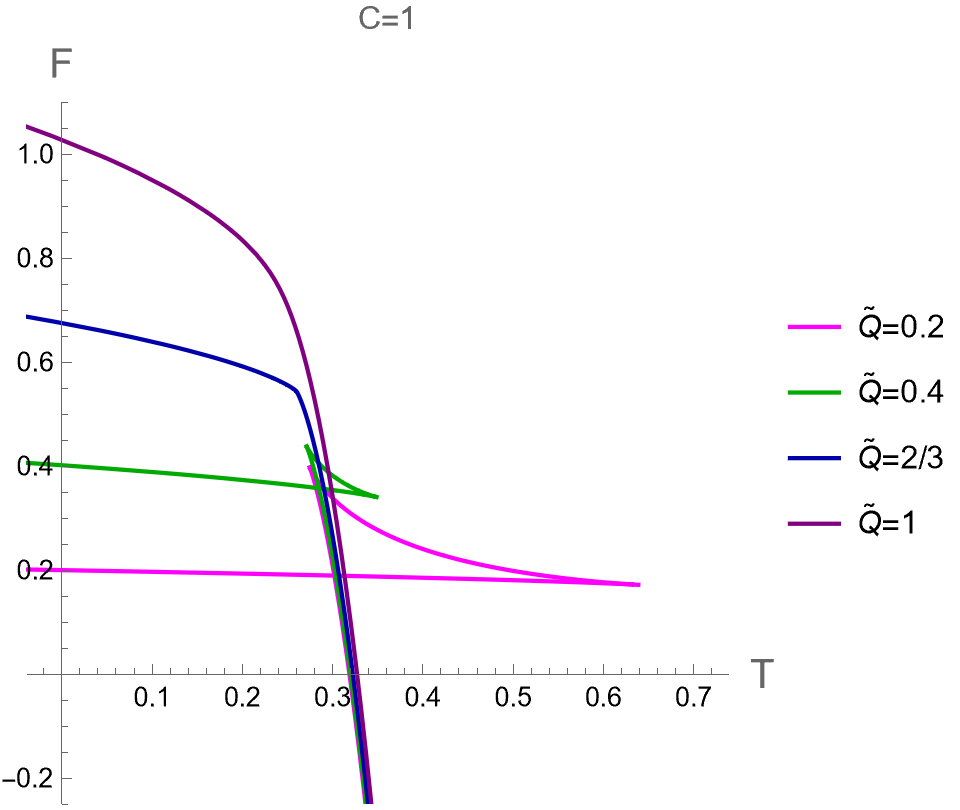}\hspace{2 cm}
    \includegraphics[width = 5cm]{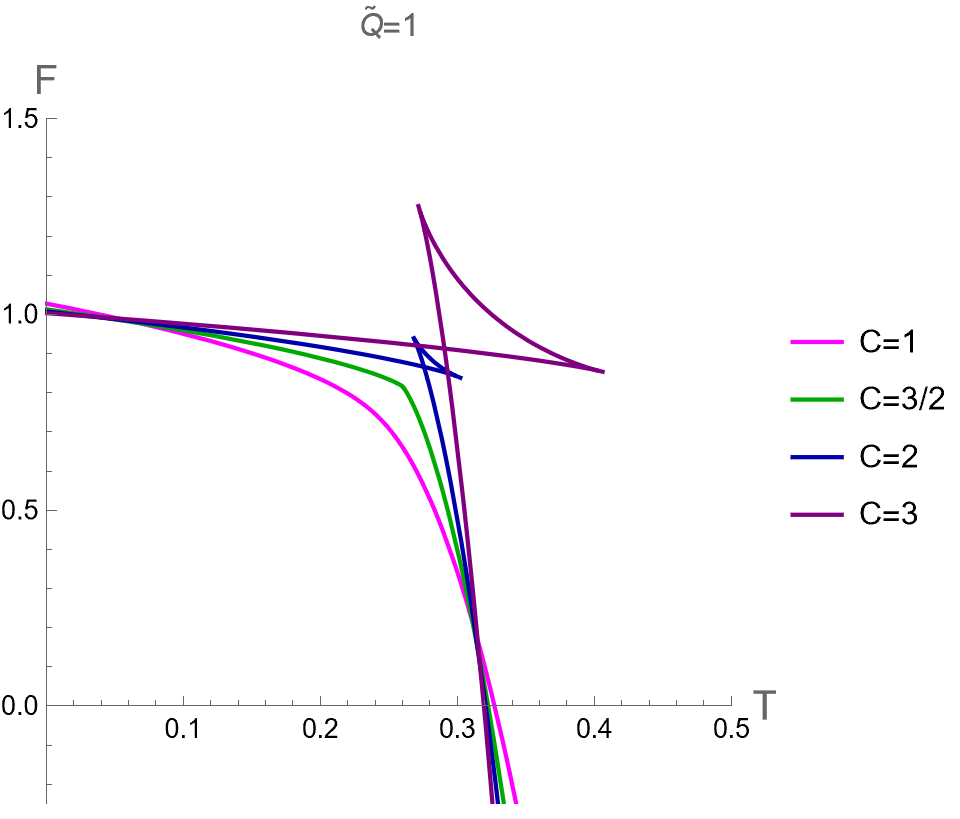}
    \caption{Plots of $F$ versus $T$ of $3$-dimensional CFT which holographically dual to $4$-dimensional charged AdS-BH.
    Note that the radius of curvature $R=1$ in these plots.
    \textbf{Left:} We fix $\mathcal{C}=1$ with different values of $\Tilde{Q}=0.2,0.4,2/3,1$ (pink, dark-green, dark-blue and purple). 
    \textbf{Right:} We fix $\Tilde{Q}=1$ and vary $\mathcal{C}=1,3/2,2,3$ (pink, dark-green, dark-blue and purple). For $\Tilde{Q}<\Tilde{Q}_c$ $(\mathcal{C}>\mathcal{C}_c)$, the free energy shows the swallowtail behavior and first-order phase transition occur. }  
    \label{fig: F vs T compare}
\end{figure}

The thermodynamic potential associated with fixed $(\Tilde{Q}, \mathcal{V}, \mathcal{C})$ ensemble is the Helmholtz free energy 
\begin{eqnarray}
    F\equiv E-TS=\frac{\mathcal{C}x^{n-2}}{R}\left(1-x^2+\frac{(2n-3)}{4\eta^2(n-1)^2\mathcal{C}^2}\frac{\Tilde{Q}^2}{x^{2n-4}}\right).\label{F holographic}
\end{eqnarray}
The variation of $F$ reads
\begin{eqnarray}
    dF&=&dE-TdS-SdT, \nonumber \\
    &=&(TdS+\Tilde{\Phi}d\Tilde{Q}-pd\mathcal{V}+\mu_\mathcal{C} d\mathcal{C})-TdS-SdT,\nonumber \\
    &=&-SdT+\Tilde{\Phi}d\Tilde{Q}-pd\mathcal{V}+\mu_\mathcal{C} d\mathcal{C},
\end{eqnarray}
where we have used Eq.~\eqref{holographic 1 st} for $dE$, this yields
\begin{eqnarray}
    \left(\frac{\partial F}{\partial T}\right)_{\Tilde{Q},\mathcal{V},\mathcal{C}}=-S, \ \ \ 
    \left(\frac{\partial F}{\partial \Tilde{Q}}\right)_{T,\mathcal{V},\mathcal{C}}=\Tilde{\Phi}, \ \ \ \left(\frac{\partial F}{\partial \mathcal{V}}\right)_{T,\Tilde{Q},\mathcal{C}}=-p, \ \ \ 
    \left(\frac{\partial F}{\partial \mathcal{C}}\right)_{T,\Tilde{Q},\mathcal{V}}=\mu_\mathcal{C}.\label{df relation}
\end{eqnarray}
As pointed out in \cite{Cong:2021jgb}, there is no critical point found in the $p-\mathcal{V}$ plane in the dual CFT.
This result from the holographic thermodynamics is different from the $P-V$ criticality behavior from the extended phase space approach as discussed in last subsection.
Therefore, we will consider $F$ as a function of $T$ instead of $p$ with constant $\Tilde{Q}$ or $\mathcal{C}$ to investigate the global stability of dual CFT.
By considering $x$ as a parameter, we parametrically plot between $F$ in Eq.\eqref{F holographic} and $T$ in Eq.\eqref{Hawking T} for dual CFT in the case (I) and (II) as shown in the left and right panels of Fig~\ref{fig: F vs T compare}, respectively. 
There are cusp points in $F(T)$ curve where $C_{\Tilde{Q},\mathcal{V},\mathcal{C}}$ diverge.
At these points, there are the second-order phase transitions occuring, manifested from the fact that $\left( \frac{\partial^2F}{\partial T^2}\right)_{\Tilde{Q},\mathcal{V},\mathcal{C}}=-\frac{C_{\Tilde{Q},\mathcal{V},\mathcal{C}}}{T}$, which can be derived via the first relation in Eq.~\eqref{df relation}.

\begin{figure}[t]
    \centering
    \subfigure[]{\includegraphics[width = 3.5cm]{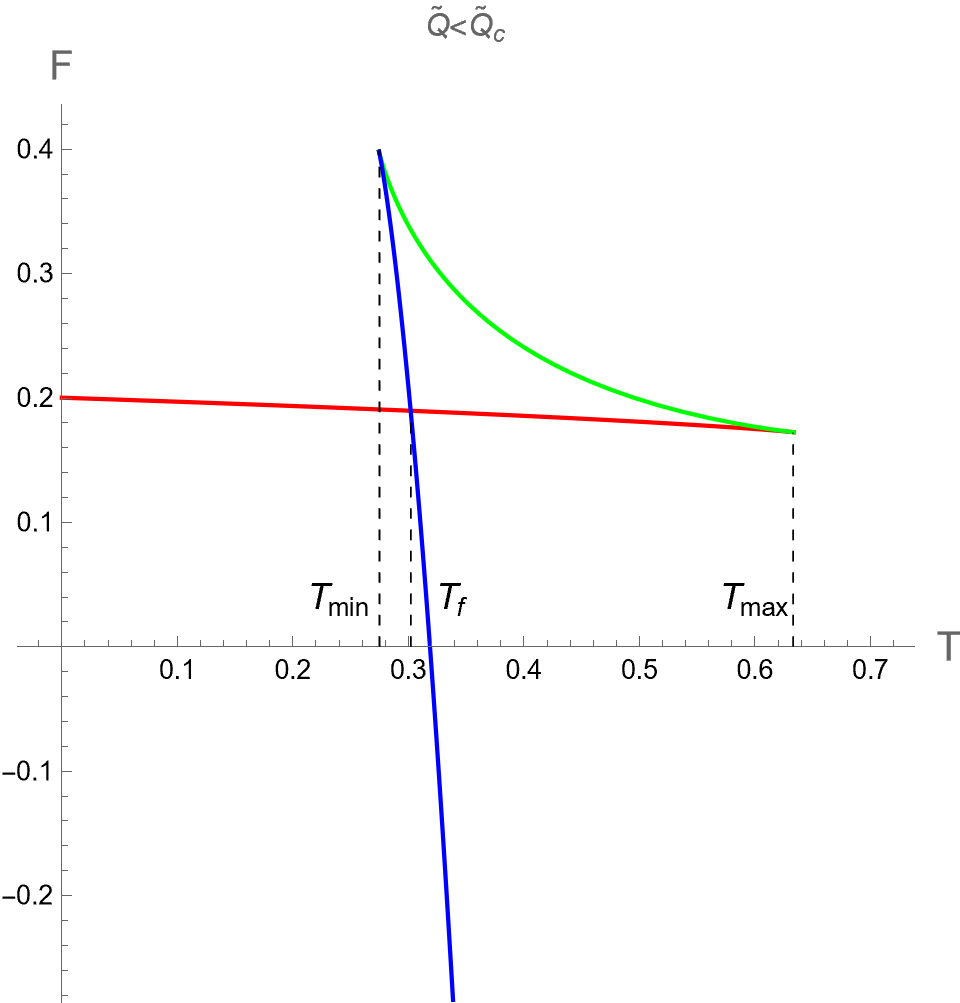}}\hspace{2 cm}
    \subfigure[]{\includegraphics[width = 3.5cm]{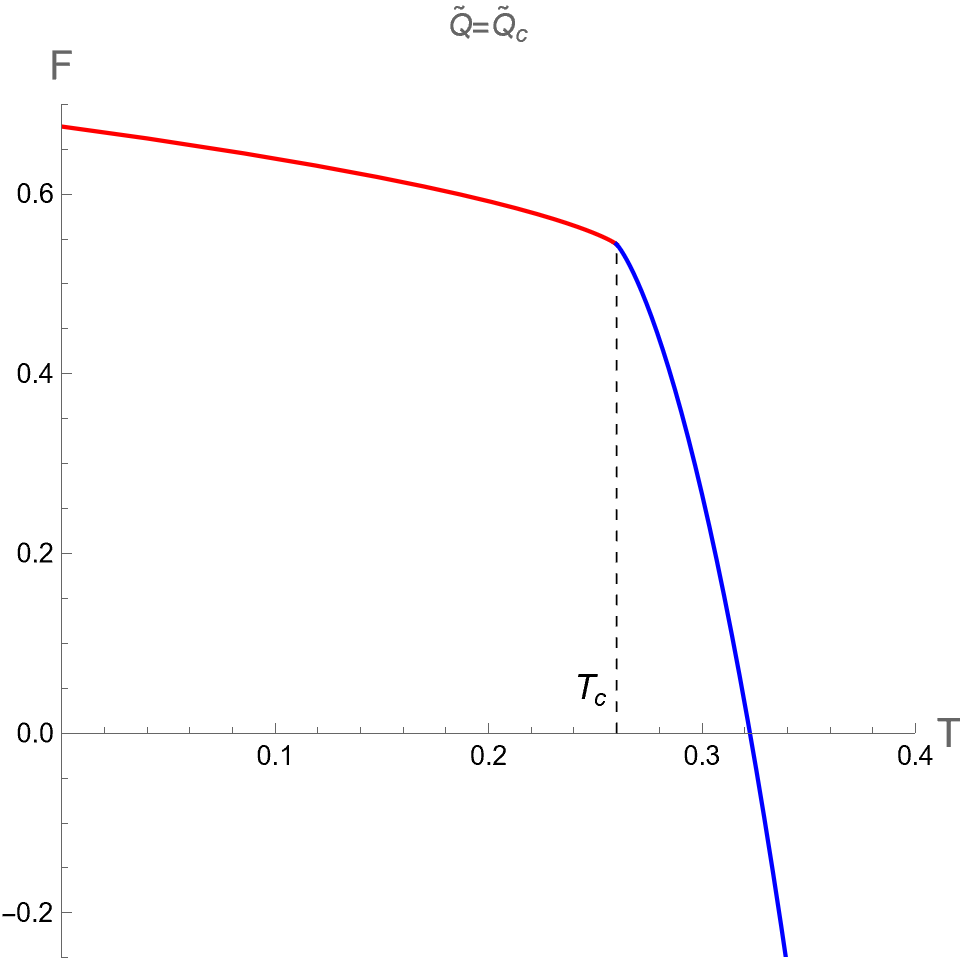}}\hspace{2cm}
    \subfigure[]{\includegraphics[width = 3.5cm]{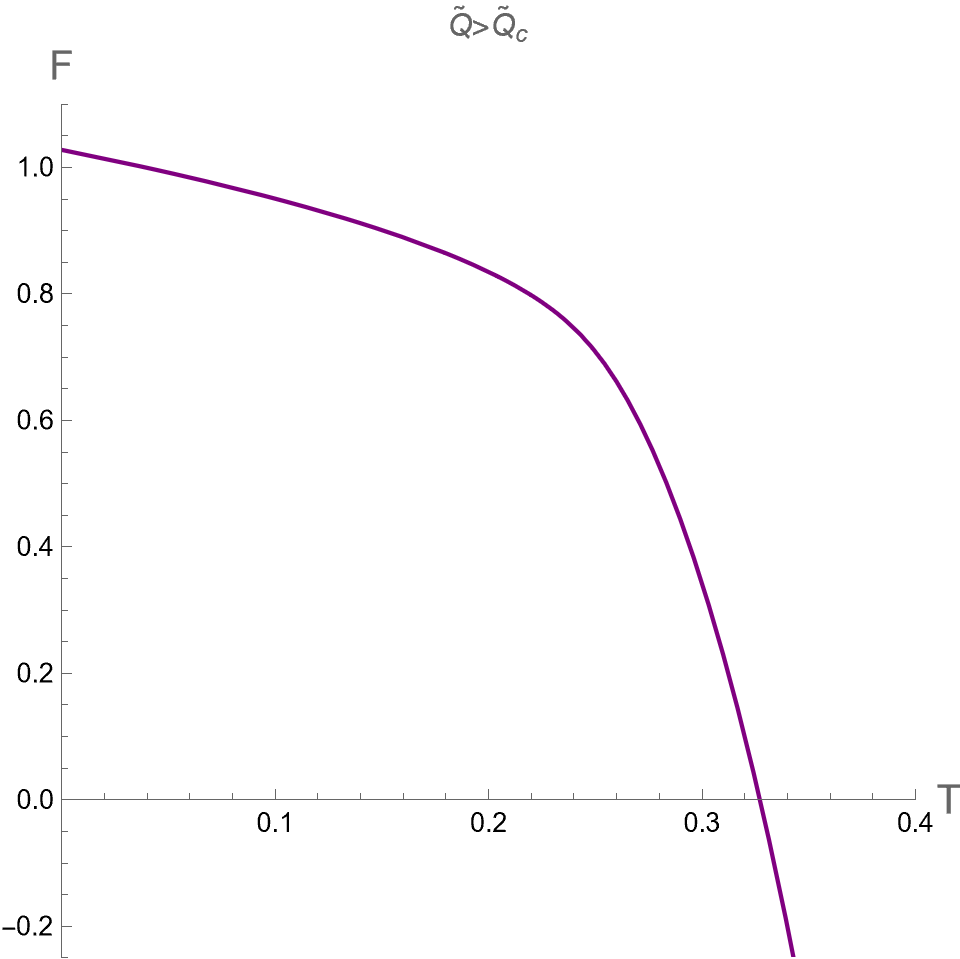}}\\
    \subfigure[]{\includegraphics[width = 3.5cm]{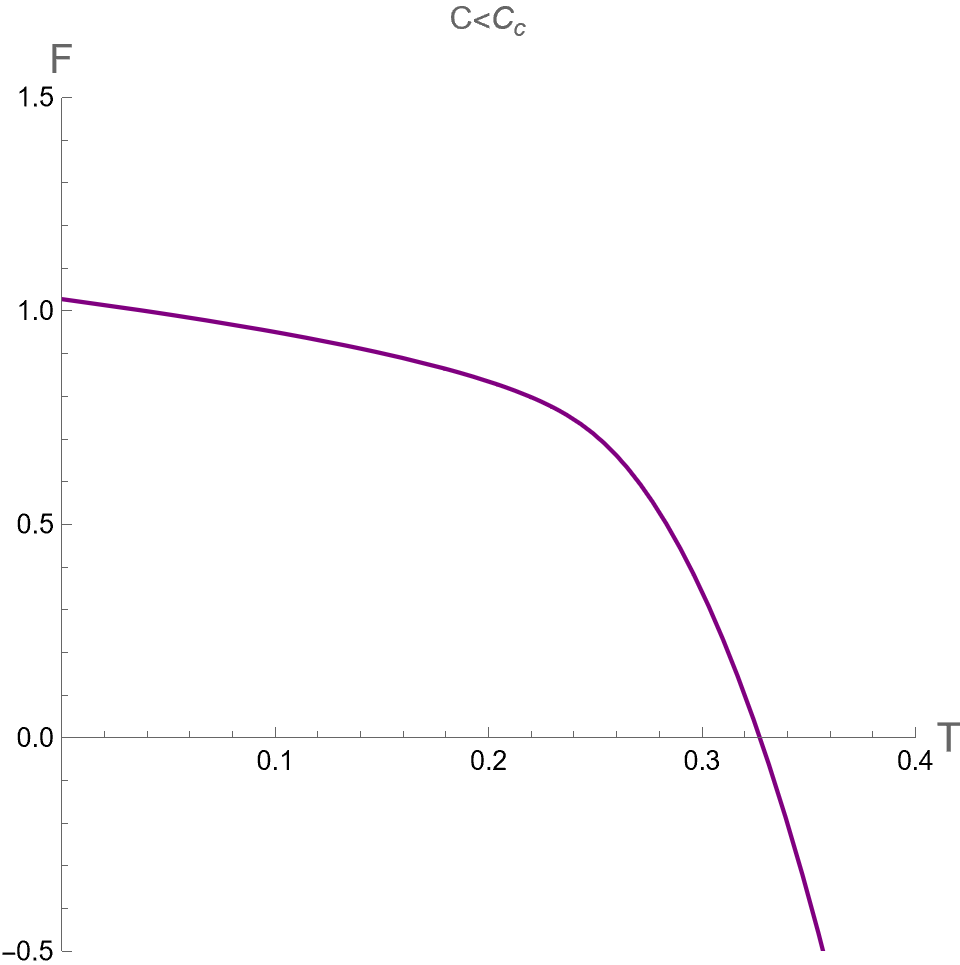}}\hspace{2cm}
    \subfigure[]{\includegraphics[width = 3.5cm]{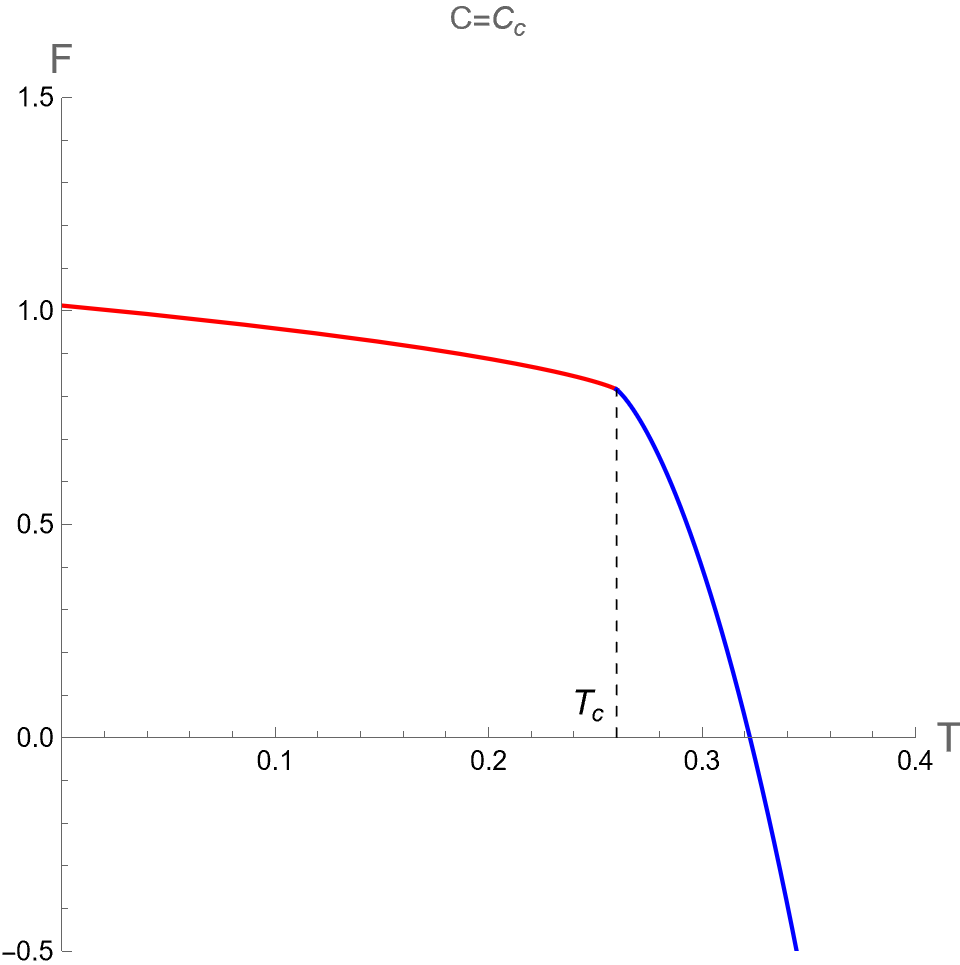}}\hspace{2cm}
    \subfigure[]{\includegraphics[width = 3.5cm]{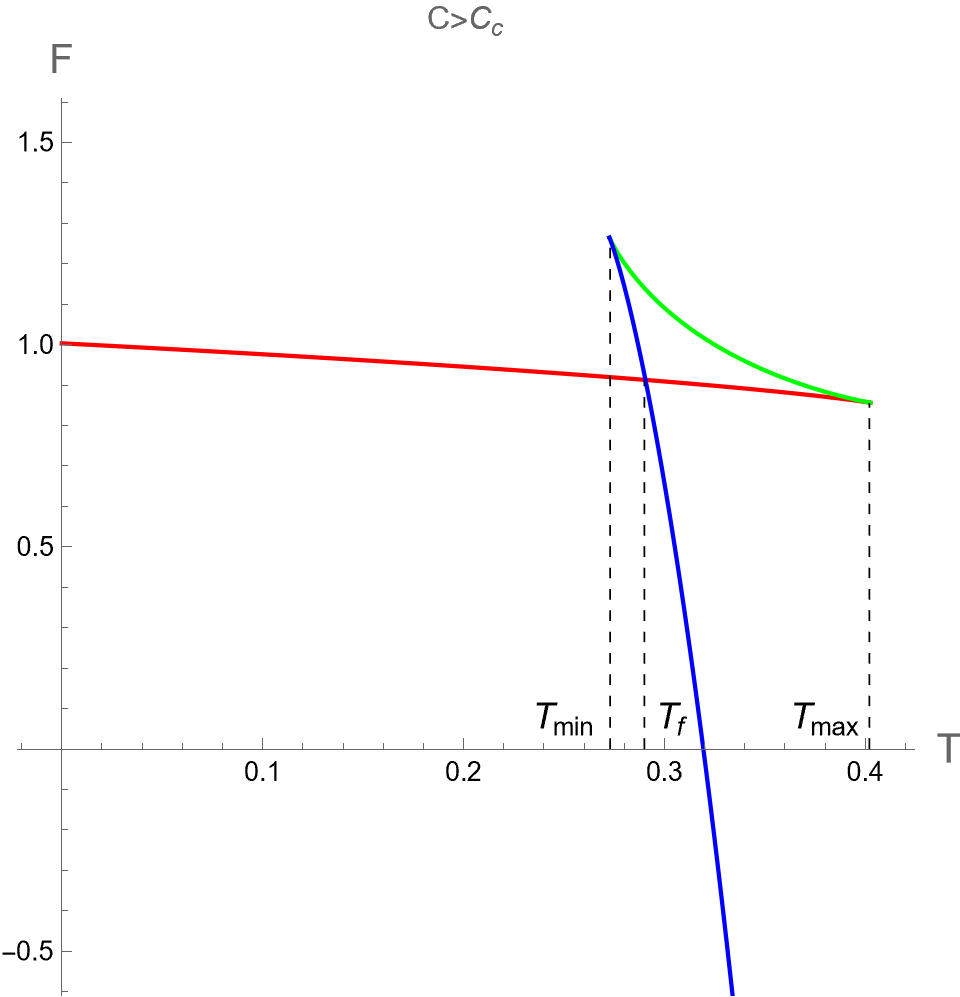}}
    \caption{The behaviors of $F(T)$ curves in the case (I) and (II) are shown in the first row and second rows, respectively.
    The swallowtail behavior of $F(T)$ occurs when $\Tilde{Q}<\Tilde{Q}_c$ and $\mathcal{C}>\mathcal{C}_c$ for case (I) and (II) are shown in figures (a) and (f), respectively, where have three thermal states, namely pCFT1, nCFT and pCFT2, represented in red, green and blue curves, respectively.
    }
    \label{fig: First F vs T fix Q}
\end{figure}

We will discuss about the phase transition of dual CFT via its $F(T)$ curve in more details as follows.
First, let us consider in the case (I).
We depicted the $F(T)$ curves when $\Tilde{Q}<\Tilde{Q}_c$, $\Tilde{Q}=\Tilde{Q}_c$ and $\Tilde{Q}>\Tilde{Q}_c$ in figures (a), (b) and (c) of Fig~\ref{fig: First F vs T fix Q}.
The free energy shows the swallowtail shape for $\Tilde{Q}<\Tilde{Q}_c$.  
In this case, the pCFT1 phase is thermodynamically preferred in low temperatures, i.e. both in the region of  $T<T_\text{min}$ and $T_\text{min}\leq T< T_f$.  There is a first-order phase transition from pCFT1 to pCFT2 occurring at $T_f$.
At slightly higher than $T_f$, the pCFT2 phase becomes the most thermodynamically preferred since it has the lowest free energy.
The swallowtail shape of $F(T)$ curve with $\Tilde{Q}=\Tilde{Q}_c$ shrink to appear as a kink at $T_c$, where the transition between pCFT1 and pCFT2 phases become the second-order type of phase transition. 
For $\Tilde{Q}>\Tilde{Q}_c$, there is only one phase with continuous curves with positive heat capacity for any $T$, thus no phase transition occurs. 

We plot $F$ versus $T$ in the case (II) in Fig~\ref{fig: First F vs T fix Q} (d), (e) and (f) from small to large values of $\mathcal{C}$.
The results indicate that phase structures of dual CFT in case (II) are qualitatively similar to the case (I).
Namely, $F(T)$ displays the swallowtail shape of three phases for $\mathcal{C}>\mathcal{C}_c$, the pCFT1-pCFT2 first-order phase transition occurs at $T=T_f$.
At the critical value of $\mathcal{C}$, the nCFT disappears and the transition between pCFT1 and pCFT2 becomes the second-order phase transition at $T_c$.
For the small values of $\mathcal{C}$, there is only one thermal state of CFT with positive heat capacity at any value of $T$.

\section{Critical Parameters in Photon Ring Region} \label{section 3}

The photon trajectories play a crucial role in determining the images of BH surrounded by emitting matter. Computing such null geodesics has become particularly interesting since the first image of a black hole was published by the Event Horizon Telescope Collaboration~\cite{ETH1,*ETH2,*ETH3,*ETH4,*ETH5,*ETH6,*EventHorizonTelescope:2022wkp, *EventHorizonTelescope:2022apq, *EventHorizonTelescope:2022wok, *EventHorizonTelescope:2022exc, *EventHorizonTelescope:2022urf, *EventHorizonTelescope:2022xqj}. In this section, we will explore null geodesics in the spacetime of charged AdS black holes. This analysis aims to provide a clear understanding of three critical parameters in the photon ring region: the orbital half-period \( \tau \), the angular-Lyapunov exponent \( \lambda_L \), and the temporal-Lyapunov exponent \( \gamma_L \). These parameters play an important role in investigating black hole phase transitions through the BH's optical characteristics.

The Lagrangian of test particle moving in the curved spacetime of the metric in Eq.~\eqref{metric} is given by
\begin{eqnarray}
    2\mathcal{L}=-f(r)\left(\frac{dt}{ds}\right)^2+\frac{1}{f(r)}\left(\frac{dr}{ds}\right)^2+r^2\left(\frac{d\phi}{ds}\right)^2,
\end{eqnarray}
where $s$ denotes the affine parameter for a null geodesic.
Since the spacetime is static and spherical symmetric, so the motion of test particle is confined in a plane that we can choose $\theta =\pi /2$ (equatorial plane) without loss of generality.
The coordinates $t$ and $\phi$ are cyclic coordinates due to the symmetry of the background, resulting in the energy $\omega$ and angular momentum $\ell$ of test particles can take the form
\begin{eqnarray}
    \omega =f(r)\frac{dt}{ds}, \ \ \ \text{and} \ \ \ \ell =r^2\frac{d\phi}{ds}, \label{const motion}
\end{eqnarray}
where they are the constant of motion.
The equation of motion in the radial direction reads
\begin{eqnarray}
    \left(\frac{dr}{ds}\right)^2+V_\text{eff}=\omega^2, \label{eom of r}
\end{eqnarray}
where the effective potential $V_\text{eff}$ is given by
\begin{eqnarray}
    V_\text{eff}=f(r)\left( \frac{\ell^2}{r^2}+\delta_1\right).
\end{eqnarray}
Note that $\delta_1=0$ and $1$ correspond to null-like and time-like geodesics, respectively.
In the following formulas, we only consider the case of photon trajectories, i.e., $\delta_1=0$. 
The unstable circular orbit can be determined by
\begin{eqnarray}
    V_\text{eff}(r_u)=\omega^2, \ \ \ \text{and} \ \ \ V^{\prime}_\text{eff}(r_u)=0, \label{ru con}
\end{eqnarray}
where $r_u$ is the \textit{radius of unstable photon sphere}.
Using the first condition in \eqref{ru con} together with \eqref{eom of r}, we obtain
\begin{eqnarray}
    b_c^2=\left( \frac{\ell}{\omega}\right)^2=\frac{r_u^2}{f(r_u)}, \label{critimpact}
\end{eqnarray}
where $b_c$ denotes a critical impact parameter.
Solving the second condition in Eq.~\eqref{ru con}, we obtain
\begin{eqnarray}
    r_u&=&\frac{1}{4}\left( 3m\pm \sqrt{9m^2-32q^2}\right), \nonumber \\
    &=&\frac{1}{4}\left[ 3r_+\left(1+\frac{r_+^2}{L^2}+\frac{q^2}{r_+^2}\right)\pm \sqrt{9r_+^2\left(1+\frac{r_+^2}{L^2}+\frac{q^2}{r_+^2}\right)^2-32q^2}\right] \label{unstable ph radius}
\end{eqnarray}
which is independent of angular momentum $\ell$.

\subsection{Orbital half-period}

The incoming photons with impact parameter $b$ larger than and close to $b_c$ can wind several times around the BH before scattering back to the asymptotic region of spacetime.
This effect can cause the light to take different paths around the BH, leading to the formation of multiple images of the source as seen by the distant observer.

Remarkably, the travel time before photons arrive at the observer screen along their null geodesics in different images is not generally equal because of the strong gravity caused by BH.
This phenomenon is called the gravitational time delay, which could be measured via the gravitational lensing observation.

It is found that the time delay between successive images in the photon ring region is governed by the time-lapse $\tau$ over each half-orbit of a photon sphere~\cite{PhysRevD.101.044031,Bozza_2004,Hsieh_2021}. 
The half-time period $\tau$ depends only on the metric spacetime.
In other words, $\tau$ turns out to be independent of the distance of the light source from observers.
This implies that $\tau$ is the parameter that is appropriate to encode the important feature about the region of spacetime near the event horizon. 

To obtain the orbital half-period $\tau$ of the bound photon orbit, we can start by calculating the angular velocity of incoming photons $\Omega =d\phi /dt$ measured by a distant observer.
Using Eq.~\eqref{const motion}, we obtain
\begin{eqnarray}
    \frac{d\phi}{ds} &=& \frac{\ell}{r^2}, \nonumber \\
    \frac{dt}{ds}\frac{d\phi}{dt}&=&\frac{\ell}{r^2}, \nonumber \\
    \Omega &=&b\frac{f(r)}{r^2},
\end{eqnarray}
where we have used $b=\ell /\omega$.
At the photon sphere radius $r_u$, the angular velocity $\Omega$ is given by
\begin{eqnarray}
    \Omega_u =b_c\frac{f(r_u)}{r_u^2}=\frac{\sqrt{f(r_u)}}{r_u}.
\end{eqnarray}
Since the time lapse over a full orbit is $T=2\pi /\Omega$, therefore the orbital half-period $\tau$ reads
\begin{eqnarray}
    \tau =\frac{T}{2}=\frac{\pi r_u}{\sqrt{f(r_u)}}. \label{half period}
\end{eqnarray}

\subsection{Angular-Lyapunov exponent}

As evident from the fact that $V^{\prime}_\text{eff}(r_u)=0$ and $V^{\prime \prime}_\text{eff}(r_u)<0$, the orbiting photons at $r_u$ are unstable.
The Lyapunov exponents measure the sensitivity of the system to changes in initial conditions, which in this work refers to slight alterations in the photon trajectories near \( r_u \) from one to a neighboring one.
In this subsection and the following one, we investigate the sub-rings structure of photon trajectories in the region very close to the unstable photon sphere by using two Lyapunov exponents, namely the angular-Lyapunov exponent $\lambda_L$ and temporal-Lyapunov exponent $\gamma_L$~\cite{cardoso2009geodesic,deich2024lyapunov}.
Studying BH's images of some given emitting light sources indicates that the ratio of photon flux received between adjacent sub-rings is determined by $\lambda_L$~\cite{Johnson:2019ljv,Broderick:2023jfl,deich2024lyapunov}. 

To obtain $\lambda_L$, we linearize Eq.~\eqref{eom of r} near the unstable shell of photon sphere in $\delta r=r-r_u$.
The effective potential becomes $V_\text{eff}(r)=V_\text{eff}(r_u+\delta r)$.
Consequently, one finds that
\begin{eqnarray}
    \frac{d \delta r}{ds}=\sqrt{\omega^2-V_\text{eff}(r_u+\delta r)},
\end{eqnarray}
Expanding $V_\text{eff}$ around $r_u$ and using two conditions in Eq.~\eqref{ru con}, we obtain
\begin{eqnarray}
    \frac{d \delta r}{ds}=\sqrt{-\frac{1}{2}V^{\prime \prime} _\text{eff}(r_u)}\delta r. \label{delta r}
\end{eqnarray}
Applying the second relation of Eq.~\eqref{const motion} into Eq.~\eqref{delta r}, which leads to the formula describing the angular $\phi$ dependence of the deviation $\delta r$ from $r_u$ as follows:
\begin{eqnarray}
    \pi \frac{d~\delta r}{d\phi}= \lambda_L ~\delta r,
\end{eqnarray}
where we have defined the angular Lyapunov exponent as
\begin{eqnarray}
    \lambda_L \equiv \pi r_u^2\sqrt{-\frac{1}{2}V_\text{eff}^{\prime \prime}(r_u)}. \label{ang Lya}
\end{eqnarray}
The angular dependence of the deviation $\delta r(\phi)$ is given by
\begin{eqnarray}
    \delta r(\phi)= \delta r_0e^{\lambda_L \phi /\pi}, \label{perturb phi}
\end{eqnarray}
where $\delta r_0$ is the initial deviation of a geodesic from the critical circular orbit.

\subsection{Temporal-Lyapunov exponent}

Gravitational waves emitted by BHs during the ringdown phase are characterized by QNMs. 
However, determining the QNMs from the field perturbation of BH can be accomplished by examining the unstable circular geodesics of photons. 
Specifically, the temporal-Lyapunov exponent $\gamma_L$ is related to the imaginary part of the complex quasinormal frequency. Namely, we have $1/\gamma_L$ as the timescale for the instability of the ringdown amplitude. 
Consequently, $\gamma_L$ associated with the photon ring region serves as an important parameter for studying BHs under various scenarios.

We calculate $\gamma_L$, which represents the deviation rate in time $t$ of photon trajectory from the unstable photon sphere.
With the first relation of Eq.~\eqref{const motion}, we obtain the rate from Eq.~\eqref{delta r} as follows:
\begin{eqnarray}
    \frac{d \delta r}{dt}= \gamma_L \delta r, 
\end{eqnarray}
where we have defined the temporal-Lyapunov exponent
\begin{eqnarray}
    \gamma_L \equiv \sqrt{-\frac{r_u^2f(r_u)}{2\ell^2}V^{\prime \prime} _\text{eff}(r_u)}. \label{time Lya}
\end{eqnarray}
Thus, the time evolution of $\delta r(t)$ is
\begin{eqnarray}
    \delta r(t)=\delta r_0 e^{\gamma_Lt},
\end{eqnarray}
where $\delta r_0$ is the initial deviation of a geodesic from the photon sphere.

\section{Probing Black Hole Phase Transitions through Optical Features} \label{section 4}

In this section, we will investigate BHs undergoing thermal phase transitions by exploring three critical parameters involving the optical features of BHs introduced in the previous section. As discussed in the Introduction, section~\ref{Intro}, it is worthwhile to investigate thermal phase transitions through optical probing in the extended phase space approach and holographic thermodynamics. As will be shown later in this section, the results from both approaches indicate that these parameters, treated as functions of either \(P\) or \(T\), can express discontinuities as well as behaviors of multi-valued functions. This is in a similar way to exploring thermodynamic phase transitions of conventional substances by observing abrupt changes in their free energy and response functions as \(P\) or \(T\) are varied. For convenience, we use \(\mathcal{O}_i\) where \(i = 1, 2, 3\) to represent the three critical parameters: \(\mathcal{O}_1\), \(\mathcal{O}_2\), and \(\mathcal{O}_3\) represent \(\tau\), \(\lambda_L\), and \(\gamma_L\), respectively.  These three critical parameters can potentially serve as order parameters in the consideration of phase transitions, as we will discuss later in Section~\ref{sec 5_orderparameter}.

\subsection{Probing the phase transition in the extended phase space approach}

The thermal profiles of $\mathcal{O}_i$ in the extended phase space
can be considered as a function of either $P$ or $T$ separately, i.e., $\mathcal{O}_i(P)$ or $\mathcal{O}_i(T)$.
This is because $r_u$ in Eq.~\eqref{unstable ph radius} can be written as $r_u=r_u(r_+,P)$ and $r_u=r_u(r_+,T)$ as follows:
\begin{eqnarray}
    r_u(r_+,P)&=&\frac{1}{4} \left[3 \left(\frac{8}{3} \pi  P r_+^3+\frac{q^2}{r_+}+r_+\right)+\sqrt{9 \left(\frac{8}{3} \pi  P r_+^3+\frac{q^2}{r_+}+r_+\right)^2-32 q^2} \,\right],  \label{r_u_P} \\
    r_u(r_+,T)&=&\frac{q^2}{r_+}+\frac{1}{2}\left[ r_+ +2\pi r_+^2T+\sqrt{r_+^2(1+2\pi r_+T)^2+(8\pi r_+T-4)q^2+\frac{4q^4}{r_+^2}} \, \right]. \label{r_u_T}
\end{eqnarray}
\begin{figure}[t]
    \centering
    \includegraphics[width = 4.cm]{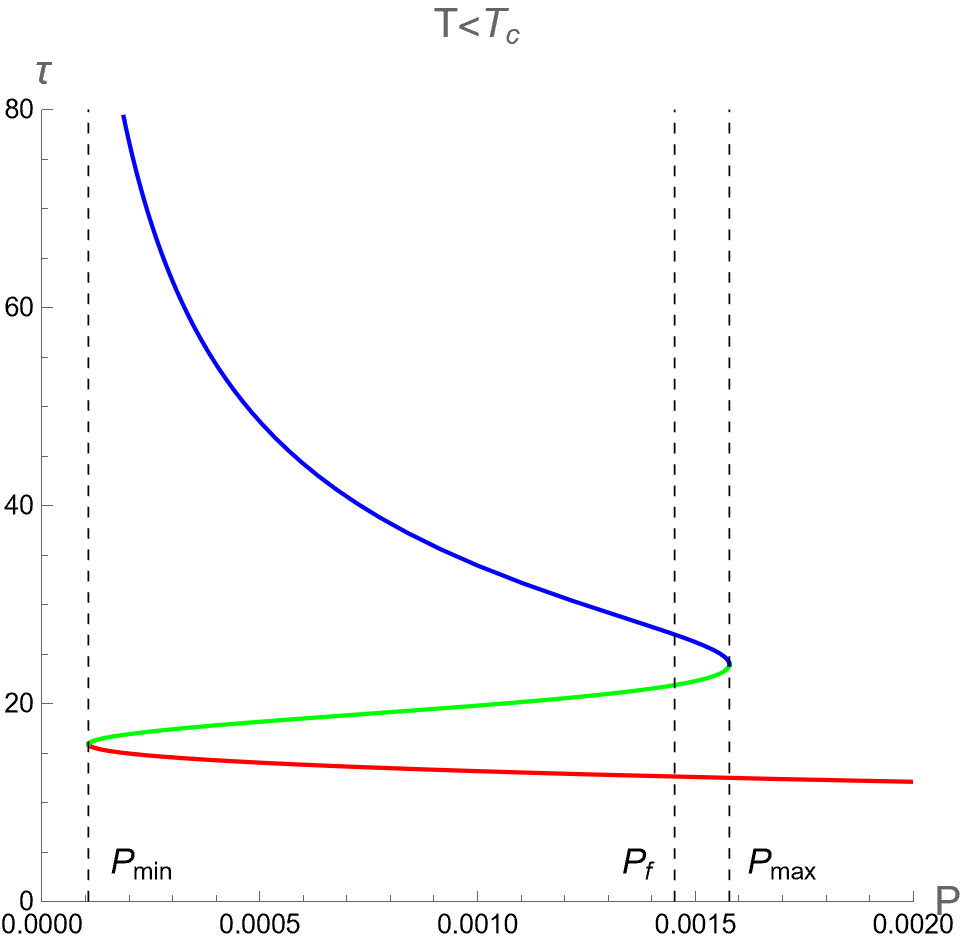}\hspace{1cm}
    \includegraphics[width = 4.cm]{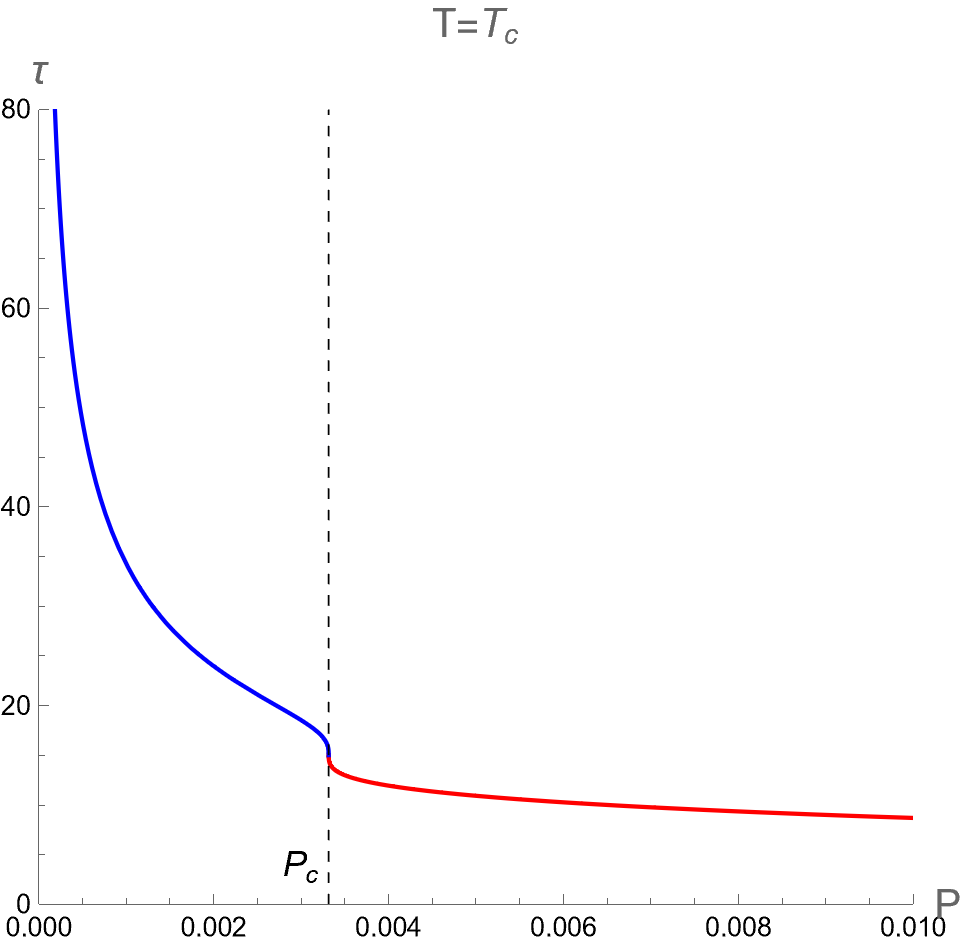}\hspace{1cm}
    \includegraphics[width = 4.cm]{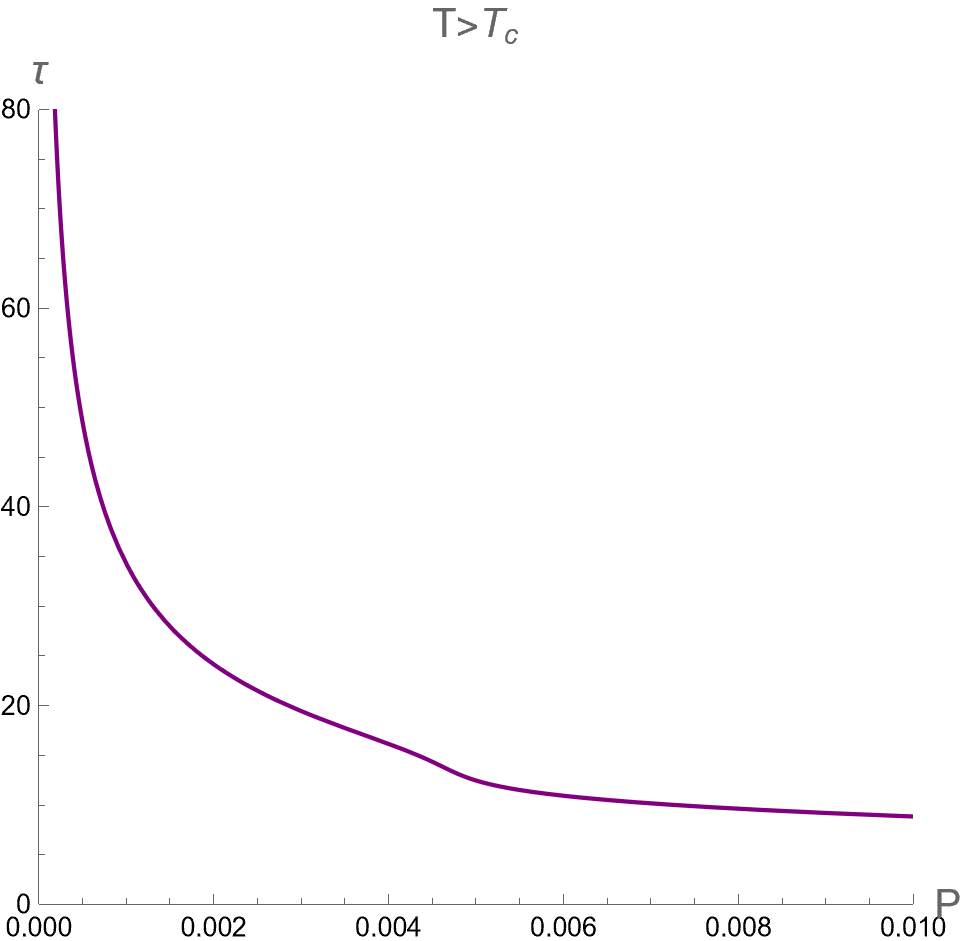}\\ 
    \includegraphics[width = 4.cm]{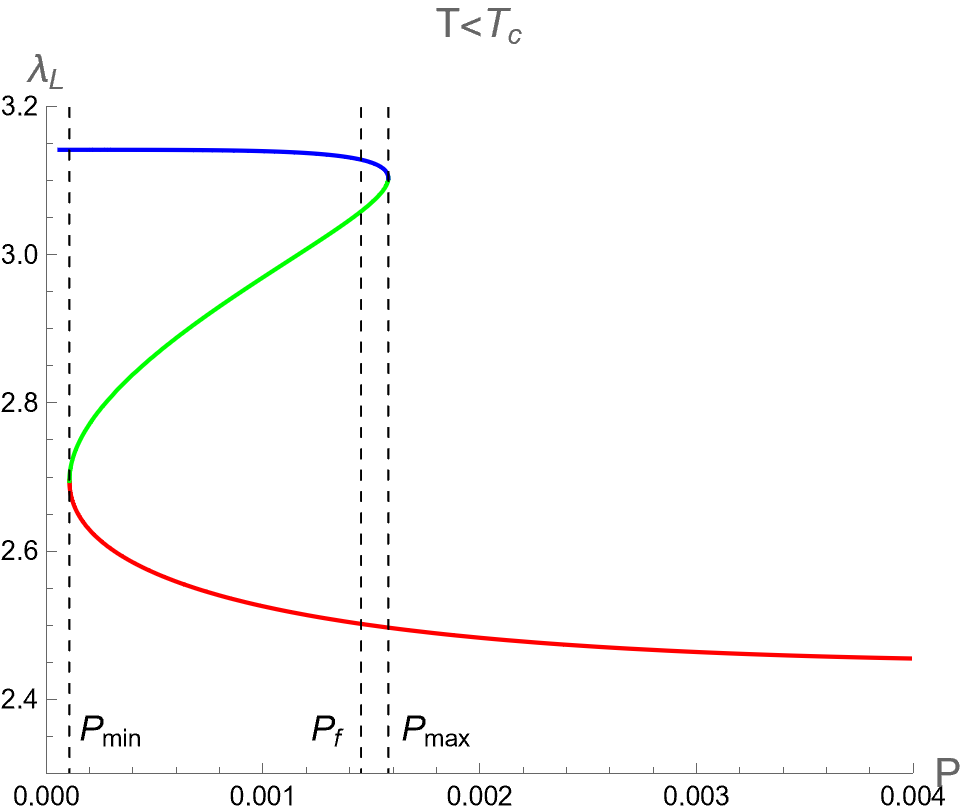}\hspace{1cm}
    \includegraphics[width = 4.cm]{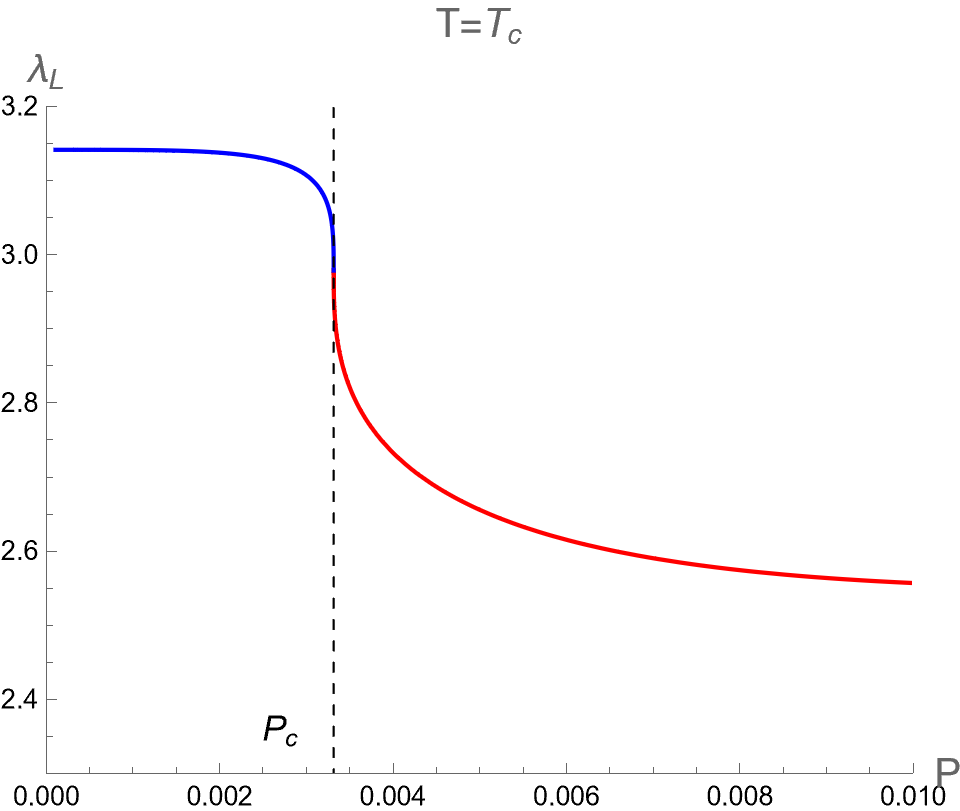}\hspace{1cm}
    \includegraphics[width = 4.cm]{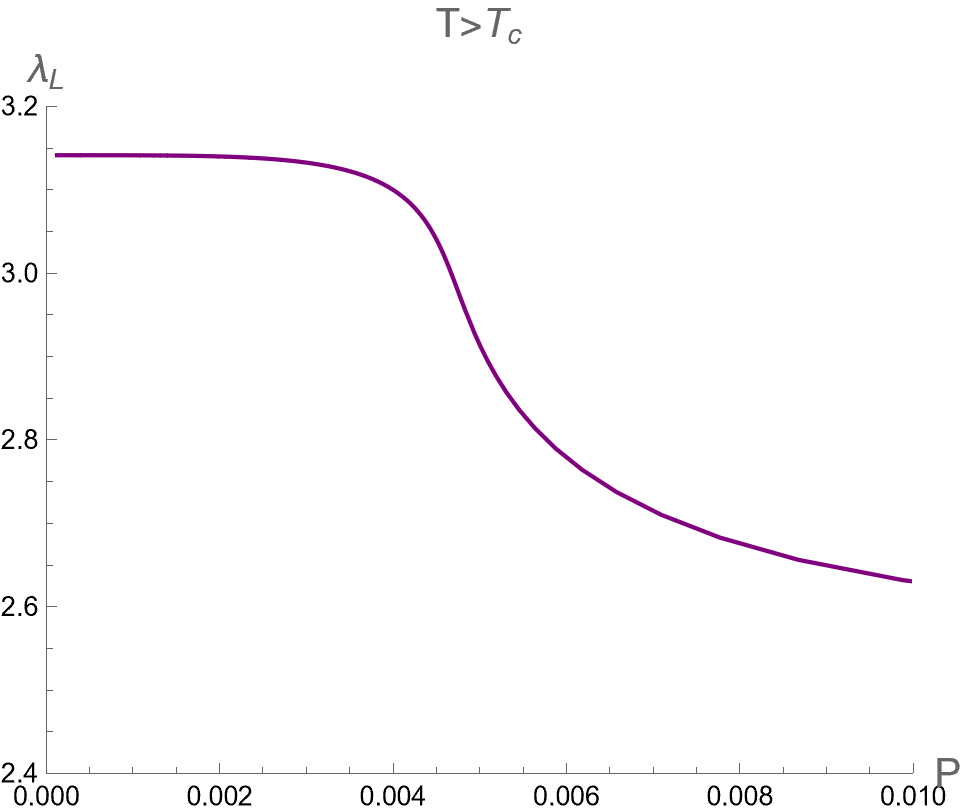}\\
    \includegraphics[width = 4.cm]{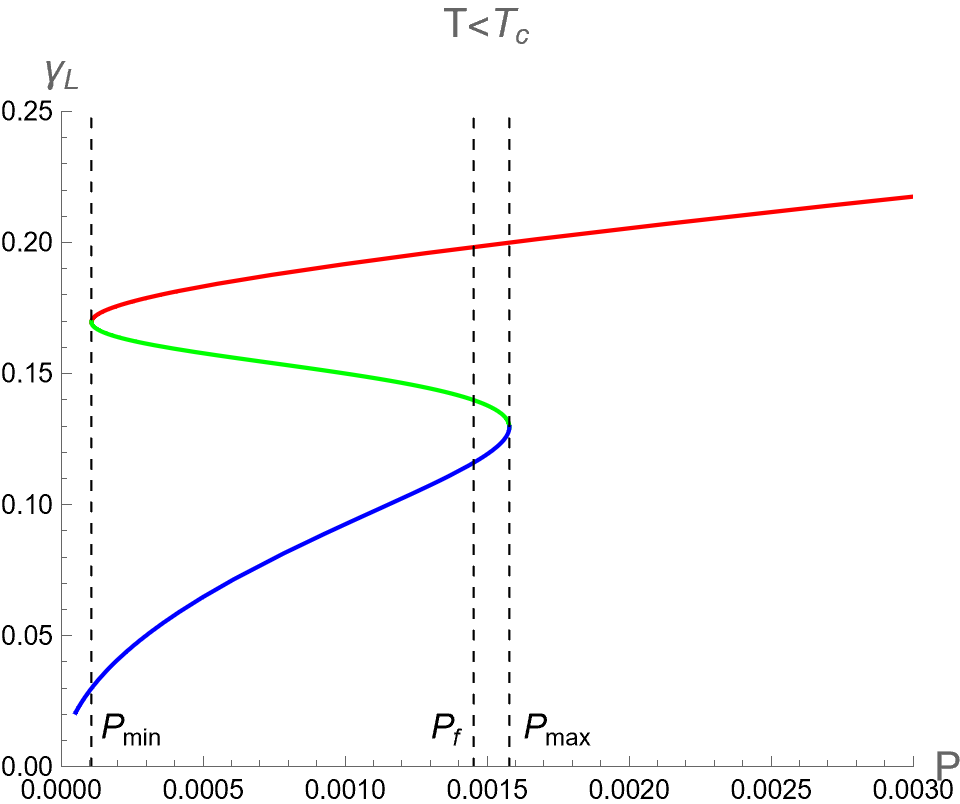}\hspace{1cm}
    \includegraphics[width = 4.cm]{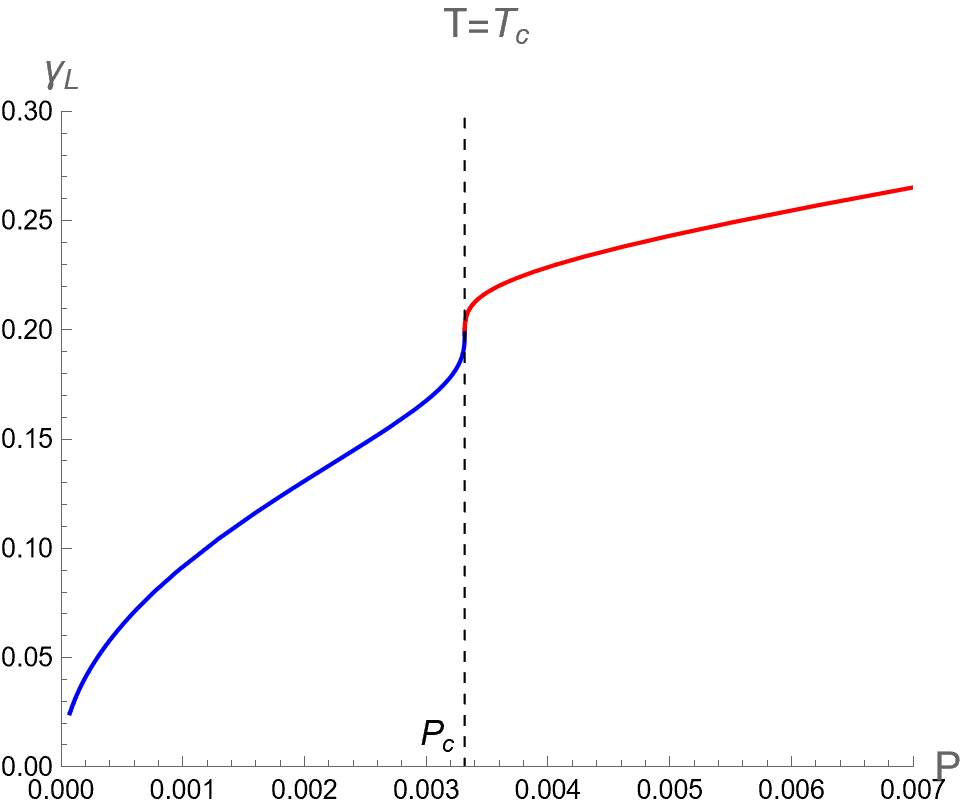}\hspace{1cm}
    \includegraphics[width = 4.cm]{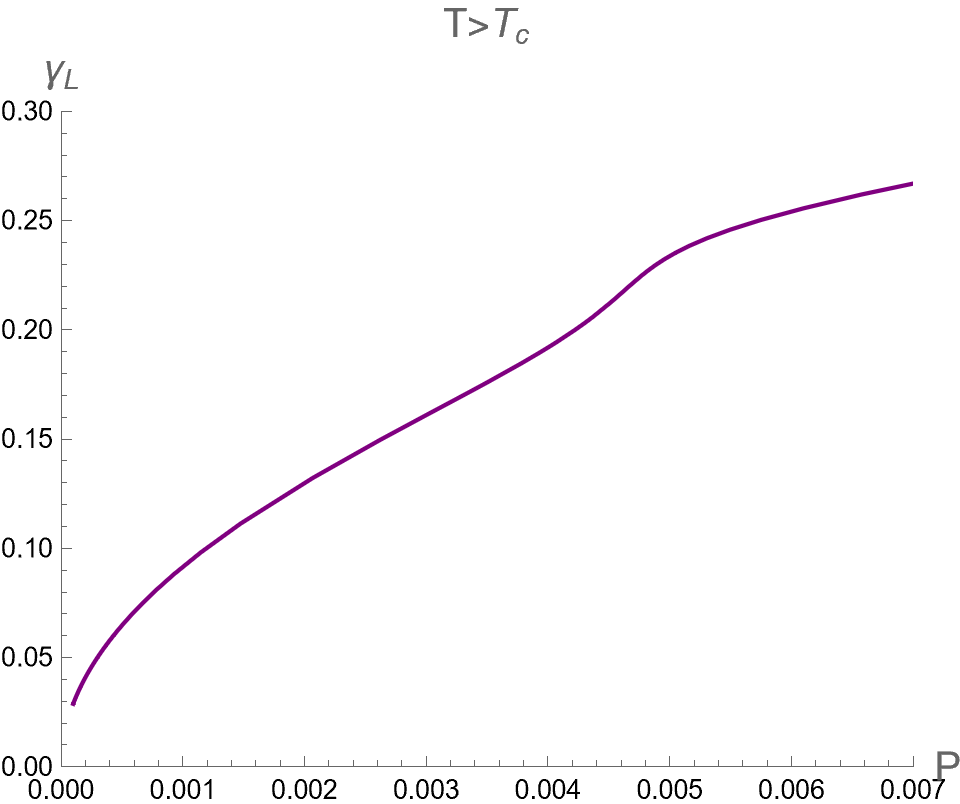}
    \caption{Isotherm curves of $\tau$, $\lambda_L$ and $\gamma_L$ as a function of $P$ are shown in the first, second and third rows, respectively. In the left, middle and right columns correspond to the temperature $T<T_c$, $T=T_c$ and $T>T_c$,  respectively.} 
    \label{fig: Optic vs P}
\end{figure}
In the context of the extended phase space, the present paper focuses on the effect of $P$ on three critical parameters during BH phase transitions in the isothermal process.
Namely, we analyze the discontinuities and multi-value behavior of $\mathcal{O}_i$ as functions of $P$ while keeping $T$ constant.
By substituting $r_u(r_+,T)$ as described in Eq.~\eqref{r_u_T} into Eqs.~\eqref{half period}, \eqref{ang Lya} and \eqref{time Lya}, one can use $r_+$ as a parameter running in the parametric plot as the isotherm curves, i.e. $\mathcal{O}_i$ versus the bulk pressure $P$. Note that the bulk pressure $P$ has been described in Eq.~\eqref{eqf1}.
The isotherm curves of $\tau ,\lambda_L$ and $\gamma_L$, as a function of $P$, are illustrated in the first, second and third rows of Fig.~\ref{fig: Optic vs P}, respectively.
The left, middle and right columns of Fig.~\ref{fig: Optic vs P} depict the cases with $T<T_c, T=T_c$ and $T>T_c$, respectively.

When $T<T_c$, the three critical parameters exhibit a multi-value function within the range $P_\text{min}\leq P\leq P_\text{max}$, as shown in the left column of Fig.~\ref{fig: Optic vs P}.
In other words, at the values of $P$ in this range, there are three values for each $\mathcal{O}_i$ corresponding with the Small (red), Intermediate (green) and Large-BHs (blue) branches. 
Outside this range of pressure, there exists only one branch, namely the Large-BH branch for $P<P_\text{min}$ and the Small-BH branch for $P>P_\text{max}$.
Interestingly, the isotherm curves of three critical parameters at $T<T_c$ expressed itself as a multi-value function has the boundary at $P_\text{min}$ and $P_\text{max}$, which correspond to the positions of the cusps of  $G$ (see Fig.~\ref{fig: F vs P}) with the divergent value of $\kappa_T$ (see Fig.~\ref{fig: P vs rh}).
Therefore, considering the parameters $\mathcal{O}_i$ can exhibit the second-order phase transition of charged AdS-BH from observing the divergence of their derivative with respect to $P$.
Remarkably, when a Small-Large BHs first-order phase transition occurs at $P=P_f$, the values of $\mathcal{O}_i$ will discontinuously change between Small and Large BH branches due to discontinuity jumping of event horizon radius caused by the Maxwell equal area law.
This is an important result that will be applied in our study about the $\mathcal{O}_i-P$ criticality, as discussed in the next section.

As the temperature $T$ is larger, the pressures $P_\text{min}$ and $P_\text{max}$
converge and degenerate to be $P_c$ at $T=T_c$, namely the Intermediate-BH phase becomes absent, as shown in the middle colums of Fig.~\ref{fig: Optic vs P}.
The values of three critical parameters at the critical point of the second-order phase transition between Small (red) and Large (blue) BHs can be expressed as follows
\begin{eqnarray}
    \tau_c=3\pi \sqrt{\frac{12 \sqrt{6}+29}{23}}, \ \ \ \lambda_c=\frac{2 \pi \ell}{\sqrt{2+\sqrt{6}}}, \ \ \ \gamma_c=\frac{2}{3} \sqrt{65-53 \sqrt{\frac{3}{2}}}.
\end{eqnarray}
Note that we use $q=1$. For $T$ is larger than $T_c$, it turns out that only one BH solution exists and no phase transition occurs.

\subsection{Probing the phase transition in holographic thermodynamics}

In this subsection, we probe phase structures of charged AdS-BH in the holographic thermodynamics approach via three critical parameters, i.e. $\tau$, $\lambda$ and $\gamma$.
The expression for the photon sphere radius $r_u$ in Eq.~\eqref{unstable ph radius} can be written in term of the bulk parameters $x$ and $y$, as defined in Eq.~\eqref{xy}, in the form
\begin{eqnarray}
    r_u=\frac{L}{4}\left[ 3x\left( 1+x^2+\frac{y^2}{x^2}\right)+\sqrt{9x^2\left( 1+x^2+\frac{y^2}{x^2}\right)^2-32y^2}\right].
\end{eqnarray}
To investigate the influence of $\Tilde{Q}$ and $\mathcal{C}$ of the boundary CFT on the behaviors of null geodesics around critical photon orbits in bulk spacetime, $y$ can be substituted using Eq.~\eqref{Q vs y} into the above equation in terms of $\Tilde{Q}$ and $\mathcal{C}$ as following
\begin{eqnarray}
    r_u=\frac{L}{4}\left[ 3x\left( 1+x^2+\frac{1}{16\mathcal{C}^2}\frac{\Tilde{Q}^2}{x^2}\right)+\sqrt{9x^2\left( 1+x^2+\frac{1}{16\mathcal{C}^2}\frac{\Tilde{Q}^2}{x^2}\right)^2-\frac{2\Tilde{Q}^2}{\mathcal{C}^2}}\right]. \label{photon sphere holo}
\end{eqnarray}
Since the $p-\mathcal{V}$ criticality is absent in the boundary thermodynamics and no $p\mathcal{V}$ term appears in the Smarr formula, we focus on the behavior of $\mathcal{O}_i$ as a function of $T$ instead $p$ for the holographic thermodynamics.
Substituting $r_u$ into Eqs.~\eqref{half period}, \eqref{ang Lya} and \eqref{time Lya}, one can express $\tau$, $\lambda_L$ and $\gamma_L$ versus $T$ for study the influence of $\Tilde{Q}$ and $\mathcal{C}$ on  $\mathcal{O}_i(T)$ in the case I and case II as discussed before, respectively.

\begin{figure}[t]
    \centering
    \includegraphics[width = 4.cm]{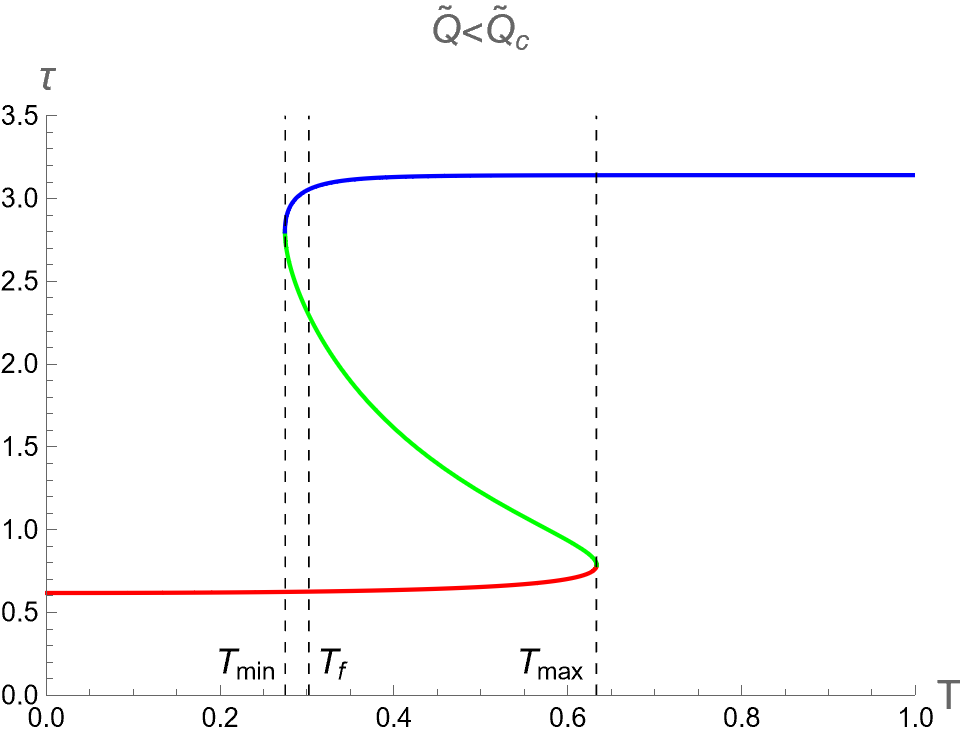}\hspace{1cm}
    \includegraphics[width = 4.cm]{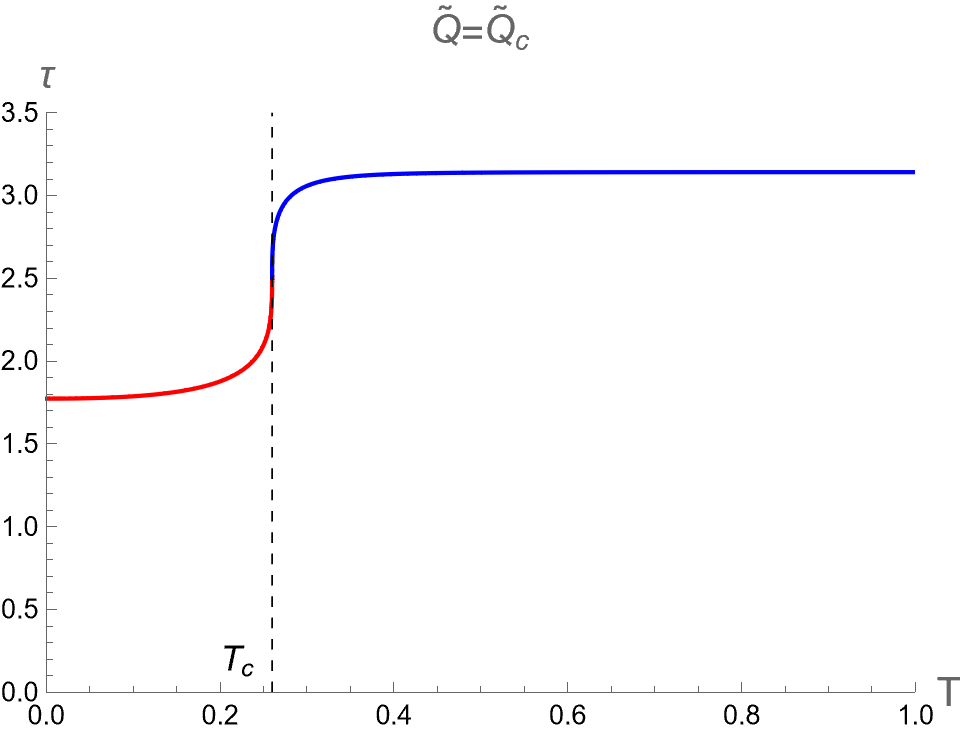}\hspace{1cm}
    \includegraphics[width = 4.cm]{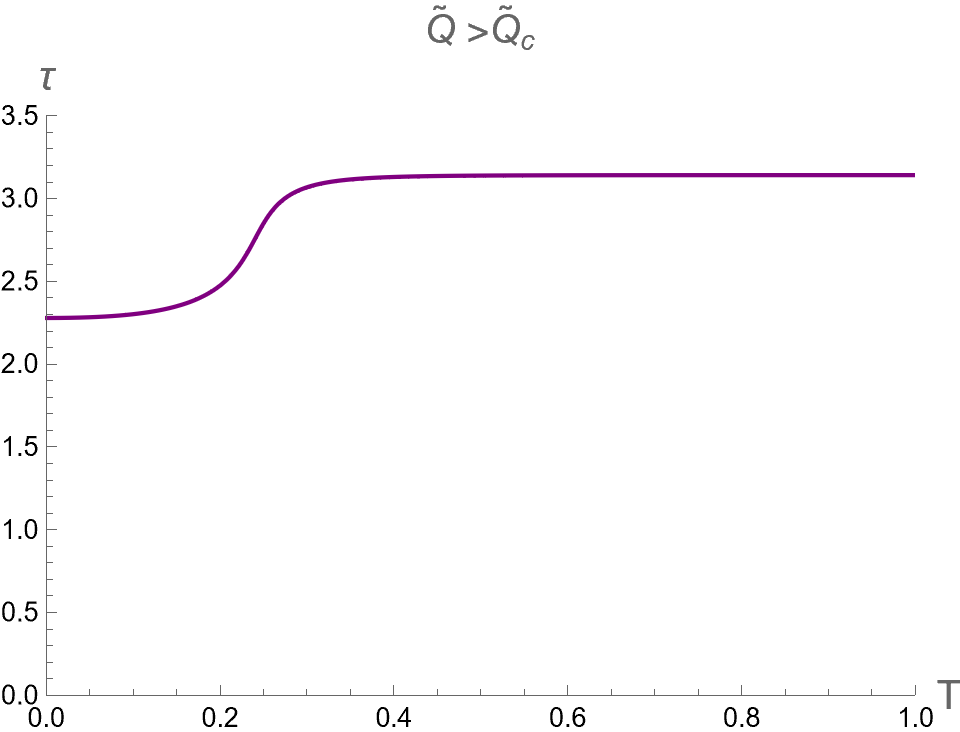}\\
    \includegraphics[width = 4.cm]{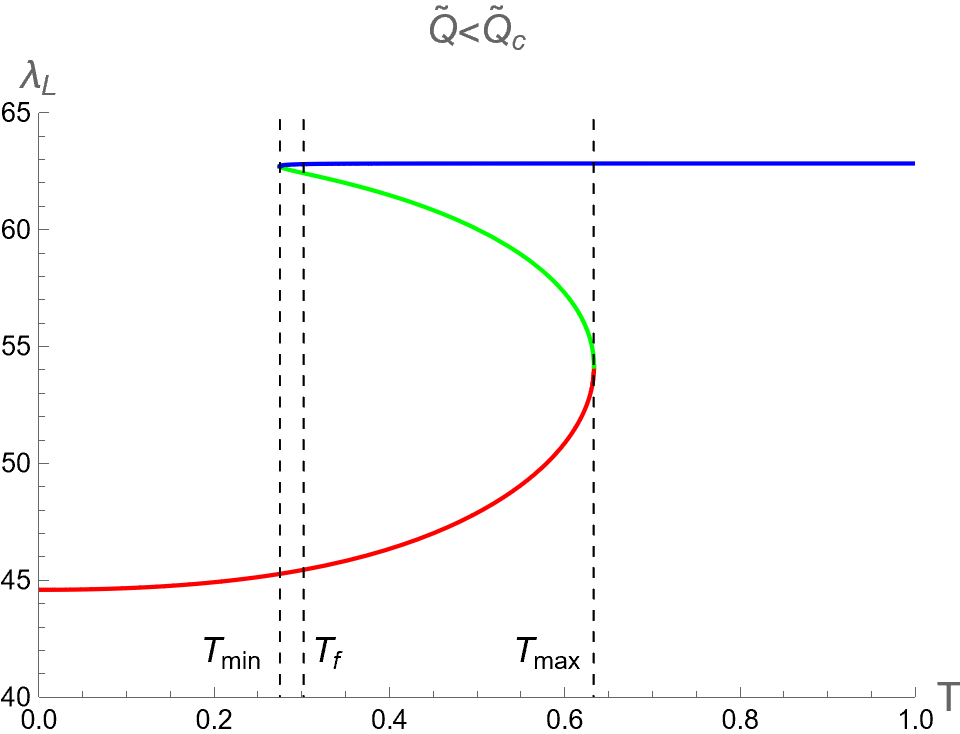}\hspace{1cm}
    \includegraphics[width = 4.cm]{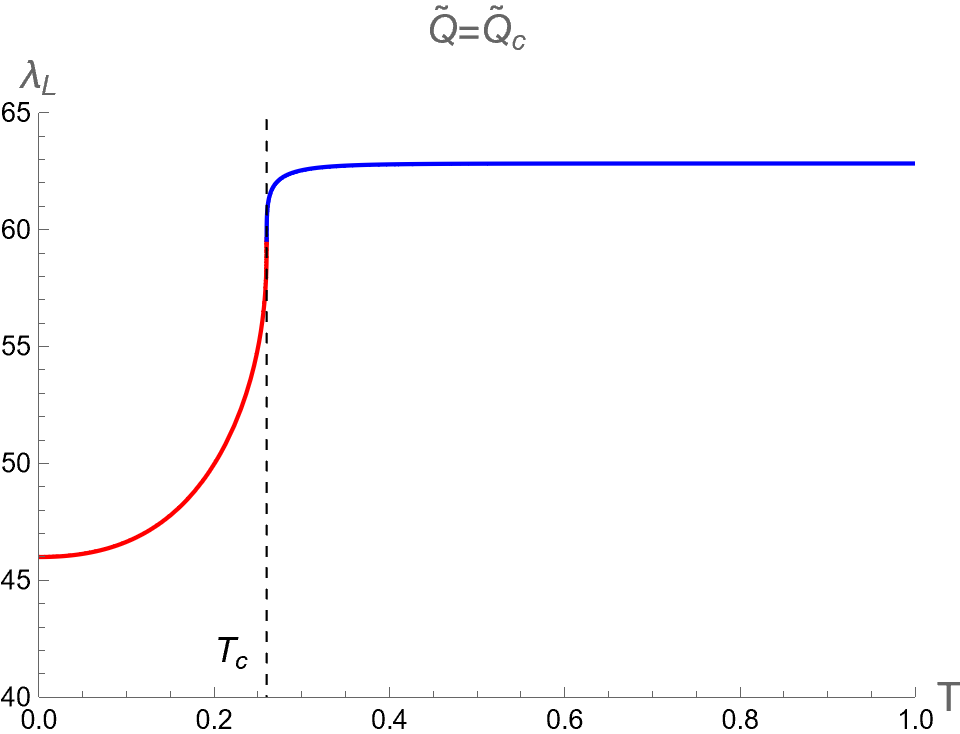}\hspace{1cm}
    \includegraphics[width = 4.cm]{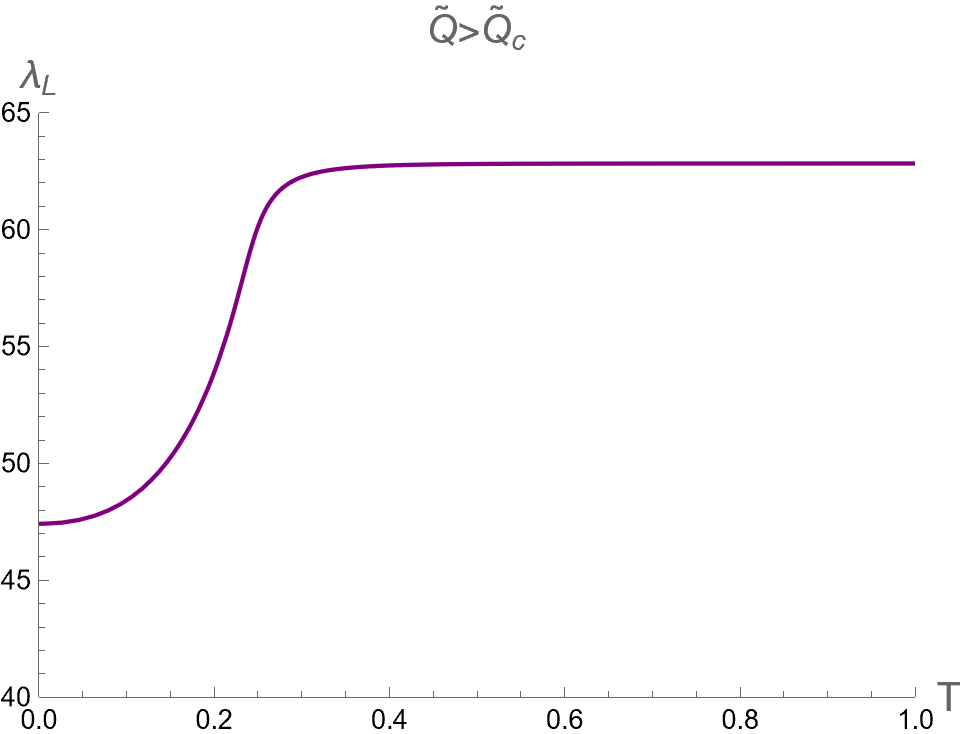}\\
    \includegraphics[width = 4.cm]{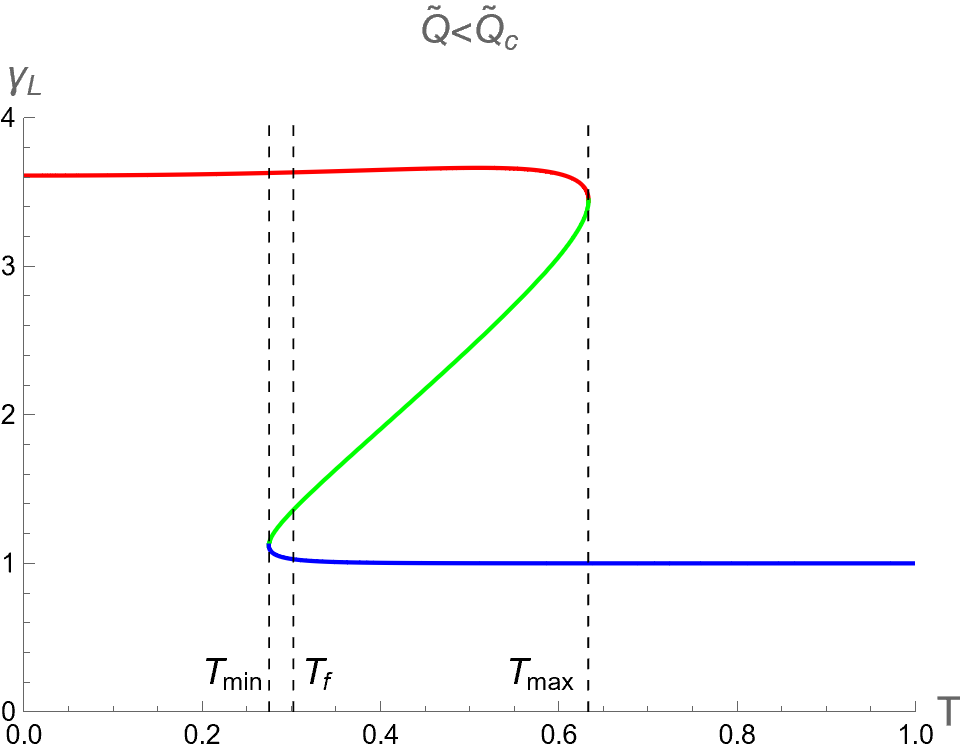}\hspace{1cm}
    \includegraphics[width = 4.cm]{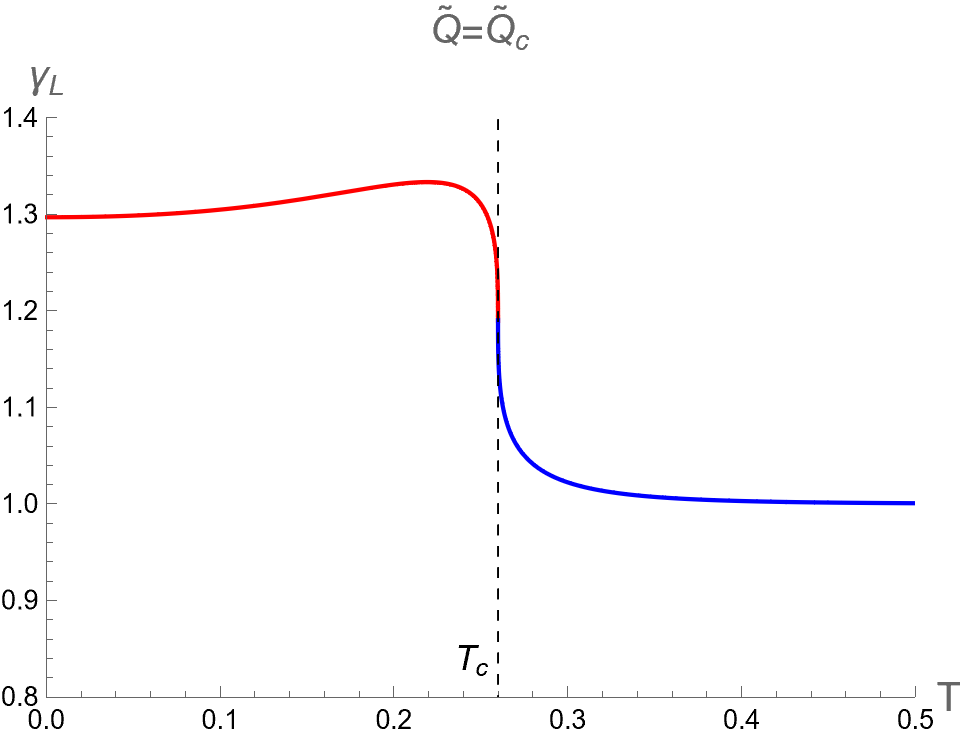}\hspace{1cm}
    \includegraphics[width = 4.cm]{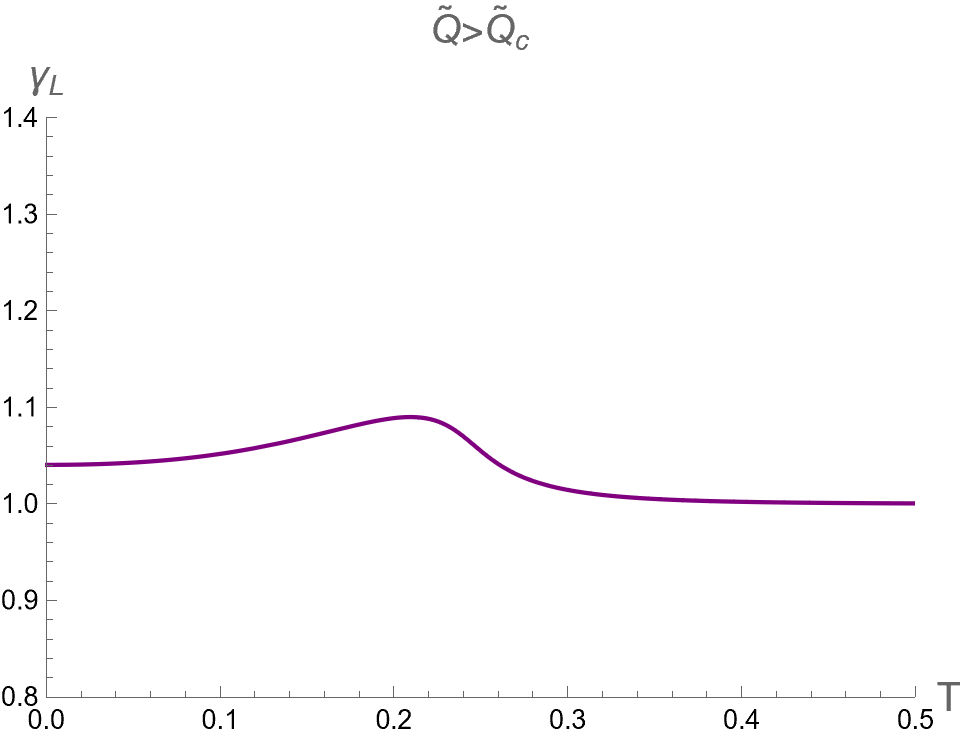}
    \caption{Isocharge curves of $\tau$, $\lambda_L$ and $\gamma_L$ as a function of $T$ are shown in the first, second and third rows, respectively. In the left, middle and right columns correspond to the charge $\Tilde{Q}<\Tilde{Q}_c$, $\Tilde{Q}=\Tilde{Q}_c$ and $\Tilde{Q}>\Tilde{Q}_c$,  respectively.} 
    \label{fig: Photon ring vs T fix Q massless}
\end{figure}

In the case I, the behaviors of $\mathcal{O}_i$ versus $T$ for $\Tilde{Q}<\Tilde{Q}_c$, $\Tilde{Q}=\Tilde{Q}_c$ and $\Tilde{Q}>\Tilde{Q}_c$ are illustrated in the left, middle and right columns in Fig.~\ref{fig: Photon ring vs T fix Q massless}, respectively.
From the graph of $F(T)$ with $\Tilde{Q}<\Tilde{Q}_c$ for dual CFT (see the figure (a) of Fig~\ref{fig: First F vs T fix Q}) reveals that three phases of CFT, namely pCFT1, nCFT and pCFT2 coexist when $T_\text{min}<T<T_\text{max}$ implying that the Small, Intermediate and Large-BH phases also coexist in this range of $T$ in the gravity picture.
We find that the parameters $\mathcal{O}_i$ associated with the critical curve in $T_\text{min}<T<T_\text{max}$ exhibits a multi-value function, i.e. three phases of BH have three different values of $\mathcal{O}_i$, as shown in the left column in Fig.~\ref{fig: Photon ring vs T fix Q massless}.
Outside this range of $T$, there exists only one phase, namely the Small-BH phase for $T<T_\text{min}$ and the Large-BH phase for $T>T_\text{max}$.
Moreover, the slope of isocharge curves in the $\mathcal{O}_i-T$ plane diverges at $T_\text{min}$ and $T_\text{max}$, which correspond to the positions of cusp of $F$ with the divergent of $C_{\Tilde{Q},\mathcal{V},\mathcal{C}}$.
Remarkably, when pCFT1-pCFT2 first-order phase transition occur at $T=T_f$, the values of $\mathcal{O}_i$ will discontinuously change between Small-BH and Large-BH phases due to discontinuos jumping of entropy. This can be exhibited by the Maxwell equal area law as illustrated in Appendix \ref{App B}.
This remarkable results from the present study could reveal the behaviors of $\mathcal{O}_i-T$ criticality, as will be discussed later in the next section.

The middle column in Fig.~\ref{fig: Photon ring vs T fix Q massless} illustrates the case of $\Tilde{Q}=\Tilde{Q}_c$. 
Notably, $\mathcal{O}_i$ becomes a single-valued function, and Intermediate-BH phase disappears. 
Two distinct configuration of BHs, i.e. Small and Large-BHs, exist without coexistence. 
Furthermore, Small-Large BHs second-order phase transition takes place at $T_c$, where the critical values of $\tau$, $\lambda_L$ and $\gamma_L$  can be expressed as follows:
\begin{eqnarray}
    \tau_c=\pi L\sqrt{\frac{29+12\sqrt{6}}{92}}, \ \ \ \lambda_c=\frac{2\pi \ell}{\sqrt{2+\sqrt{6}}},  \ \ \ \gamma_c=\frac{4}{L}\sqrt{65-53\sqrt{\frac{3}{2}}}.\label{parameter at crit}
\end{eqnarray}

In the right column of Fig.~\ref{fig: Photon ring vs T fix Q massless}, we examine the scenario where $\Tilde{Q}>\Tilde{Q}_c$. 
In this case, the BH is characterized by a single phase (purple curve).

\begin{figure}[H]
    \centering
    \includegraphics[width = 4.cm]{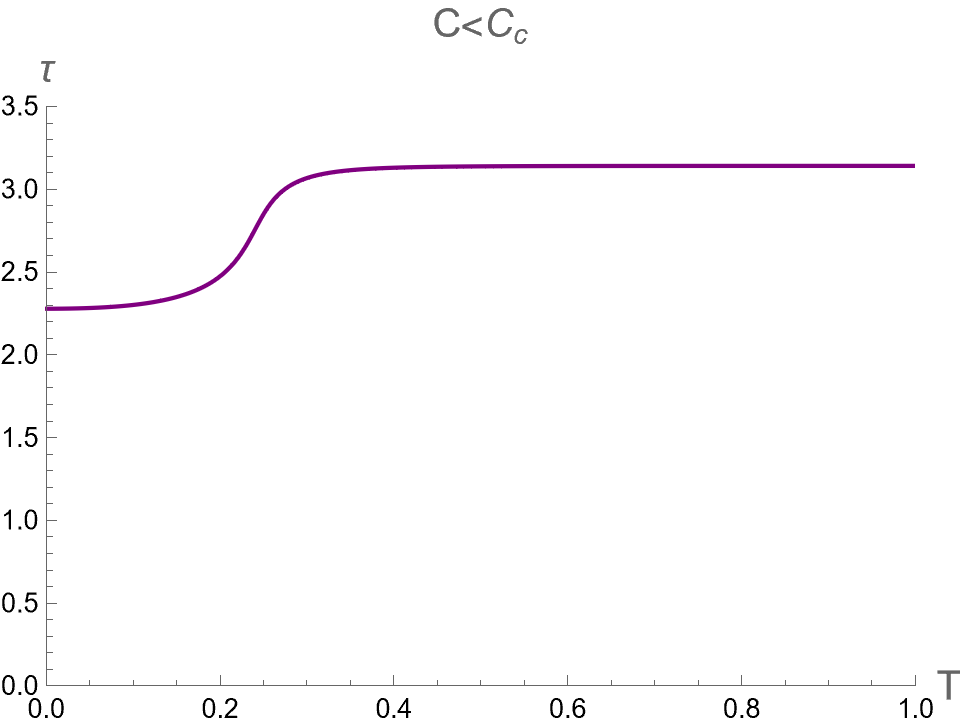}\hspace{1cm}
    \includegraphics[width = 4.cm]{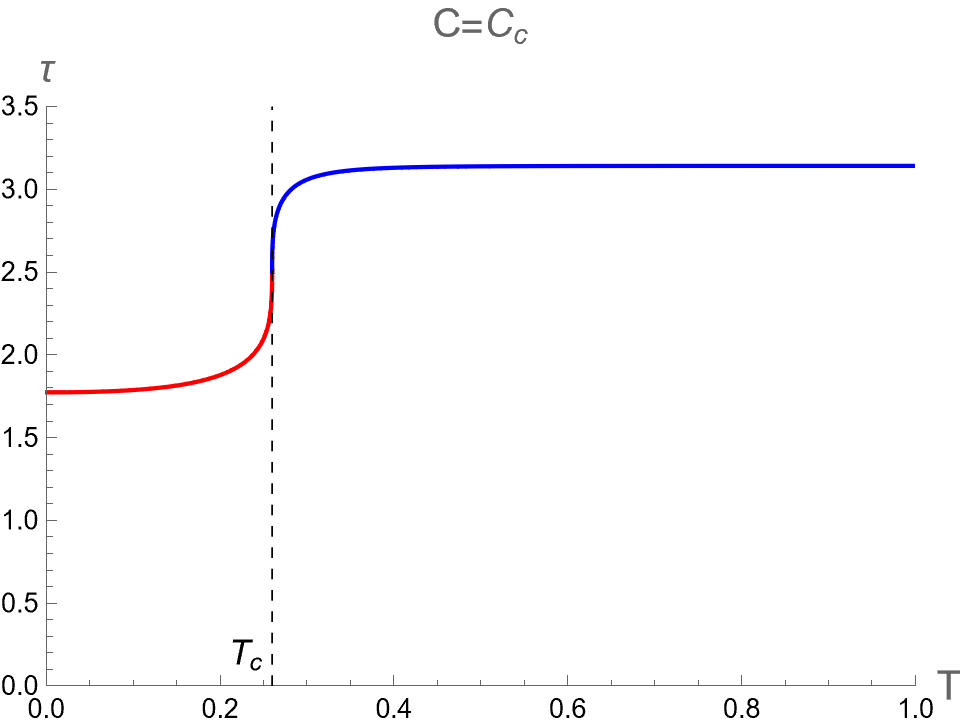}\hspace{1cm}
    \includegraphics[width = 4.cm]{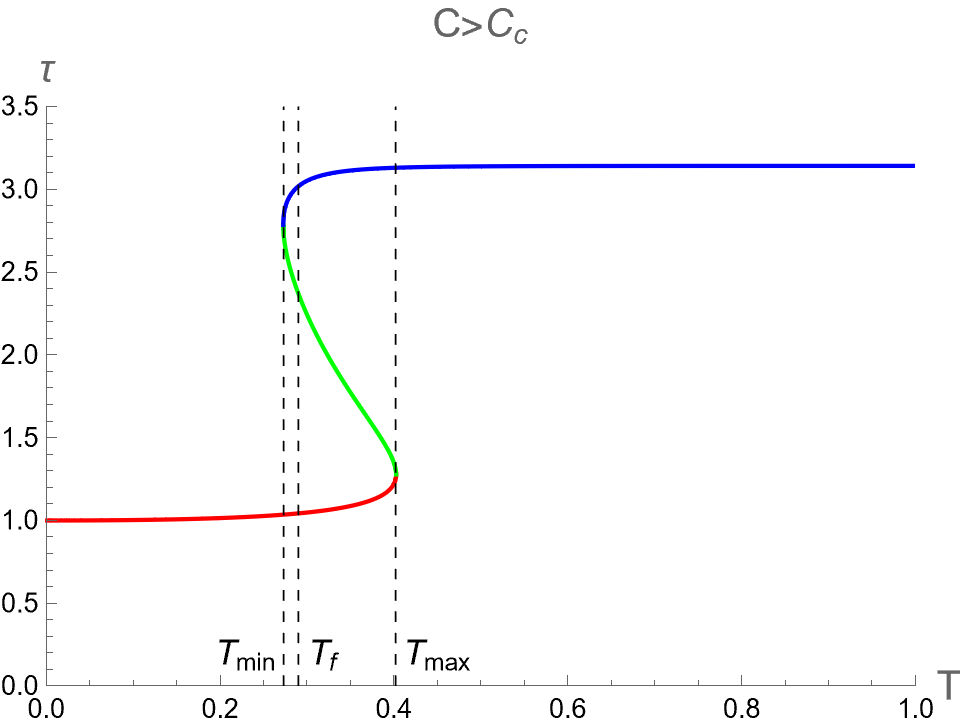}\\
    \includegraphics[width = 4.cm]
    {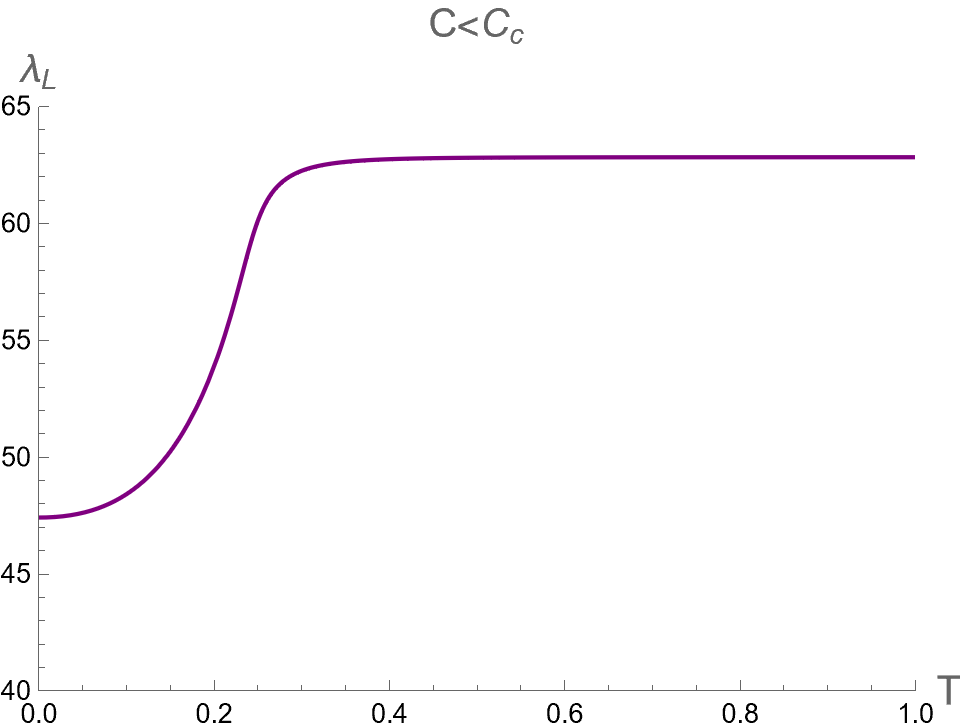}\hspace{1cm}
    \includegraphics[width = 4.cm]{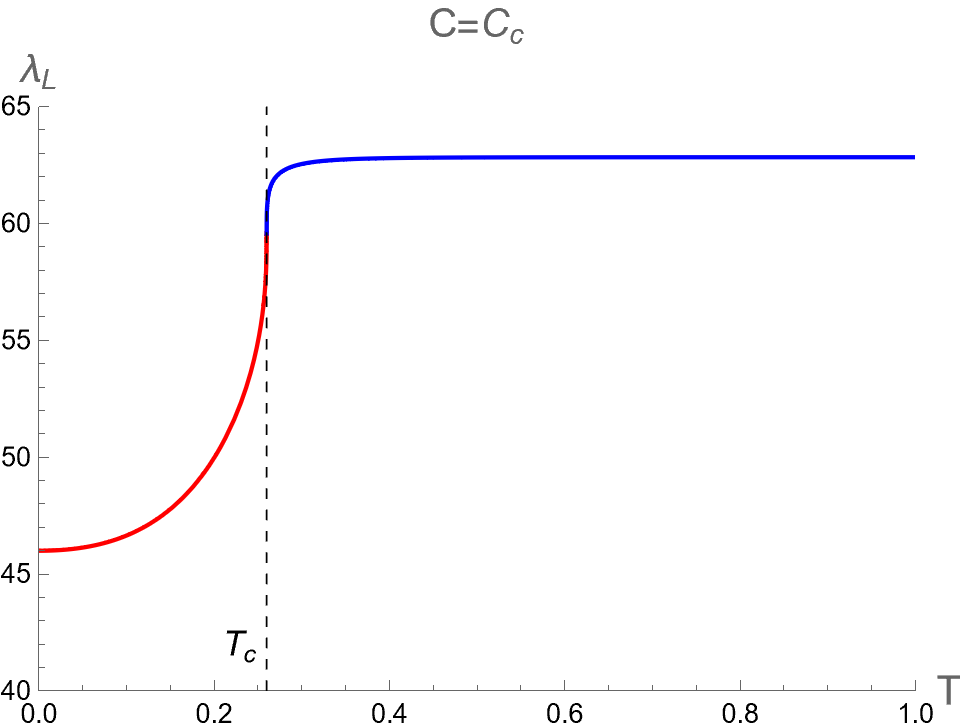}\hspace{1cm}
    \includegraphics[width = 4.cm]{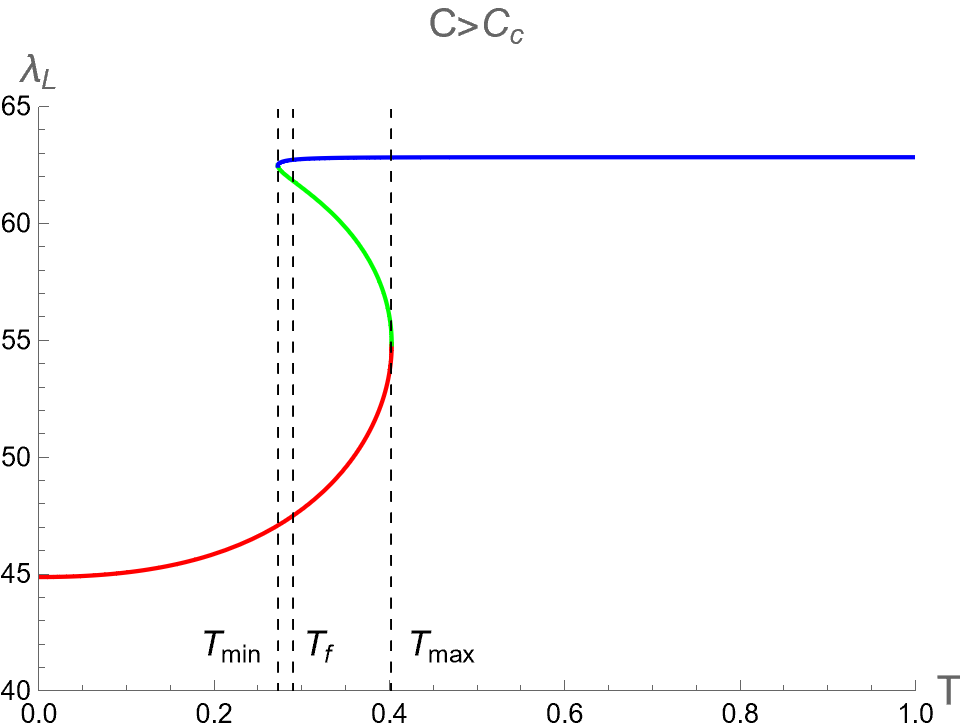}\\
    \includegraphics[width = 4.cm]{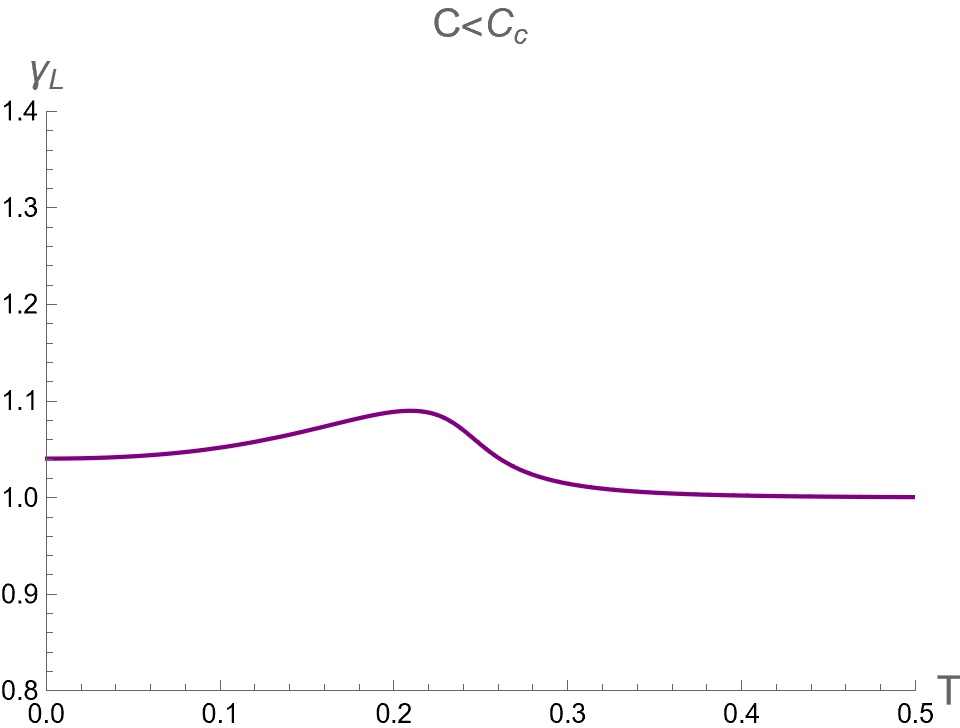}\hspace{1cm}
    \includegraphics[width = 4.cm]{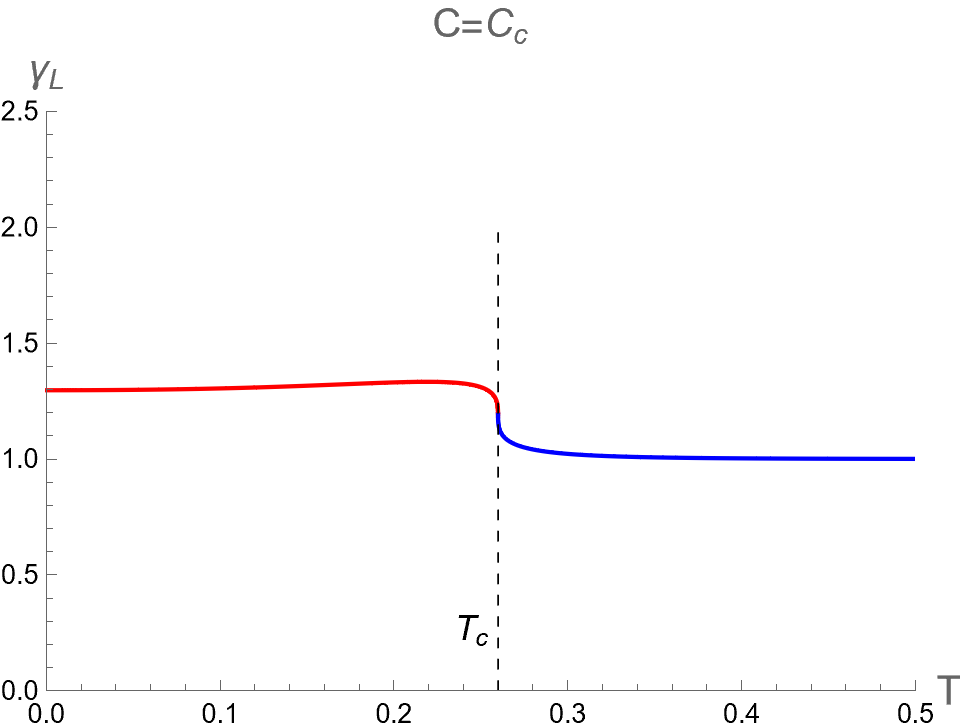}\hspace{1cm}
    \includegraphics[width = 4.cm]{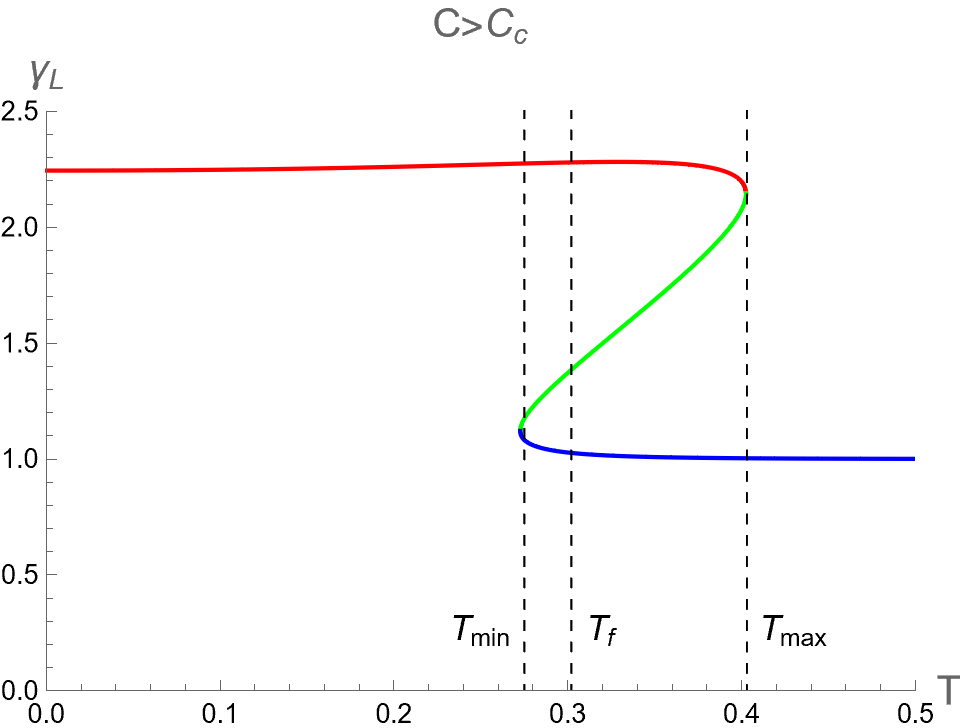}
    \caption{The curves with constant central charge for $\tau$, $\lambda_L$ and $\gamma_L$ as a function of $T$ are shown in the first, second and third rows, respectively. In the left, middle and right columns correspond to the central charge $\mathcal{C}<\mathcal{C}_c$, $\mathcal{C}=\mathcal{C}_c$ and $\mathcal{C}>\mathcal{C}_c$,  respectively.} 
    \label{fig: Photon ring vs T fix C massless}
\end{figure}

By introducing $\mathcal{C}$ as a new thermodynamic variable, we investigate its influence on $\mathcal{O}_i(T)$ in the case II as shown in Fig.~\ref{fig: Photon ring vs T fix C massless}.
We examine the curves of $\mathcal{O}_i$ versus $T$ in $\mathcal{C}<\mathcal{C}_c$ scenario in the left column in Fig.~\ref{fig: Photon ring vs T fix C massless}. 
Since the BH and its dual CFT on the boundary exist in a single phase without any phase transition, the $\mathcal{O}_i(T)$ curves exhibit no multi-value function and discontinuity.  

In the case of $\mathcal{C}=\mathcal{C}_c$, the curves $\mathcal{O}_i(T)$ display a single-value function, whose derivative with respect to $T$ diverges at $T_c$.
The BH configurations below and above the critical temperature $T_c$ correspond to the Small and Large BH phases, respectively.
Note that the values of $\tau$, $\lambda_L$ and $\gamma_L$ at the critical point of case II is the same as those of case I, as shown in Eq.\eqref{parameter at crit}.

For $\mathcal{C}>\mathcal{C}_c$, we find that the parameters $\mathcal{O}_i$ in $T_\text{min}<T<T_\text{max}$ range exhibits a multi-value function.
Namely, three different values of $\mathcal{O}_i$ correspond to three phases of BH, as shown in the right column in Fig.~\ref{fig: Photon ring vs T fix C massless}.
Outside this range of $T$, there exists only one phase, namely the Small-BH phase for $T<T_\text{min}$ and the Large-BH phase for $T>T_\text{max}$.
Moreover, the slope of constant-$\mathcal{C}$ curves in the $\mathcal{O}_i-T$ plane diverges at $T_\text{min}$ and $T_\text{max}$, which correspond to the positions of cusp of $F$ with the divergent value of $C_{\Tilde{Q},\mathcal{V},\mathcal{C}}$.

\section{Scaling Behavior of the optical parameters} \label{sec 5_orderparameter}

In this section, we investigate the scaling behavior of the BH's optical appearance parameters \(O_i\). In phase transition studies, scaling behavior typically emerges near the critical transition point, where different types of matter can exhibit similar scaling laws for thermodynamic quantities.
The scaling behavior with identical critical exponents across different types of matter suggests that these matters belong to the same universality class, independent of their particle composition and interactions. Notably, the critical exponents in the scaling law are crucial for understanding the nature of matter near the critical point of phase transition.

To investigate the system near the critical point, we focus on \textit{order parameter} that vanishes at this point, while maintaining non-zero values prior to reaching it.
In examining the dependence of \(\tau\), \(\lambda_L\), and \(\gamma_L\) on \(P\) and \(T\) within the extended phase space approach and holographic thermodynamics, there is a potential that \(\Delta \mathcal{O}_i = \left|\mathcal{O}_{iL} - \mathcal{O}_{iS}\right|\) can serve as an order parameter, where subscripts \(S\) and \(L\) denote Small and Large BHs, respectively.
At \(P_f\) and \(T_f\), where the Small-Large BH phase transition occurs, \(\mathcal{O}_i\) experiences a discontinuous jump, indicating \(\Delta \mathcal{O}_i \neq 0\). 
As the value of \(P_f\) or \(T_f\) changes along the phase boundary, the values of \(\mathcal{O}_i\) for Small and Large BHs become increasingly similar, causing \(\Delta \mathcal{O}_i\) to decrease, eventually reaching zero at the critical point.

Here, we will categorize the phase transition of the charged AdS-BH in the extended phase space and holographic thermodynamics description via investigating the scaling law of $\Delta \mathcal{O}_i$.
In particular, our analysis in this section will show that the scaling law can be written in the following form
\begin{eqnarray}
    \frac{ \mathcal{O}_i}{\mathcal{O}_{ic}}\sim a_i(1-p^*)^{\alpha_i}
\end{eqnarray}
in the extended phase space approach, whereas in the form 
\begin{eqnarray}
    \frac{\Delta \mathcal{O}_i}{\mathcal{O}_{ic}}\sim b_i(t^*-1)^{\beta_i} 
\end{eqnarray}
in holographic thermodynamics approach. Note that $\mathcal{O}_{ic}$ denotes the value of $\mathcal{O}_i$ at the critical point. Moreover, $a_i$ and $b_i$ are the proportionality constants, $\alpha_i$ and $\beta_i$ are the critical exponents. Recall that $p^*=P_f/P_c$ and $t^*=T_f/T_c$, as previously defined in section~\ref{section 2}.

\subsection{Extended phase space approach}
\begin{figure}[t]
    \centering
    \includegraphics[width=0.3\textwidth]{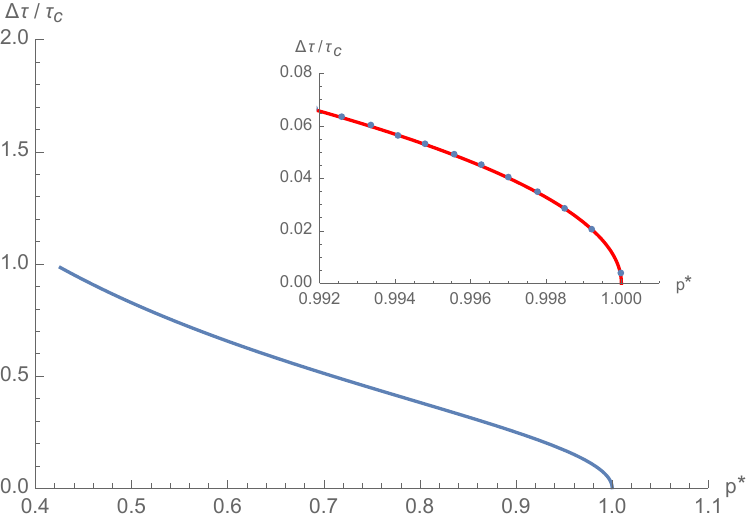} \hfill 
    \includegraphics[width=0.3\textwidth]{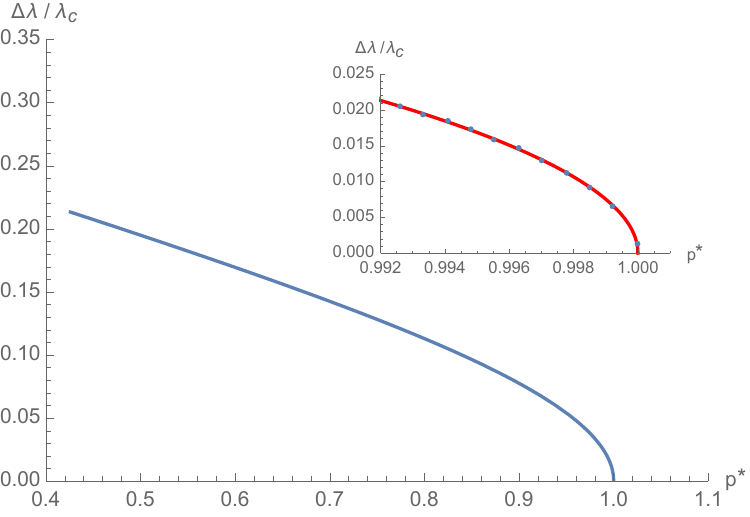}
    \hfill 
    \includegraphics[width=0.3\textwidth]{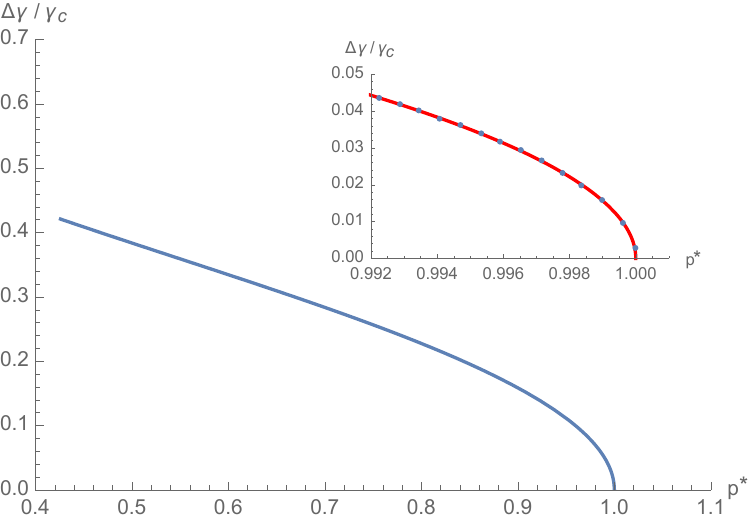} \hfill 
    \\
    \caption{Plots of the reduced order parameters $\Delta \mathcal{O}_i/\mathcal{O}_{ic}$ versus $p^*$ at the Small-Large BH first-order phase transition in the extended phase space approach are shown as the blue curve, while they are shown as blue points in the zoomed panel with $p^*\sim 1$ in order to compare with the scaling laws obtained from our analysis (red).}
    \label{fig: Scaling Extended}
\end{figure}

Let us consider the system near the critical point, one can apply the power series expansion to express $\mathcal{O}_{iS}(r_S)$ and $\mathcal{O}_{iL}(r_L)$ as follows
\begin{eqnarray}
    \mathcal{O}_{iS}(r_S)=\mathcal{O}_{ic}+\left(\frac{\partial \mathcal{O}_i}{\partial r_+}\right)_c(r_S-r_c)+\dots , \\
    \mathcal{O}_{iL}(r_L)=\mathcal{O}_{ic}+\left(\frac{\partial \mathcal{O}_i}{\partial r_+}\right)_c(r_L-r_c)+\dots ,
\end{eqnarray}
where $r_S$ and $r_L$ are the horizon radii of Small and Large BHs at Hawking-Page pressure.
The order parameter is then
\begin{eqnarray}
    \frac{\Delta \mathcal{O}_i}{\mathcal{O}_{ic}}= \frac{\left|\mathcal{O}_{iL} - \mathcal{O}_{iS}\right|}{\mathcal{O}_{ic}}=\frac{1}{\mathcal{O}_{ic}}\left(\frac{\partial \mathcal{O}_i}{\partial r_+}\right)_c\left(r_L-r_S\right). \label{scaling ep}
\end{eqnarray}
From Eq.~\eqref{small r+} and \eqref{large r+}, the leading order of $r_S$ and $r_L$ around the critical point is given by
\begin{eqnarray}
    r_S&=&\sqrt{6}-2\sqrt{3}\sqrt{-\bar{t}}+\cdots ,\nonumber \\
    r_L&=&\sqrt{6}+2\sqrt{3}\sqrt{-\bar{t}}+\cdots ,\nonumber
\end{eqnarray}
where  
\begin{eqnarray}
    \bar{t}=\frac{T_f-T_{c}}{T_{c}}.
\end{eqnarray}
Thus
\begin{eqnarray}
    r_L-r_S&=&4\sqrt{3}\sqrt{-\bar{t}}. \label{scaleExtended}
\end{eqnarray}
Expanding the reduced Hawking-Page pressure $p^*$ as expressed in Eq.~\eqref{Maxwell 1} near the critical temperature $t=1$, we obtain
\begin{eqnarray}
    p^*&=&1+\frac{8}{3}\bar{t}+\cdots .
\end{eqnarray}
It is important to emphasize that we are considering the system near the critical point, allowing us to express the above equation up to the first order of \(\bar{t}\), thereby neglecting higher-order terms. Consequently, we can use this first-order approximation to eliminate \(\bar{t}\) in Eq.~\eqref{scaleExtended}. By substituting the result into Eq.~\eqref{scaling ep}, we obtain
\begin{eqnarray}
    \frac{\Delta\mathcal{O}_{i}}{\mathcal{O}_{ic}}
    &=&a_{i}\left(1-p^*\right)^{1/2}, \label{cri_1sub2}
\end{eqnarray}
where $a_{i}$ could be expressed as
\begin{equation}
    a_{i}=\frac{3\sqrt{2}}{\mathcal{O}_{ic}}\left(\frac{\partial \mathcal{O}_i}{\partial r_+}\right)_c .
\end{equation}
From Eq.~\eqref{cri_1sub2}, we observe that the critical exponent $\alpha_i=1/2$, which suggests that its behaviors are in the same way as the vdW fluid type.
The scaling behavior for three optical order parameters are
\begin{eqnarray}
    \frac{\Delta \tau}{\tau_c} &\sim & 0.7337 \left(1-p^*\right)^{1/2} , \label{scaling ep tau} \\ 
    \frac{\Delta \lambda_L}{\lambda_c} &\sim & 0.2384\left(1-p^*\right)^{1/2} , \label{scaling ep lamb}\\ 
    \frac{\Delta \gamma_L}{\gamma_c} &\sim & 0.4953 \left(1-p^*\right)^{1/2} . \label{scaling ep gam}
\end{eqnarray}
Using Eqs.~\eqref{Maxwell 1}, \eqref{small r+}, \eqref{large r+} and \eqref{r_u_T}, we can determine each pair of the reduced order parameters $\displaystyle \frac{\Delta \mathcal{O}_i}{\mathcal{O}_{ic}}$ and  the reduced Hawking-Page pressure $p^*$ by running the temperature increasing from the minimum value $T_0$ to $T_c$. Recall that we have shown the formula of $T_0$ in Eq.~\eqref{T_0}. In other words, by writing in the form of reduced temperature $t=T/T_c$, we run $t$ from $t=t_0=T_0/T_c=1/\sqrt{2}$ to the critical value $t=1$, and then we obtain the plot of the reduced order parameters versus $p^*$, as shown in Fig~\ref{fig: Scaling Extended}.
Notably, we find that three reduced order parameters decrease as $p^*$ increase and become vanish at $p^*=1$. 
In the zoomed panels, the blue points represent the numerical values of the exact results of the reduced order parameter near the critical point.

Interestingly, the expressions of the scaling laws, as shown in Eqs.~\eqref{scaling ep tau}, \eqref{scaling ep lamb} and \eqref{scaling ep gam},  provide the results that fit the blue points remarkably well, as illustrated as the red curves.
These indicate that the three order parameters exhibit a discontinuous jump, resulting in non-vanishing \(\Delta \mathcal{O}_i\) within the range \(p^* < 1\). This behavior signifies the occurrence of a second-order phase transition at \(p^* = 1\). 
At this critical point, all order parameters \(\Delta \mathcal{O}_i\) vanish as \(\mathcal{O}_{iS}\) and \(\mathcal{O}_{iL}\) converge to the same value. Beyond the critical point, the order parameters no longer exhibit multiple values, indicating the presence of a single BH phase.

\subsection{Holographic thermodynamics approach}
In the case of the holographic thermodynamics approach, we consider the power series expansion of $\mathcal{O}_i(x)$ around the critical point up to the first order as
\begin{eqnarray}
    \mathcal{O}_{iS}(x)=\mathcal{O}_{ic}+\left(\frac{\partial \mathcal{O}_i}{\partial x}\right)_c(x_S-x_c)+\dots , \\
    \mathcal{O}_{iL}(x)=\mathcal{O}_{ic}+\left(\frac{\partial \mathcal{O}_i}{\partial x}\right)_c(x_L-x_c)+\dots ,
\end{eqnarray}
where $x_S$ and $x_L$ refer to the horizon radii for Small and Large BHs at the first-order phase transition, respectively.
The reduced order parameters near the critical point are then
\begin{eqnarray}
    \frac{\Delta \mathcal{O}_i}{\mathcal{O}_{ic}}= \frac{\left|\mathcal{O}_{iL} - \mathcal{O}_{iS}\right|}{\mathcal{O}_{ic}}=\frac{1}{\mathcal{O}_{ic}}\left(\frac{\partial \mathcal{O}_i}{\partial x}\right)_c\left(x_L-x_S\right). \label{delta O}
\end{eqnarray}
Let us consider the phase transition in the case I first.  As $\mathcal{C}$ is fixed while $\Tilde{Q}$ can be varied, we define the parameter $\Bar{q}$ as
\begin{eqnarray}
    \Bar{q}=\frac{\Tilde{Q}-\Tilde{Q}_c}{\Tilde{Q}_c}.
\end{eqnarray}
The expression of $x_S$ and $x_L$ in Eqs.~\eqref{Small r holo} and \eqref{Large r holo} can be expanded around  $x_c$ as follows
\begin{eqnarray}
    x_S&=&x_c-\frac{1}{2}\sqrt{-\Bar{q}}-\dots , \nonumber \\
    x_L&=&x_c+\frac{1}{2}\sqrt{-\Bar{q}}-\dots . \nonumber
\end{eqnarray}
By keeping up to only the leading order, the difference between these two horizon radii is given by
\begin{eqnarray}
    x_L-x_S=\sqrt{-\Bar{q}}. \label{diff x 1}
\end{eqnarray}
Since we consider the behavior of reduced order parameters whose values are calculated at the Hawking-Page temperature $T_f$ near the critical temperature $T_c$, let us write $t^*$, Eq.~\eqref{Tf of fix C}, in the form of the series expansion as
\begin{eqnarray}
    t^*&=&1-\frac{1}{4}\Bar{q}+\dots. \nonumber 
\end{eqnarray}
Keeping up to only the leading order, we can have
\begin{eqnarray}
    -\Bar{q}&=&4(t^*-1). \label{delta t in q bar}
\end{eqnarray}
Consequently, using Eqs.~\eqref{diff x 1} and \eqref{delta t in q bar},   Eqs.~\eqref{delta O} can be rewritten in the form of the scaling law of reduced order parameters as
\begin{eqnarray}
    \frac{\Delta \mathcal{O}_i}{\mathcal{O}_{ic}}=b_i(t^*-1)^{1/2}, \label{critical_case I}
\end{eqnarray} 
where the proportionality coefficient is 
\begin{eqnarray}
    b_i=\frac{2}{\mathcal{O}_{ic}}\left(\frac{\partial \mathcal{O}_i}{\partial x}\right)_c.
\end{eqnarray}
As obviously shown in Eq.~\eqref{critical_case I}, the critical exponents of three order parameters $\beta_i=1/2$ for $i=1,2$ and  $3$.

\begin{figure}[t]
    \centering
    \includegraphics[width=0.3\textwidth]{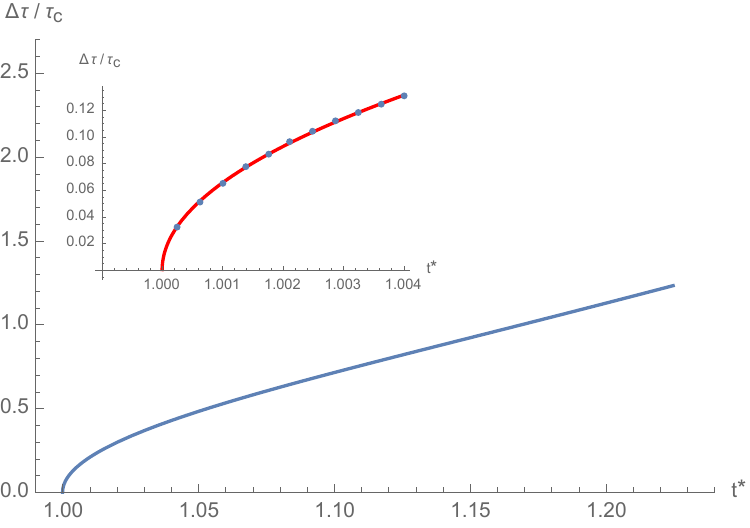} \hfill 
    \includegraphics[width=0.3\textwidth]{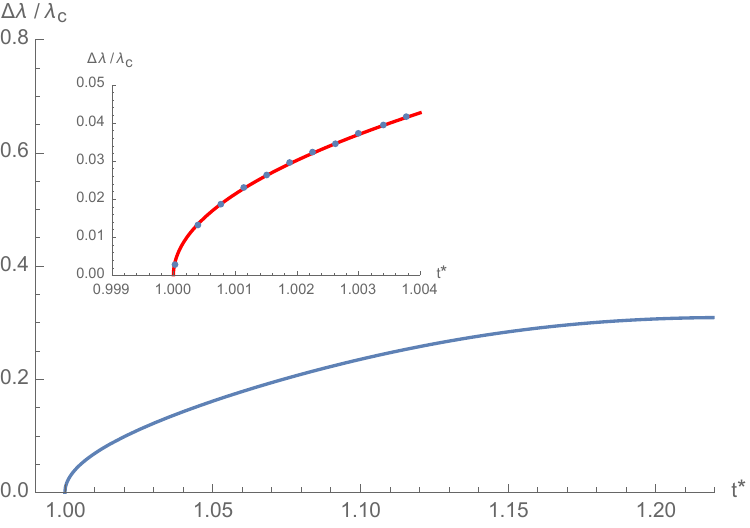}
    \hfill 
    \includegraphics[width=0.3\textwidth]{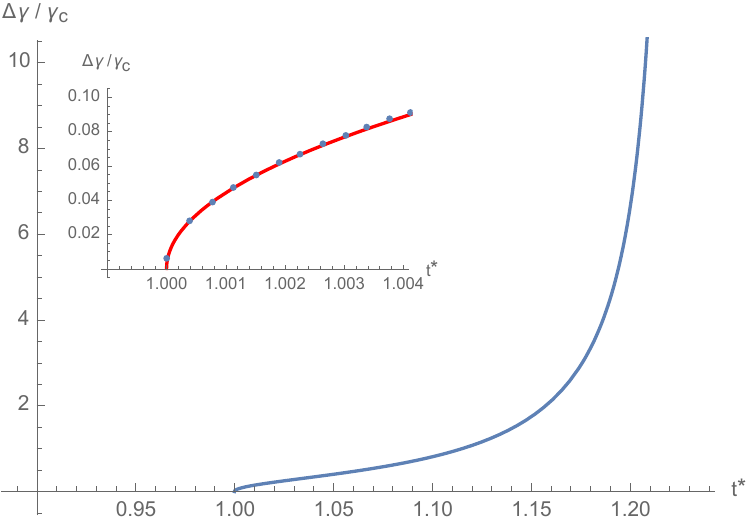} \hfill 
    \\
    \caption{Plots of the reduced order parameters $\Delta \mathcal{O}_i/\mathcal{O}_{ic}$ versus $t^*$ at the Small-Large BH first-order phase transition in the holographic thermodynamics approach are shown in blue curve, while they are shown as blue points in the zoomed panel with $t^*\sim 1$. Note that the red curve follows the scaling laws as obtained from our analysis.}
    \label{fig: Scaling holographic}
\end{figure}

For the case II of phase transition in the holographic thermodynamics approach, we introduce
\begin{eqnarray}
    \bar{c}=\frac{\mathcal{C}-\mathcal{C}_c}{\mathcal{C}_c}.
\end{eqnarray}
According to the previous approximation, we find that the scaling law for three reduced order parameters near the critical point of case II is the same as case I as follows
\begin{eqnarray}
    \frac{\Delta \tau}{\tau_c} &\sim &2.0751 \left( t^*-1\right)^{1/2} , \label{scaling tau fixed C} \\
    \frac{\Delta \lambda_L}{\lambda_c} &\sim &0.6742\left( t^*-1\right)^{1/2} , \label{scaling lambda fixed C} \\
    \frac{\Delta \gamma_L}{\gamma_c} &\sim &1.4008 \left( t^*-1\right)^{1/2} . \label{scaling gamma fixed C}
\end{eqnarray}
Since the scaling behavior near the critical point of reduced order parameters in case II is similar to case I, we will display the plot of three reduced order parameters $\displaystyle \frac{\Delta \mathcal{O}_i}{\mathcal{O}_{ic}}$ versus the reduced Hawking-Page temperature $t^*$ only in case I, as illustrated in Fig~\ref{fig: Scaling holographic}.
Using Eqs.~\eqref{Tf of fix C},  \eqref{Small r holo}, \eqref{Large r holo} and \eqref{photon sphere holo}, $\displaystyle \frac{\Delta \mathcal{O}_i}{\mathcal{O}_{ic}}$ can be determined at fixed $c$ versus $t^*$,  with $q$ running from 1 to 0 in case I, whereas it is determined at fixed $q$ versus $t^*$ with $c^{-1}$ running from 1 to $0$ in case II.
On the other hand, the figure indicates that three reduced order parameters are more than zero starting at $t^*=1$ and terminating at $t^*\approx 1.22$, beyond which a Hawking-Page phase transition does not exist.  This can also be seen from Fig.~\ref{fig: Maxwell cano fix C} (b) in Appendix~\ref{App B}.    
In the zoomed panels, the blue points represent the numerical values of the exact results of the reduced order parameter near the critical point. 
Interestingly, the expressions of the scaling laws, as shown in Eqs.~\eqref{scaling tau fixed C}, \eqref{scaling lambda fixed C} and \eqref{scaling gamma fixed C},  provide the results that fit the blue points very well, as illustrated as the red curves.
These indicate that the three order parameters exhibit a discontinuous jump, resulting in non-vanishing \(\Delta \mathcal{O}_i\) within the range $1<t^*<1.22$. 
This behavior signifies the occurrence of a second-order phase transition at \(t^* = 1\). 
At this critical point, all order parameters \(\Delta \mathcal{O}_i\) vanish as \(\mathcal{O}_{iS}\) and \(\mathcal{O}_{iL}\) converge to the same value. 
For $T<T_c$, the order parameters no longer exhibit multiple values, indicating the presence of a single BH phase.

It is worth emphasizing that in the extended phase space approach, the scaling law can be found near the critical point for these three optical order parameters as a function of 
$p^*$, where all exhibit the critical exponent with the value of $1/2$. On the other hand, in holographic thermodynamics where the concepts of bulk pressure and volume are absent, the scaling law behavior can also be found near the critical point for these parameters as a function of $t^*$ with the critical exponent equal to $1/2$.
The results in the holographic thermodynamics context are particularly noteworthy because they do not rely on bulk pressure and volume. 
This suggests that the critical behavior of BHs can be understood only in terms of boundary field theory quantities.
\section{Conclusions} \label{sec 6_conclusion}

In this study, we have thoroughly examined the intricate phase structures of charged AdS-BHs within the extended phase space and holographic thermodynamics approaches by analyzing null geodesics near their critical curves. 
By utilizing horizon-scale observation of BHs, namely the orbital half-period $\tau$, angular Lyapunov exponent $\lambda_L$, and temporal Lyapunov exponent $\gamma_L$, we can characterize BH phase transitions in these two frameworks.

In particular, for the extended phase space approach, when $T<T_c$, these three critical parameters are multi-valued functions of bulk pressure $P$, indicating the presence of three coexisting phases of BH.
When $T>T_c$, these parameters display the singled-valued function versus $P$ corresponding with a single BH phase.

In the holographic thermodynamics approach, the phase transition of charged AdS-BH in the bulk is holographically dual to the CFT phase transition in the boundary with including the central charge $\mathcal{C}$ and its conjugate as a new pair of thermodynamic variable. 
We investigate the phase structures of BH dual to the fixed $(\Tilde{Q}, \mathcal{V}, \mathcal{C})$ ensemble of CFT. 
For $\Tilde{Q} < \Tilde{Q}_c$ or $\mathcal{C} > \mathcal{C}_c$, the three critical parameters exhibit a multi-valued function of temperature $T$, indicating the presence of three coexisting BH phases. 
Conversely, for $\Tilde{Q} > \Tilde{Q}_c$ or $\mathcal{C} < \mathcal{C}_c$, these parameters display a single-valued function with respect to $T$, corresponding to a single BH phase.

In many fields, including condensed matter physics, there are numerous examples of phase transitions that often lack a clearly defined order parameter or the concept may not be entirely appropriate. In this study, we propose the possibility of an order parameter based on the difference between each optical parameter of the Small and that of Large BH phases at the Hawking-Page phase transition. 
This approach is advantageous because upcoming space-based VLBI missions aim at studying the photon subring structure of BH, thereby providing a practical and observable method to characterize BH phase transitions.
Remarkably, the critical exponents for these parameters near the critical point, from both extended phase space and holographic thermodynamics approaches, are found to be 1/2, indicating a significant thermodynamic similarity to van der Waals fluids.
It is important to emphasize that, in our work, the scaling law near the critical point can be derived through precise theoretical calculations, avoiding any ambiguous methods.

Evidently, the BH optical appearance parameters from both approached are effective in characterizing BH phase transitions in AdS space. 
It is important to highlight that these critical parameters may be linked to some phenomena in strongly coupled systems through holographic duality. 
In the context of the AdS/CFT correspondence, a thermal state in the CFT is dual to a BH in the bulk AdS space. 
A perturbation in the dual field theory, induced by a local operator, corresponds to a perturbation of the BH by turning on the scalar fields propagating in the bulk. 
The process of returning to thermal equilibrium in the field theory side can be dual to the black hole returning to equilibrium after a perturbation, which can be observed through imaginary part of quasinormal frequencies describing the rate at which a perturbed BH returns to equilibrium. 
Consequently, the QNM spectrum of a BH can be interpreted holographically as the quantum Ruelle resonances in boundary theory, which are complex frequency modes that describe the late-time decay of perturbations in chaotic systems~\cite{Jahnke:2018off}.
Moreover, in the eikonal limit, the decay of QNMs can be governed by the temporal Lyapunov exponent $\gamma_L$ of unstable null geodesics \cite{cardoso2009geodesic,hadar2022,Hashimoto:2023buz,riojas2023photon}.
In this eikonal regime in the AdS-BHs background, there exist the long-lived modes, which dominate the late-time behavior of BH while returning to equilibrium~\cite{Festuccia:2008zx,Morgan:2009vg}. 
The existence of long-lived QNMs in AdS suggests that certain perturbations in the dual CFT decay very slowly, implying that the system takes a long time to return to thermal equilibrium. 
In our results, we find that at the Small-Large BHs phase transitions, the behavior of $\gamma_L$ and hence the late-time behavior of long-lived modes change discontinuously due to the Maxwell equal area law.  
Holographically, this behavior presumably reflects a sudden change in the spectrum of Ruelle resonances across the first-order phase transition on the field theory side. The order parameter we propose could be valuable for extending studies on holographic thermalization in a field theory, particularly concerning phase transitions, in future research.


In particular, we focus on optically probing the phase structures of charged BH in AdS space, as they exhibit rich phase structures and undergo intriguing phase transitions, such as the Hawking-Page phase transition and critical behavior. 
However, it is also worthwhile to consider applying these methods to study BH phase transitions in more realistic scenarios, such as BHs in asymptotically flat or de Sitter (dS) spacetimes. 
Additionally, this approach could be extended to investigate BH phase transitions in the context of modified entropies \cite{Czinner:2017tjq,*Promsiri:2020jga,*ElMoumni:2022chi,*Tannukij:2020njz,*Nakarachinda:2021jxd,*Barzi:2024bbj,*MEJRHIT201945,*barrow2020area,*Jawad:2022bmi}, providing a broader framework for understanding BH thermodynamics across different theoretical models.

\section*{Acknowledgement}

EH  is supported by Thailand Science Research and Innovation (TSRI) Basic Research Fund: Fiscal year 2023 under project number FRB660073/0164. 
WH is supported by the National Science and Technology Development Agency under the Junior Science Talent Project (JSTP) scholarship, grant number SCA-CO-2565-16992-TH.  This research has received funding support from the NSRF via the Program Management Unit for Human Resources \& Institutional Development, Research and Innovation [grant number B13F660066].

\appendix
\section{Eulerian scaling argument} \label{App A}
In the following discussions, we review how the first law of thermodynamics is obtained from the Smarr formula by using Euler's theorem for a homogeneous function as detailed in
\cite{kubizvnak2017black,2014Galax...2...89A}.

A function $f(x_1, \dots , x_n)$ is a homogeneous function of degree $q$ if      
\begin{eqnarray}
f(\alpha^{p_1} x_1, \alpha^{p_2} x_2, \dots , \alpha^{p_n} x_n) = \alpha^q f(x_1, x_2, \dots , x_n).
\end{eqnarray}
The Euler's theorem states that if $f(x_1, \dots , x_n)$ is a homogeneous function of degree $q$, then it satisfy the partial differential equation:
\begin{eqnarray}
qf(x_1, \dots , x_n) = p_1\left(\frac{\partial f}{\partial x_1}\right)x_1 + p_2\left(\frac{\partial f}{\partial x_2}\right)x_2 + \cdots + p_n\left(\frac{\partial f}{\partial x_n}\right)x_n, \label{Euler}
\end{eqnarray}
then an infinitesimal change of $f$ is given by
\begin{eqnarray}
df = \left(\frac{\partial f}{\partial x_1}\right)dx_1 + \left(\frac{\partial f}{\partial x_2}\right)dx_2 + \dots + \left(\frac{\partial f}{\partial x_n}\right)dx_n. \label{df}
\end{eqnarray}
It is noticed that the mass of non-rotating charged AdS-BH can be expressed in terms of the area $A$, the electric charge $Q$ and the cosmological constant $\Lambda$ as follows
\begin{eqnarray}
    M(A,Q,\Lambda)=\frac{(n-1)\omega_{n-1}^{\frac{1}{n-1}}}{16\pi G_N}A^{\frac{n-2}{n-1}}-\frac{1}{8\pi G_N\omega_{n-1}^{\frac{1}{n-1}}n}\Lambda A^{\frac{n}{n-1}}+\frac{4\pi G_N}{(n-1)\omega_{n-1}^{\frac{1}{n-1}}\eta^2}\frac{Q^2}{A^{\frac{n-2}{n-1}}}, \label{M(A,Q,Lambda)}
\end{eqnarray}
which is indeed the homogeneous function degree $n-2$ because $M(\alpha^{n-1} A,\alpha^{n-2} Q,\alpha^{-2}\Lambda)=\alpha^{n-2} M(A,Q,\Lambda)$.
From Eq.~\eqref{Euler}, we have
\begin{eqnarray}
    (n-2)M=(n-1)\frac{\partial M}{\partial A}A+(n-2)\frac{\partial M}{\partial Q}Q+(-2)\frac{\partial M}{\partial \Lambda}\Lambda .
\end{eqnarray}
Using Eq.~\eqref{M(A,Q,Lambda)}, the partial derivative terms in the above equation can be derived as
\begin{eqnarray}
    \frac{\partial M}{\partial A}A=\frac{\partial M}{\partial S}S=TS, \ \ \
    \frac{\partial M}{\partial Q}Q=\Phi Q \ \ \ \text{and} \ \  \
    \frac{\partial M}{\partial \Lambda}\Lambda =\frac{\Theta \Lambda}{{8\pi G_N}},
\end{eqnarray}
and the Smarr formula is then
\begin{eqnarray}
    M=\frac{n-1}{n-2}\frac{\kappa A}{8\pi G_N}+\Phi Q-\frac{1}{n-2}\frac{\Theta \Lambda}{4\pi G_N}. \label{Smarr_ex}
\end{eqnarray}
The formula in Eq.~\eqref{df} give the variation of $M$ as follow
\begin{eqnarray}
    dM=\frac{\kappa}{8\pi G_N}dA +\Phi dQ+\frac{\Theta}{8\pi G_N}d\Lambda . \label{dm_ex}
\end{eqnarray}
Identifying respectively, $\Lambda$ and its conjugate variable $\Theta$ as the bulk pressure $P$ and volume $V$ via the relation in Eq.~\eqref{extended phase} in Eqs.~\eqref{Smarr_ex} and \eqref{dm_ex}, we will obtain Eqs.~\eqref{MMM} and \eqref{dM extended}.

\section{Holographic Maxwell's equal area law in $T-S$ plane} \label{App B}
\begin{figure}[t]
    \centering
    \subfigure[]{\includegraphics[width = 5 cm]{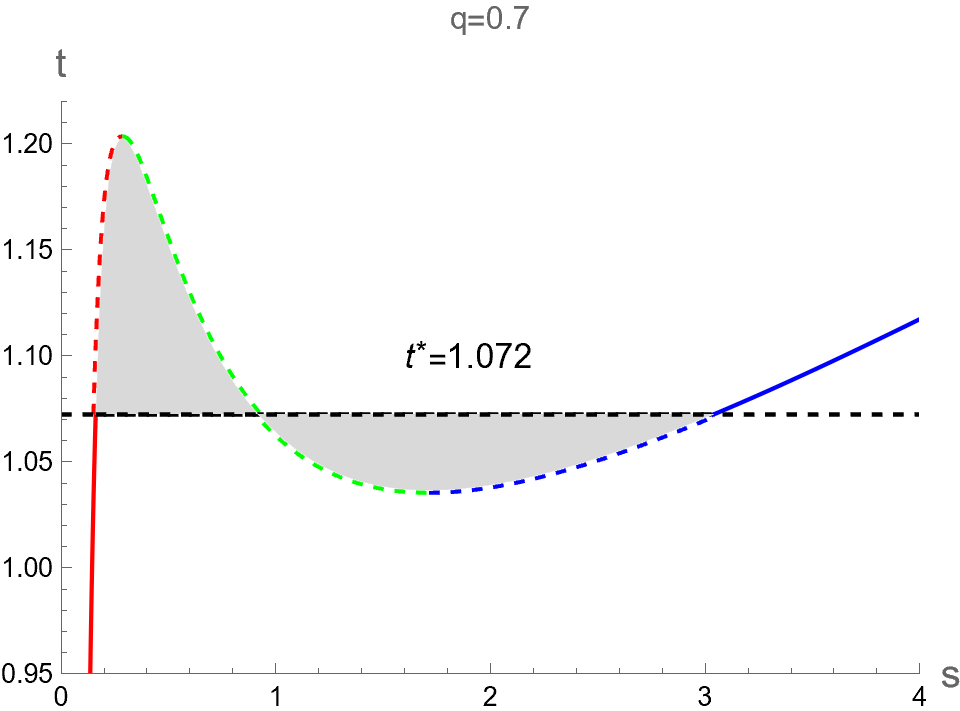}}\hspace{2cm}
    \subfigure[]{\includegraphics[width = 5 cm]{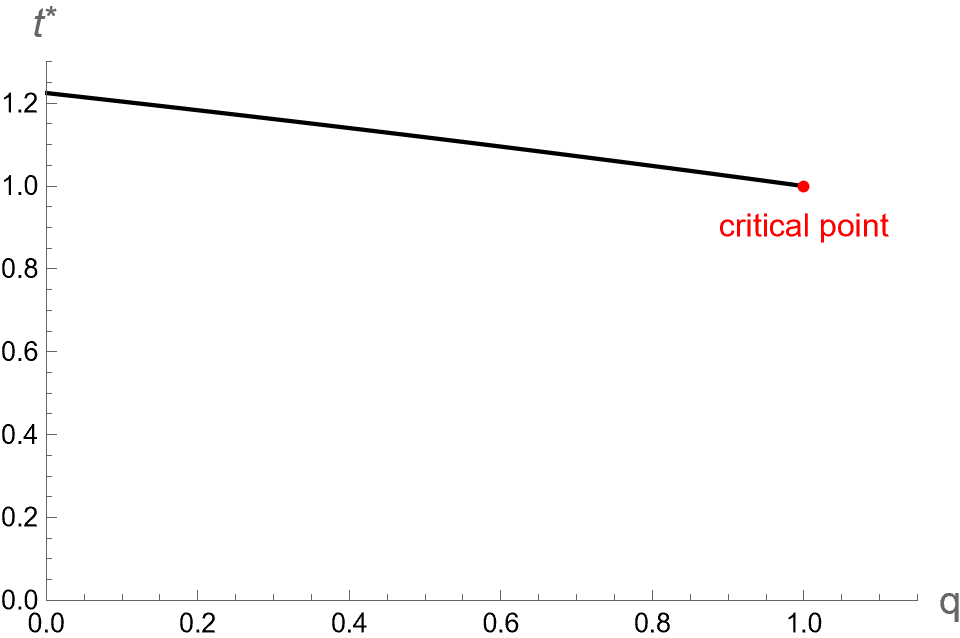}}
    \subfigure[]{\includegraphics[width = 5 cm]{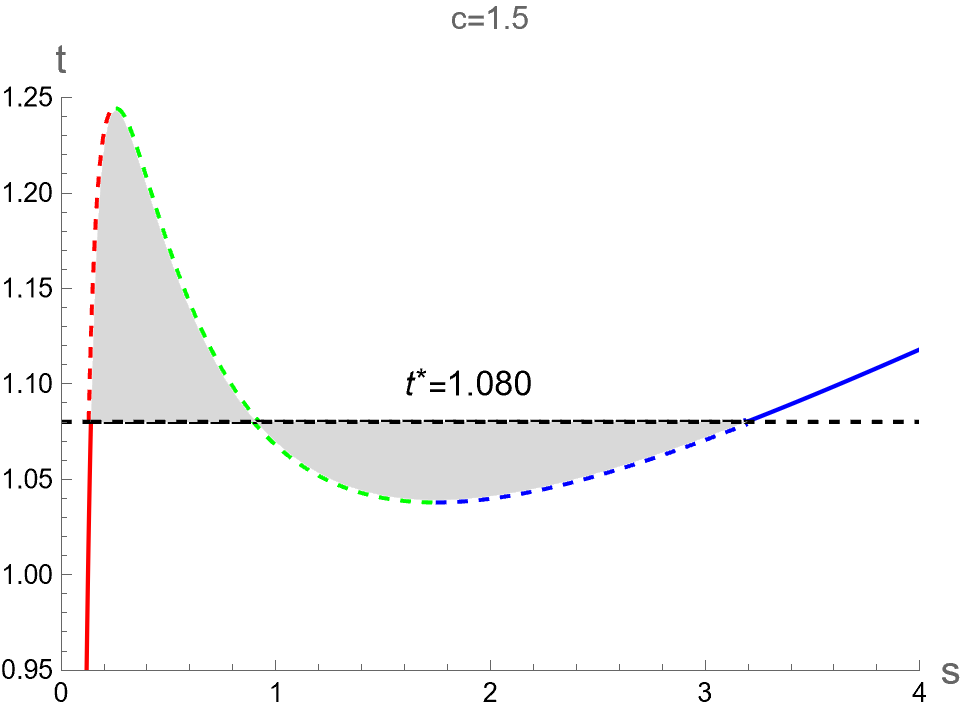}}\hspace{2cm}
    \subfigure[]{\includegraphics[width = 5 cm]{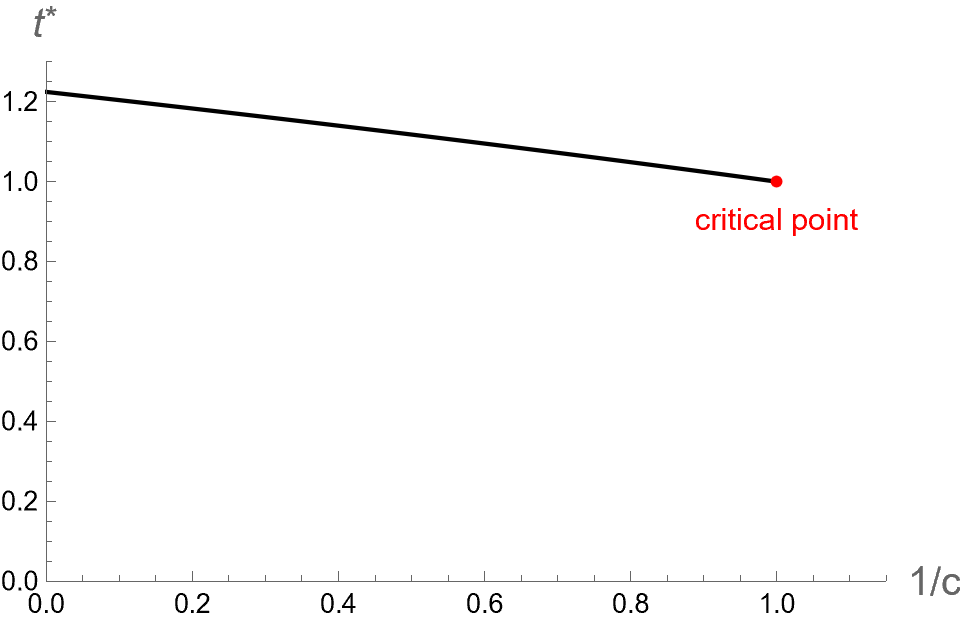}}
    \caption{The curves of the reduced BH temperature $t$ versus the reduced entropy $s$ with the Maxwell construction for finding the first-order phase transition $t^*$ are plotted for case I with $q=0.7$ (a) and case II with $c=1.5$ (c). Resulting from the Maxwell construction, we obtain the values of $t^*$ at different values of $q$ presented as the line terminated at the critical point $q=1$ (b), whereas $t^*$ versus $1/c$ presented as the line terminated at the critical point $1/c =1$ (d).}      
    \label{fig: Maxwell cano fix C}
\end{figure}
Here, we consider the calculations of pCFT1-pCFT2 phase transition temperature $T_f$ for CFT, which dual to the Small-Large BHs phase transition of charged AdS-BH in the holographic description.
Substituting the horizon radius $x$ into Eq.~\eqref{temp 4D} and then writing $x$ in the form of the entropy $S$ using the area law, we obtain $T=T(S, \Tilde{Q}, \mathcal{C})$ as follows
\begin{eqnarray}
    T=\frac{1}{2}\sqrt{\frac{\mathcal{C}}{\pi S}}\left(1+\frac{3S}{4\pi \mathcal{C}}-\frac{\pi \Tilde{Q}^2}{4\mathcal{C}S}\right). \label{T in S}
\end{eqnarray}
We can define the reduced parameters as follows:
\begin{eqnarray}
    t=\frac{T}{Tc}, \ \ \ s=\frac{S}{S_c}, \ \ \  q=\frac{\Tilde{Q}}{\Tilde{Q}_c} \ \ \ \text{and} \ \ \ c=\frac{\mathcal{C}}{\mathcal{C}_c}.
\end{eqnarray}
Consider case I as introduced in section~\ref{section 2}, we can express the reduced BH temperature in Eq.~\eqref{T in S} in terms of $s$ and $q$ as 
\begin{eqnarray}
    t=\frac{3}{4}\left(\frac{1}{\sqrt{s}}+\frac{\sqrt{s}}{2}-\frac{q^2}{6s^{\frac{3}{2}}}\right). \label{t fix q}
\end{eqnarray}
Interestingly, it does not explicitly depend on central charge $\mathcal{C}$.
As shown in Fig.~\ref{fig: Maxwell cano fix C} (a), the reduced Hawking-Page phase transition temperature $t^*=T_f/T_c$ is represented in dashed-black horizontal line.
This line is positioned such that the areas above and below the isocharge curve are equal, in accordance with Maxwell's equal area law.
With $s_1$ and $s_3$ representing the reduced entropy associated with pCFT1 and pCFT2 at the $t^*$, respectively, the equal area law is expressed as
\begin{eqnarray}
    t^*(s_3-s_1)=\int_{s_1}^{s_3}t(s)ds. \label{Maxwell area}
\end{eqnarray}
Then, we obtain
\begin{eqnarray}
    t^*=\frac{1}{\sqrt{s_1}+\sqrt{s_3}}\left( \frac{3}{2}-\frac{q^{2}}{4\sqrt{s_1s_3}}+\frac{1}{4}(s_1+s_3+\sqrt{s_1s_3})\right). \label{Maxwell}
\end{eqnarray}
Note that the first-order phase transition appear when $q<1$ ($\Tilde{Q}<\Tilde{Q}_c$).
Substituting $x=\sqrt{s_1}$ and $y=\sqrt{s_3}$ into Eq.~\eqref{t fix q}  together with Eq.~\eqref{Maxwell}, we obtain the system of equations as follows
\begin{eqnarray}
    t^*&=&\frac{3}{4}\left( \frac{1}{x}+\frac{x}{2}-\frac{q^{2}}{6x^3}\right), \label{t1} \\ 
    t^*&=&\frac{3}{4}\left( \frac{1}{y}+\frac{y}{2}-\frac{q^{2}}{6y^3}\right), \label{t2} \\
    t^*&=&\frac{1}{x+y}\left( \frac{3}{2}-\frac{q^{2}}{4xy}+\frac{1}{4}(x^2+y^2+xy)\right). \label{t3}
\end{eqnarray}
These equations can be solved analytically by following the procedure as suggested in \cite{Lan:2015bia}. 
First, using the fact the RHS of Eq.~\eqref{t1} equal to Eq.~\eqref{t2}, we obtain
\begin{eqnarray}
    x^2+y^2=\frac{6}{q^{2}}z^2\left(1-\frac{z}{2}\right)-z, \label{a1}
\end{eqnarray}
where $z=xy$.
Then, using $2\eqref{t3}=\eqref{t1}+\eqref{t2}$, we have
\begin{eqnarray}
    3z-\frac{q^{2}}{2}+\frac{z}{2}(x^2+y^2+z)=\frac{3}{4}(x^2+y^2+2z)\left( 1+\frac{z}{2}-\frac{q^{2}}{6z^2}(x^2+y^2-z) \right).
\end{eqnarray}
By eliminating term $x^2+y^2$ via the relation in Eq.~\eqref{a1}, we have the quartic equation in variable $z$ as follows
\begin{eqnarray}
    z^4-2z^3+2q^{2}z-q^{4}=0.
\end{eqnarray}
The nontrivial solution is $z=xy=q$.
To obtain $x$ and $y$, we substitute $z=q$ into Eq.~\eqref{a1} and solve following equations
\begin{eqnarray}
    x^2+y^2&=&6-4q, \\
    xy&=&q.
\end{eqnarray}
simultaneously. Since $x<y$, therefore the solutions for $x$ and $y$ are
\begin{eqnarray}
    x&=&\frac{q}{\sqrt{3+\sqrt{3(q-3)(q-1)}-2q}}, \label{x for small BH} \\
    y&=&\sqrt{3+\sqrt{3(q-3)(q-1)}-2q}. \label{x for large BH}
\end{eqnarray}
Thus, $t^*$ becomes
\begin{eqnarray}
    t^*=\frac{3+\sqrt{3(q-3)(q-1)}-q}{2\sqrt{3+\sqrt{3(q-3)(q-1)}-2q}}. \label{HP temp}
\end{eqnarray}

For case II,  we express the BH temperature in Eq.~\eqref{T in S} by substituting $S=sS_c$ and $\mathcal{C}=c\mathcal{C}_c$.
The resulting $t$ does not depend on electric charge $\Tilde{Q}$ as following
\begin{eqnarray}
    t=\frac{3}{4}\left( \frac{1}{\sqrt{s}}+\frac{\sqrt{s}}{2}-\frac{1}{6c^2s^{\frac{3}{2}}}\right). \label{t fix c}
\end{eqnarray}
Note that the first-order phase transition appears when $c>1$ ($\mathcal{C}>\mathcal{C}_c$).
Since Eqs.~\eqref{t fix q} and \eqref{t fix c} is symmetric under $q^2\rightarrow 1/c^2$, the solutions of variables $x, y$ and $t^*$ as a function of $c^*$ in this case can be written as follows
\begin{eqnarray}
    x&=&\frac{1}{\sqrt{3c^{2}+c\sqrt{3(1-3c)(1-c)}-2c}}, \\
    y&=&\frac{1}{c}\sqrt{3c^{2}+c\sqrt{3(1-3c)(1-c)}-2c}, \\
    t^*&=&\frac{3c+\sqrt{3(1-3c)(1-c)}-1}{2\sqrt{3c^{2}+c\sqrt{3(1-3c)(1-c)}-2c}}.
\end{eqnarray}

\bibliography{ref}

\end{document}